\documentclass{cup-elements}

\usepackage{blindtext}
\usepackage{graphicx}

\usepackage[natbibapa]{apacite}
\usepackage{amssymb,amsmath}
\usepackage{bm}
\usepackage{xcolor}
\usepackage{enumitem} 

\usepackage{listings}
\definecolor{codegreen}{rgb}{0,0.6,0}
\definecolor{codegray}{rgb}{0.5,0.5,0.5}
\definecolor{codepurple}{rgb}{0.58,0,0.82}
\definecolor{backcolour}{rgb}{0.95,0.95,0.92}

\lstdefinestyle{mystyle}{
    backgroundcolor=\color{backcolour},   
    commentstyle=\color{codegreen},
    keywordstyle=\color{magenta},
    numberstyle=\tiny\color{codegray},
    stringstyle=\color{codepurple},
    basicstyle=\ttfamily\scriptsize,
    breakatwhitespace=false,         
    breaklines=true,                 
    captionpos=b,                    
    keepspaces=true,                 
    numbers=left,                    
    numbersep=5pt,                  
    showspaces=false,                
    showstringspaces=false,
    showtabs=false,                  
    tabsize=2
}

\DeclareCaptionType{floatbox}[Box]
\newcommand{\numpy}{NumPy}
\newcommand{\networkx}{NetworkX}

\renewcommand{\O}{O}

\CUPseries{The Structure and Dynamics of Complex Networks}
\CUPelements{Gillespie Algorithms: Tutorial}

\title{Gillespie algorithms for stochastic multiagent dynamics in populations and networks}

\author{Naoki Masuda}
\affil{Department of Mathematics, State University of New York at Buffalo, NY, USA}
\affil{Computational and Data-Enabled Science and Engineering Program, State University of New York at Buffalo, NY, USA\\
naokimas@buffalo.edu}

\author{Christian L. Vestergaard}
\affil{Decision and Bayesian Computation, Department of Neuroscience, CNRS UMR 3571, Department of Computational Biology, CNRS USR 3756, Institut Pasteur, Paris, France\\
cvestergaard@gmail.com} 

\begin{document}

\frontmatter  
\maketitle

\begin{abstract}
Many multiagent dynamics, including various collective dynamics occurring on networks, can be modeled as a stochastic process in which the agents in the system change their state over time in interaction with each other. 
The Gillespie algorithms are popular algorithms that exactly simulate such stochastic multiagent dynamics when each state change is driven by a discrete event, the dynamics is defined in
continuous time, and the stochastic law of event 
occurrence is governed by independent Poisson processes.
In the first main part of this volume, 
we provide a tutorial on the Gillespie algorithms focusing on simulation of social multiagent dynamics occurring in populations and networks.  
We do not assume advanced knowledge of mathematics (or computer science or physics). We clarify why one should use the continuous-time models and the Gillespie algorithms in many cases, instead of easier-to-understand discrete-time models. In the remainder of this volume, we review recent extensions of the Gillespie algorithms aiming to add more reality to the model (i.e., non-Poissonian cases) or to speed up the simulations.
\end{abstract}



\copyrightauthor{Naoki Masuda, Christian L. Vestergaard, 2021}

\mainmatter  



\section{Introduction}

We are compelled to understand and intervene in the dynamics of various complex systems in which different elements, such as human individuals, interact with each other. Such complex systems are often modeled by multi-agent or network-based models that explicitly dictate how each individual behaves and influences other individuals. 
Stochastic processes are popular models for the dynamics of multiagent systems when it is realistic to assume random elements in how agents behave or in dynamical processes taking place in the system. 
For example, random walks have been successfully applied to describe locomotion and foraging of animals \citep{Okubo2001book,Codling2008JRSocInterface}, dynamics of neuronal firing \citep{Tuckwell1988book2,Gabbiani2010book}, and financial market dynamics \citep{CampbellLoMackinlay1997book,Mantegna2000book} to name a few (see \cite{Masuda2017PhysRep} for a review). 
Branching processes are another major type of stochastic processes that have been applied to describe, for example, information spread \citep{Eugster2004Computer,Gleeson2021JCompNetw}, spread of infectious diseases \citep{Farrington2003Biostatistics,Britton2010MathBiosci}, cell proliferation \citep{Jagers1975book},
and the abundance of species in a community
\citep{Mcgill2007EcolLett} as well as other ecological dynamics \citep{Black2012TrendsEcolEvol}.

Stochastic processes in which the state of the system changes via discrete events that occur at given points in time are a major class of models for dynamics of complex systems~\citep{Singer1976AmJSociol,Daley1999book,Andersson2000book,Barrat2008book,Liggett2010book,Shelton2014JArtifIntelResearch,Vanmieghem2014book,KissMillerSimon2017book,Dearruda2018PhysRep}. For example, in typical models for infectious disease spread, each infection event occurs at a given time $t$ such that an individual transitions instantaneously from a healthy to an infectious state. Such processes are called \emph{Markov jump processes}\index{Markov jump process} when they satisfy certain independence conditions~\citep{Hanson2007book}, which we will briefly discuss in Section~\ref{sub:independence}. A jump is equivalent to a discrete event. In Markov jump processes, jumps occur according to \emph{Poisson processes}\index{Poisson process}. In this volume, we focus on how to simulate Markov jump processes. Specifically, we will introduce a set of exact and computationally efficient simulation algorithms collectively known as Gillespie algorithms. In the last technical section of this volume (i.e., Section~\ref{sec:renewal_and_temporal}), we will also consider more general, non-Markov, jump processes\index{jump processes}, in which the events are generated in more complicated manners than by Poisson processes. In the following text, we refer collectively to Markov jump processes and non-Markov jump processes  as jump processes.

The Gillespie algorithms\index{Gillespie algorithm} were originally proposed in their general forms by Daniel Gillespie in 1976 for simulating systems of chemical reactions~\citep{Gillespie1976JComputPhys}, whereas several specialized variants had been proposed earlier; see Section~\ref{sub:history} for a brief history review. 
Gillespie proposed two different variants of the simulation algorithm, the \emph{direct method}\index{direct method}, also known as Gillespie's \emph{stochastic simulation algorithm} (SSA), or often simply \emph{the Gillespie algorithm}, and the \emph{first reaction method}.
Both the direct and first reaction methods have found widespread use and in fields far beyond chemical physics. 
Furthermore, researchers have developed many extensions and improvements of the original Gillespie  algorithms to widen the types of processes that we can simulate with them and to improve their computational efficiency.

The Gillespie algorithms are practical algorithms to simulate coupled Poisson processes\index{coupled Poisson processes} exactly (i.e., without approximation error). 
Here ``coupled'' means that an event that occurs somewhere in the system potentially influences the likelihood of future events' occurrences in different parts of the same system. 
For example, when an individual in population, $v_i$, gets infected by a contagious disease, the likelihood that a different healthy individual in the same population, $v_j$, will get infected in the near future may increase. If interactions were absent, it would suffice to separately consider single Poisson processes, and simulating the system would be straightforward. 

We believe that the Gillespie algorithms are important tools for students and researchers that study dynamic social systems, where social dynamics is broadly construed and include both human and animal interactions, ecological systems, and even technological systems. While there already exists a large body of references on the Gillespie algorithms and their variants, most are concise, mathematically challenging for beginners, and focused on chemical reaction systems.

Given these considerations, the primary aim of this volume is to provide a detailed tutorial on the Gillespie algorithms, with specific focus on simulating dynamic social systems. We assume only limited prior knowledge of stochastic processes. 
We will realize the tutorial in the first part of the volume (Sections~\ref{sec:preliminaries} and \ref{sec:classic_Gillespie}). In this part, we assume basic knowledge of calculus and probability. Although we do introduce stochastic processes and explain the Gillespie algorithms and related concepts with much reference to networks, we do not assume prior knowledge of stochastic processes or of networks. To understand the coding section, readers will need basic knowledge of programming. The second part of this volume (Sections~\ref{sec:computational_complexity} and \ref{sec:renewal_and_temporal})
is devoted to a survey of recent advancements of Gillespie algorithms for simulating social dynamics. These advancements are concerned with accelerating simulations and/or increasing the realism of the models to be simulated.

\section{Preliminaries}
\label{sec:preliminaries}

We review in this section mathematical concepts needed to understand the Gillespie algorithms. 
In Sections \ref{sub:systems to simulate} to \ref{sub:sir}, we introduce the types of models we will be concerned with, namely \emph{jump processes}, and in particular a simple type of jump process termed \emph{Poisson processes}.
In Sections \ref{sub:waiting_time} to \ref{sub:superposition}, we derive main mathematical properties of Poisson processes.
The concepts and results presented in Sections 
\ref{sub:systems to simulate} to \ref{sub:superposition} are necessary for understanding Section~\ref{sec:classic_Gillespie}, where we derive the Gillespie algorithms. 
In Sections~\ref{sub:ODE} and \ref{sub:rejection sampling}, we review two simple methods for solving the models which predate the Gillespie algorithms and discuss some of their shortcomings. 
These two final subsections motivate the need for exact simulation algorithms such as the Gillespie algorithms.

\subsection{\label{sub:systems to simulate}Jump processes}

Before getting into the nitty-gritty of the Gillespie algorithms, we first explore which types of systems they can be used to simulate. 
First of all, with the Gillespie algorithms, we are interested in simulating a dynamic system. 
This can be, for example, epidemic dynamics in a population in which the number of infectious individuals varies over time, or the evolution of the number of crimes in a city, which also varies over time in general. Second, the Gillespie algorithms rely on a predefined and parametrized mathematical model for the system to simulate. 
Therefore, we must have the set of rules for how the system or the individuals in it change their states.
Third, Gillespie algorithms simulate stochastic processes, not deterministic systems. In other words, every time one runs the same model starting from the same initial conditions, the results will generally differ. In contrast, in a deterministic dynamical system, if we specify the model and the initial conditions, the behavior of the model will always be the same. Fourth and last, the Gillespie algorithms simulate processes in which changes in the system are primarily driven by discrete events taking place in continuous time. For example, when a chemical reaction obeying the chemical equation A $+$ B $\to$ C $+$ D happens, one unit of A and of B are consumed, and one unit of C and of D are produced. This event is discrete in that we can count the event and say when the event has happened, but it can happen at any point in time (i.e. time is not discretized but continuous). 

We refer to the class of mathematical models that satisfy these conditions and may be simulated by a Gillespie algorithm as \emph{jump processes}\index{jump process}. In the remainder of this section, we explore these processes more extensively through motivating examples. Then, we introduce some fundamental mathematical definitions and results that the Gillespie algorithms rely on.

\subsection{\label{sub:network-population}Representing a population as a network}

Networks are an extensively used abstraction for representing a structured population, and Gillespie algorithms lend themselves naturally to simulate stochastic dynamical processes taking place in networks. 
In a network representation, each individual in the population corresponds to a node in the network, and edges are drawn between pairs of individuals that directly interact. 
What constitutes an interaction generally depends on the context. 
In particular, for the simulation of dynamic processes in the population, the interaction depends on the nature of the process we wish to simulate.
For simulating the spread of an infectious disease for example, a typical type of relevant interaction is physical proximity between individuals.

Formally, we define a network\index{network} as a graph\index{graph} $G=(V,E)$, where $V = \{1, 2, \ldots, N\}$ is the set of nodes\index{node}, $E = \{(u,v): u,v\in V\}$ is the set of edges\index{edge}, and each edge $(u,v)$ defines a pair of nodes $u,v\in V$ that are directly connected. The pairs $(u,v)$ may be ordered, in which case edges are directed (by convention from $u$ to $v$), or unordered, in which case edges are undirected (i.e., $v$ connects to $u$ if and only if $u$ connects to $v$). We may additionally add weights to the edges to represent different strengths of interactions, or we may even consider graphs that evolve in time (so-called temporal networks\index{temporal networks}) to account for the dynamics of interactions in a population.

We will primarily consider simple (i.e., static, undirected, and unweighted) networks in our examples. However, the Gillespie algorithms apply to simulate jump processes in all kinds of populations and networks. 
(For temporal networks, we need to extend the classic Gillespie algorithms to cope with the time-varying network structure; see Section~\ref{sub:TGA}.)

\subsection{\label{sub:sir}Example: Stochastic SIR model in continuous time}

We introduce jump processes and explore their mathematical properties by way of a running example. We show how we can use them to model epidemic dynamics using the stochastic susceptible-infectious-recovered (SIR) model\index{susceptible-infectious-recovered (SIR) model}\footnote{The SIR model was incidentally one of the first applications of a Gillespie-type algorithm in a 1953 article~\citep{Bartlett1953JRStatSocSerCApplStat}.}.
For more examples (namely, SIR epidemic dynamics in metapopulation networks, the voter model, and the Lotka-Volterra model for predator-prey dynamics), see Section~\ref{sub:code}.

\begin{figure}
\centering
\includegraphics[width=\textwidth]{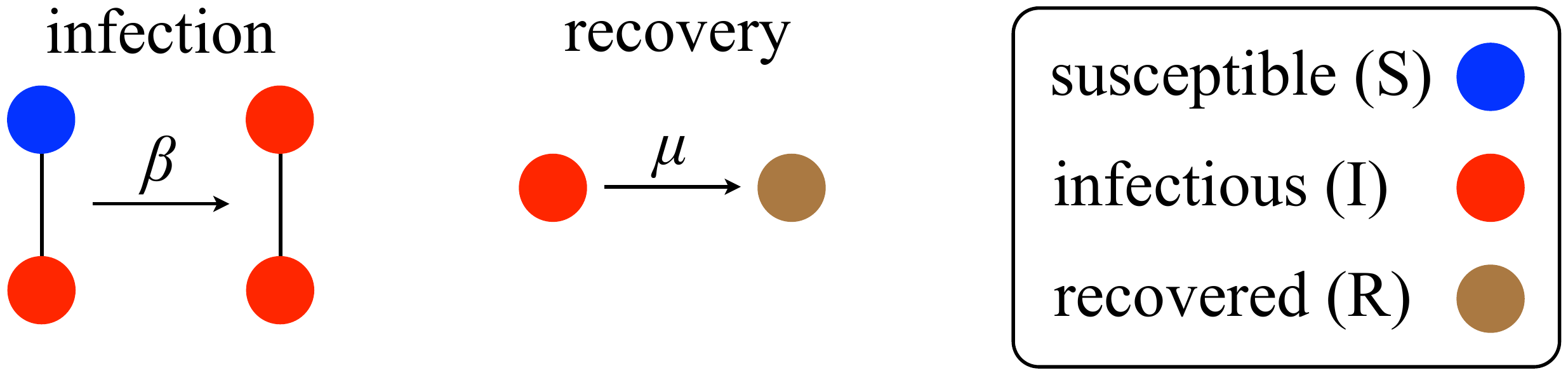}
\caption{Rules of state changes in the SIR Model. An infectious individual infects a susceptible neighbor at a rate $\beta$. Each infectious individual recovers at a rate $\mu$.}
\label{fig:SIR rule}
\end{figure}

We examine a stochastic version of the SIR model in continuous time defined as follows. We consider a constant population of $N$ individuals (nodes). At any time, each individual is in one of three states: susceptible (denoted by $S$; meaning healthy), infectious (denoted by $I$), or recovered (denoted by $R$). The rules governing how individuals change their states are shown schematically in Fig.~\ref{fig:SIR rule}. 
An infectious individual that is in contact with a susceptible individual infects the susceptible individual in a stochastic manner with a constant \emph{infection rate}\index{infection rate} $\beta$. 
Independently of the infection events, an infectious individual may recover at any point in time, with a constant \emph{recovery rate}\index{recovery rate} $\mu$. If an infection event occurs, the susceptible individual that has been infected changes its state to I. 
If an infectious individual recovers, it transits from the I to the R state. 
Nobody leaves or joins the population over the course of the dynamics.
After reaching the R state, an individual cannot be reinfected or infect others again. Therefore, R individuals do not influence the increase or decrease in the number of S or I individuals. 
Because R individuals are as if they no longer exist in the system, the R state is mathematically equivalent to having died of the infection; once dead, an individual will not be reinfected or infect others.

We typically start the stochastic SIR dynamics with a single infectious individual, which we refer to as the \emph{source} or \emph{seed}, and $N_{\rm S} = N-1$ susceptible individuals (and thus no recovered individuals). Then, various infection and recovery events may occur. The dynamics stops when no infectious individuals are left. 
In this final situation, the population is composed entirely of susceptible and/or recovered individuals. 
Since both infection and recovery involve an infectious individual, and there are no infectious individuals left, the dynamics is stuck. The final number of recovered nodes, denoted by $N_{\rm R}$, is called the \emph{epidemic size}\index{epidemic size}, also known as the \emph{final epidemic size} or simply the \emph{final size}\footnote{The fraction $N_{\rm R}/N$ is typically also referred to as the epidemic size.}. 
The epidemic size tends to increase as the infection rate $\beta$ increases or as the recovery rate $\mu$ decreases. 
Many other measures to quantify the behavior of the SIR model exist~\citep{Pastorsatorras2015RevModPhys}. For example, we may be interested in the time till the dynamics terminates or in the speed at which the number of infectious individuals grows in the initial stage of the dynamics.

\begin{figure}
\centering
\includegraphics[width=\textwidth]{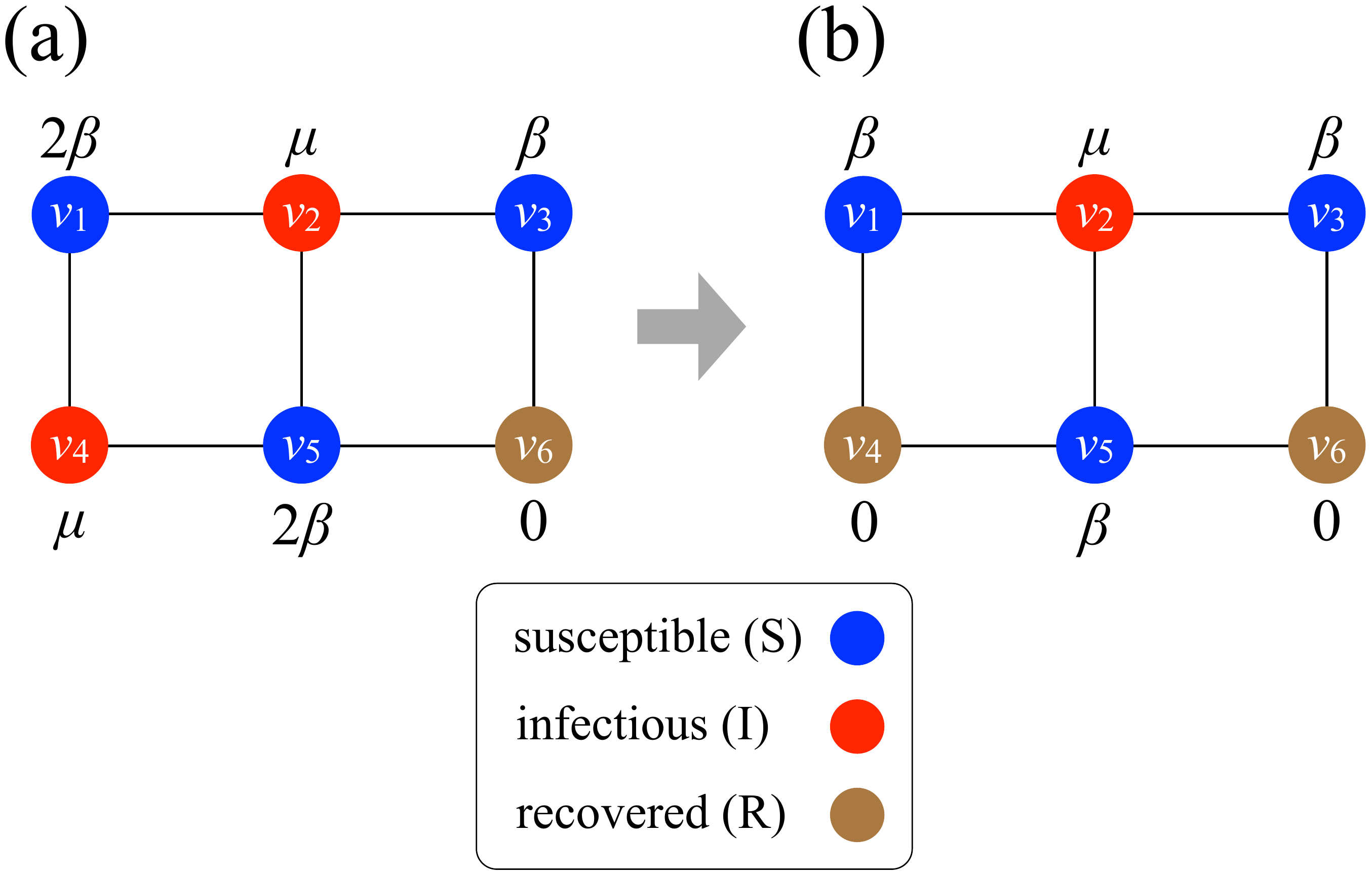}
\caption{Stochastic SIR process on a square-grid network with 6 nodes. (a) Status of the network at an arbitrary time $t$. (b) Status of the network after $v_4$ has recovered. The values attached to the nodes indicate the rates of the events that the nodes may experience next.}
\label{fig:6 node SIR}
\end{figure}

Consider Fig.~\ref{fig:6 node SIR}(a), where individuals are connected as a network. 
We generally assume that infection may only occur between pairs of individuals that are directly connected by an edge (called adjacent nodes)\index{adjacent}. 
For example, the node $v_4$ can infect $v_1$ and $v_5$ but not $v_3$. 
The network version of the SIR model is fully described by the infection rate $\beta$, the recovery rate $\mu$, the network structure, i.e., which node pairs are connected by an edge, and the choice of source node to initialize the dynamics.

Mathematically, we describe the system by a set of coupled, constant-rate jump processes; constant-rate jump processes are known as \emph{Poisson processes}\index{Poisson process} (Box~\ref{box:Poisson_process}).
Each possible event that may happen is associated to a Poisson process, i.e., the recovery of each infectious individual is described by a Poisson process, and so is each pair of infectious and susceptible individuals where the former may infect the latter. 
The Poisson processes are coupled because an event generated by one process may alter the other processes by changing their rates, generating new Poisson processes, or making existing ones disappear. For example, after a node gets infected it may in turn infect any of its susceptible neighbors, which we represent mathematically by adding new Poisson processes. This coupling implies that the set of coupled Poisson processes generally constitutes a process that is more complicated than a single Poisson process.

In the following subsections we develop main mathematical properties of Poisson processes and of sets of Poisson processes. We will rely on these properties in Section~\ref{sec:classic_Gillespie} to construct the Gillespie algorithms that can simulate systems of coupled Poisson processes exactly. Note that the restriction to Poisson (i.e., constant-rate) processes is essential for the classic Gillespie algorithms to work; see Section~\ref{sec:renewal_and_temporal} for recent extensions to the simulation of non-Poissonian processes.

\begin{figure}
\begin{boxedtext}{}
\captionof{floatbox}{\textbf{Properties of Poisson processes.}}
\label{box:Poisson_process}

A Poisson process is a jump process that generates events with a constant rate, $\lambda$.

\paragraph{\small Waiting-time distribution.}

The waiting times $\tau$ between consecutive events generated by a Poisson process are exponentially distributed. In other words, $\tau$ obeys the probability density
\begin{equation}
    \psi(\tau) = \lambda e^{-\lambda \tau} .
    \label{eq:p(tau)-Poisson}
\end{equation}

\paragraph{\small Memoryless property.}

The waiting time left until a Poisson process generates an event given that a time $t$ has already elapsed since the last event is independent of $t$. This property is called the memoryless property of Poisson processes and is shown as follows:
\begin{equation}
    \psi(t+\tau|t) = \frac{\psi(t+\tau)}{\Psi(t)} =  \frac{\lambda e^{-\lambda(t+\tau)}}{e^{-\lambda t}} =  \lambda e^{-\lambda \tau} ,
\label{eq:memoryless_exponential_dist}    
\end{equation}
where $\psi(t+\tau|t)$ represents the conditional probability density that the next event occurs a time $t+\tau$ after the last event given that time $t$ has already elapsed; $\Psi(t)=\int_t^\infty \psi(\tau)\text{d}\tau = e^{-\lambda t}$ is called the survival probability\index{survival probability} and is the probability that no event takes place for a time $t$. The first equality in Eq.~\eqref{eq:memoryless_exponential_dist} follows from the definition of the conditional probability. The second equality follows from Eq.~\eqref{eq:p(tau)-Poisson}.

\paragraph*{\small Superposition theorem.}

\index{superposition theorem}Consider a set of Poisson processes indexed by $i\in\{1, 2, \ldots, M\}$.
The superposition of the processes is a jump process that generates an event whenever any of the individual processes does. 
It is another Poisson process whose rate is given by
\begin{equation}
    \Lambda = \sum_{i=1}^M \lambda_i ,
    \label{eq:superposition-Poisson}
\end{equation}
where $\lambda_i$ is the rate of the $i$th Poisson process.

\paragraph*{\small Probability of a given process generating an event in a superposition of Poisson processes.}

Consider any given event generated by a superposition of Poisson processes. The probability $\Pi_i$ that the $i$th individual Poisson process has generated this event is proportional to the rate of the $i$th process. In other words,
\begin{equation}
    \Pi_i = \lambda_i/\Lambda .
    \label{eq:Pi_superposition_Poisson}
\end{equation}

\end{boxedtext}
\end{figure}

\subsection{\label{sub:waiting_time}Waiting-time distribution for a Poisson process}

We derive in this subsection  the \emph{waiting-time distribution}\index{waiting time} for a Poisson process, which characterizes how long one has to wait for the process to generate an event. It is often easiest to start from a discrete-time description when exploring properties of a continuous-time stochastic process. Therefore, we will follow this approach here. 
We use the recovery of a single node in the SIR model as an example in our development.

Let us partition time into short intervals of length $\delta t$. As $\delta t$ goes to zero, this becomes an exact description of the continuous-time process. An infectious individual recovers with probability $\mu \delta t$ after each interval given that it has not recovered before\footnote{To address a common misunderstanding, we emphasize that $\mu$ is a rate, not a probability, and thus can be larger than one. Note however that $\mu\delta t$ is a probability and thus cannot be greater than one.}. 

Formally, we define the SIR process in the limit $\delta t \to 0$. 
Then, you might worry that the recovery event is unlikely to ever take place because the probability with which it happens during each time step, i.e., $\mu \delta t$, goes towards 0 when the step size $\delta t$ does so. However, this is not the case; because the number of time steps in any given finite interval grows inversely proportional to $\delta t$, the probability to recover in finite time stays finite. For example, if we use a different step size $\overline{\delta t} = \delta t / 10$, which is ten times smaller than the original $\delta t$, then the probability of recovery within the short duration of time $\overline{\delta t}$ is indeed 10 times smaller than $\mu \delta t$ (i.e., $= \mu \overline{\delta t}$). However, there are ${\delta t/\overline{\delta t}} = 10$ windows of size $\overline{\delta t}$ in one time window of size $\delta t$. So, we now have 10 chances for recovery to happen instead of only one chance. The probability for recovery to occur in any of these 10 time windows is equal to 1 minus the probability that it does not occur. The probability that the individual does not recover in time $\delta t$ is equal to $(1 - \mu\overline{\delta t})^{\delta t/\overline{\delta t}}$. Therefore, the probability that the individual recovers in any of the $\delta t/\overline{\delta t}$ windows
is
\begin{equation}
p_{\mathrm{I}\to\mathrm{R}} = 1 - (1 - \mu\overline{\delta t})^{\delta t/\overline{\delta t}}.
\label{eq:prob_recover_discrete_time}
\end{equation}
Equation~\eqref{eq:prob_recover_discrete_time} does not vanish as we make $\overline{\delta t}$ small. 
In fact, the Taylor expansion of Eq.~\eqref{eq:prob_recover_discrete_time} in terms of $\overline{\delta t}$ yields $p_{\mathrm{I}\to\mathrm{R}} \approx (\delta t/\overline{\delta t}) \times \mu\overline{\delta t} = \mu\delta t$, where $\approx$ represents ``approximately equal to''. Therefore, to leading order, the recovery probabilities are the same between the case of a single time window of size $\delta t$ and the case of $\delta t/\overline{\delta t}$ time windows of size $\overline{\delta t}$.

In the limit $\delta t \to 0$, the recovery event may happen at any continuous point in time. 
We denote by $\tau$ the waiting time from the present time until the time of the recovery event. 
We want to determine the probability density function (probability density or pdf for short) of $\tau$, which we denote by $\psi_{\text{I} \to \text{R}}(\tau)$. 
By definition, $\psi_{\text{I} \to \text{R}}(\tau)\delta t$ is equal to the probability that the recovery event happens in the interval $[\tau, \tau + \delta t)$ for an infinitesimal $\delta t$ (i.e., for $\delta t \to 0$). 
To calculate $\psi_{\text{I}\to\text{R}}(\tau)$, we note that the probability that the event occurs after $r = \tau/\delta t$ time windows, denoted by $p_{\text{I} \to \text{R}}(r)$, is equal to the probability that it did not occur during the first $r$ time windows and then occurs in the $(r+1)$th window. This probability is equal to
\begin{equation}
p_{\text{I} \to \text{R}}(r) = (1-\mu \delta t)^r \times \mu \delta t =
(1- \mu \delta t)^{\tau/\delta t} \mu \delta t .
\label{eq:wait and recovery discrete time}
\end{equation}
The first factor on the right-hand side of Eq.~\eqref{eq:wait and recovery discrete time} is the probability that the event has not happened before the $(r+1)$th window; it is simply equal to the probability that the event has not happened during a single window, raised to the power of $r$. The second factor is the probability that the event happens in the $(r+1)$th window. By applying the identity $\lim_{x \to 0} (1+x)^{1/x} = e$, known from calculus (see Appendix~\ref{app:exp_identity}), with $x = - \mu \delta t$ to Eq.~\eqref{eq:wait and recovery discrete time}, we obtain the pdf of the waiting time as follows:
\begin{align}
  \psi_{\text{I} \to \text{R}}(\tau) 
  &= \lim_{\delta t \to 0} \frac{p_{\text{I} \to \text{R}}(\tau/\delta t)}{\delta t} \nonumber\\
  &= \mu \lim_{\delta t \to 0} (1- \mu \delta t)^{\tau/\delta t} \nonumber\\ 
  &= \mu \left[ \lim_{\delta t \to 0} (1- \mu \delta t)^{1/(-\mu\delta t)} \right]^{- \mu \tau} \nonumber\\
  &= \mu e^{-\mu \tau} .
   \label{eq:psi(tau) recovery}
\end{align}
Equation~\eqref{eq:psi(tau) recovery} shows the intricate connection between the Poisson process and the exponential distribution: the waiting time of a Poisson process with rate $\mu$ (here, specifically the recovery rate) follows an exponential distribution with rate $\mu$ (Box~\ref{box:Poisson_process}). This fact implies that the mean time we have to wait for the recovery event to happen is $1/\mu$.
The exponential waiting-time distribution actually completely characterizes the Poisson process. In other words, the Poisson process is the only jump process that generates events separated by waiting times that follow a fixed exponential distribution.

If we consider the infection process between a pair of S and I nodes in complete isolation from the other infection and recovery processes in the population, then exactly the same argument (Eq.~\eqref{eq:psi(tau) recovery}) holds true. 
In other words, the time until infection takes place between the two nodes is exponentially distributed with rate $\beta$, i.e.,
\begin{equation}
\psi_{\text{S} \to \text{I}}(\tau) = \beta e^{-\beta \tau}.
\label{eq:psi(tau) infection}
\end{equation}
However, in practice the infection process is more complicated than the recovery process because it is coupled to other processes. Specifically, if another process generates an event before the infection process does, then Eq.~\eqref{eq:psi(tau) infection} may no longer hold true for the infection process in question. 
For example, consider a node $v_1$ that is currently susceptible and an adjacent node $v_2$ that is infectious, as in Fig.~\ref{fig:6 node SIR}. 
For this pair of nodes, two events are possible: $v_2$ may infect $v_1$, or $v_2$ may recover. 
As long as neither of the events has yet taken place, either of the two corresponding Poisson processes may generate an event at any point in time, following Eqs.~\eqref{eq:psi(tau) infection} and \eqref{eq:psi(tau) recovery}, respectively.  
However, if $v_2$ recovers before it infects $v_1$, then the infection event is no longer possible, and so Eq.~\eqref{eq:psi(tau) infection} no longer holds. 
We explore in the following two subsections how to mathematically deal with this coupling.

\subsection{Independence and inter-dependence of jump processes\label{sub:independence}}

Most models based on jump processes and most simulation methods, including the Gillespie algorithms, implicitly assume that different concurrent jump processes are independent of each other in the sense that the internal state of one process does not influence another. 
This notion of independence may be a source of confusion because a given process may depend on the events generated earlier by other processes, i.e., the processes may be \emph{coupled}, as we saw is the case for the infection processes in the SIR model.  
In this section, we sort out the notions of independence and coupling and what they mean for the types of jump processes we want to simulate.
We will also explore another type of independence of Poisson processes, which is their independence of the past, called the \emph{memoryless} property\index{memoryless property}.

We can state the independence assumption as the condition that different processes are only allowed to influence each other by changing the state of the system.
In other words, at any point in time each process generates an event at a rate that is independent of all other processes given the current state of the system, i.e., the processes are \emph{conditionally independent}\index{conditional independence}. 
For example, the rate at which $v_2$ infects $v_1$ in Fig.~\ref{fig:6 node SIR}(a) depends on $v_2$ being infectious and $v_1$ being susceptible (corresponding to the system's current state). However, it does not depend on any internal state of $v_2$'s recovery process such as the time left till $v_2$ recovers. Given the states of all nodes, the two processes are independent. 
Poisson processes are always conditionally independent in this sense.
The conditional independence property follows directly from the fact that Poisson processes have constant rates by definition and thus are not influenced by other processes.
The conditional independence is essential for the Gillespie algorithms to work.
Even the recent extensions of the Gillespie algorithms to simulate non-Poissonian processes which we review in Section~\ref{sec:renewal_and_temporal} rely on an assumption of conditional independence between the jump processes.

We underline that the assumption of conditional independence does not imply that the different jump processes are not coupled with each other. 
Such uncoupled processes would indeed be boring. 
If the jump processes constituting a given system were all uncoupled, then they would not be able to generate any collective dynamics.
On the technical side, there would in this case be no reason to consider the set of processes as one system. 
It would suffice to analyze each process separately.
We would in particular have no need for the specialized machinery of the Gillespie algorithms since we could simply simulate each process by sampling waiting times from the corresponding exponential distribution [Box~\ref{box:Poisson_process}, Eq.~\eqref{eq:p(tau)-Poisson}]. 

In fact, the conditional independence assumption allows different processes to be coupled, as long as they only do so by changing the physical state of the system. 
This is a natural constraint in many systems. 
For example, in chemical reaction systems, the processes (i.e., chemical reactions) are coupled through discrete reaction events that use molecules of some chemical species to generate others.
Similarly, in the SIR model different processes influence each other by changing the state of the nodes, i.e., from S to I in an infection event or from I to R in a recovery event. 
In the example shown in Fig.~\ref{fig:6 node SIR}, 
when node $v_4$ recovers, it decreases the probability that its neighboring susceptible node $v_1$ gets infected within a certain time horizon compared with the scenario where $v_4$ remains infectious. As this example suggests, the probability that a susceptible node gets infected depends on the past states of its neighbors. 
Therefore, over the course of the entire simulation, the dynamics of a node's state (e.g., $v_1$) is dependent on that of its neighbors (e.g., $v_2$ and $v_4$). 

Because of the coupling between jump processes, which is present in most systems of interest, we cannot simply simulate the system by separately generating the waiting times for each process according to Eq.~\eqref{eq:p(tau)-Poisson}. Any event that occurs will alter the processes to which it is coupled, thus rendering the waiting times we drew for the affected processes invalid.
What the Gillespie algorithms do instead is to successively generate the waiting time until the next event, update the state of the system, and reiterate. 

Besides being conditionally independent of each other, Poisson processes also display a temporal independence property, the so-called \emph{memoryless property}\index{memoryless property} (Box~\ref{box:Poisson_process}).
In Poisson processes, the probability of the time to the next event, $\tau$, is independent of how long we have already waited since the last event. In this sense, we do not need to worry about what has happened in the past. The only things that matter are the present status of the population (such as $v_1$ is susceptible and $v_2$ is infectious right now) and the model parameters (such as $\beta$ and $\mu$). 
The memoryless property can be seen as a direct consequence of the exponential distribution of waiting times of Poisson processes  (Box~\ref{box:Poisson_process}, Eq.\eqref{eq:memoryless_exponential_dist}).
The direct method exploits the memoryless property.

\subsection{\label{sub:superposition}Superposition of Poisson processes}

In this section, we explain a remarkable property of Poisson processes called the superposition theorem. The direct method exploits this theorem. 
Other methods, such as the rejection sampling algorithm (see Section~\ref{sub:rejection sampling} below) and the first reaction method, can also benefit from the superposition theorem to accelerate the simulations without impacting their accuracy.

\begin{figure}
\centering
\includegraphics[width=.9\textwidth]{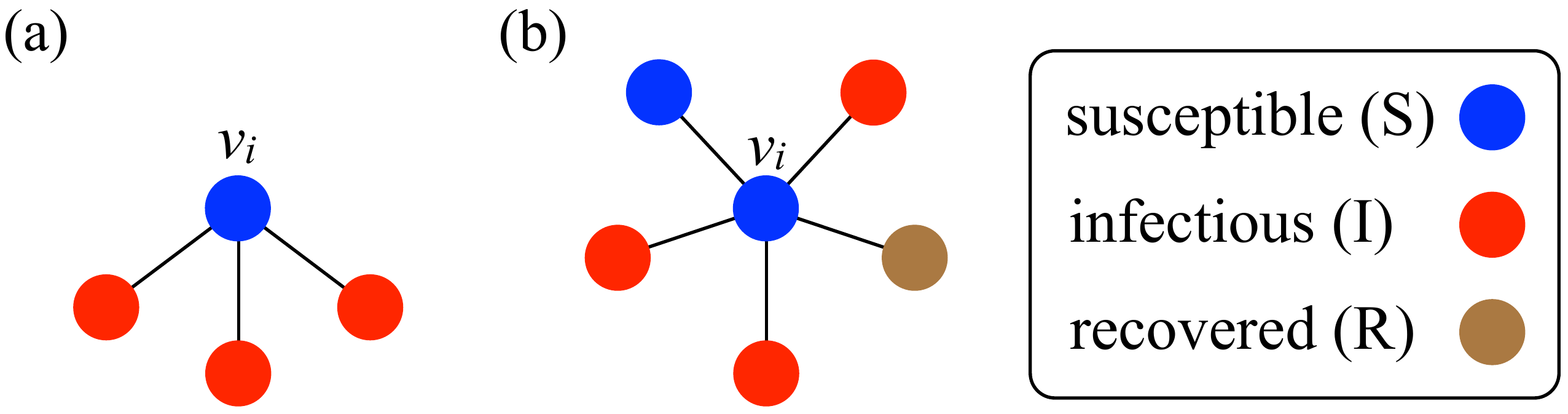}
\caption{A susceptible node and other nodes surrounding it. (a) A susceptible node $v_i$ surrounded by three infectious nodes. (b) A susceptible node $v_i$ surrounded by five nodes in different states.}
\label{fig:local net}
\end{figure}

Consider a susceptible individual $v_i$ in the SIR model that is in contact with $N_{\rm I}$ infectious individuals. 
Any of the $N_{\rm I}$ infectious individuals may infect $v_i$. Consider the case shown in Fig.~\ref{fig:local net}(a), where $N_{\rm I}=3$. 
If we focus on a single edge connecting $v_i$ to one of its neighbors and ignore the other neighbors, the probability that $v_i$ is infected via this edge exactly in time $[\tau, \tau + \delta t)$ from now, where $\delta t$ is small, is given by $\psi_{\text{S}\to\text{I}}(\tau)\delta t$
(see Eq.~\eqref{eq:psi(tau) infection}). 
Each of $v_i$'s $N_{\rm I}$ neighbors may infect $v_i$ in the same manner and independently. 
The neighbor that does infect $v_i$ is the one for which the corresponding waiting time is the shortest, provided that it does not recover before it infects $v_i$.
From this we can intuitively see that the larger $N_{\rm I}$ is, the shorter the waiting time before $v_i$ gets infected tends to be. To simulate the dynamics of this small system, we need to know, not when each of its neighbors would infect $v_i$, but rather the time until any of its neighbors infects $v_i$.  

To calculate the waiting-time distribution for the infection of  $v_i$ by any of its neighbors, we again resort to the discrete-time view of the infection processes. 
Because the infection processes are independent, the probability that $v_i$ is not infected by any of its $N_{\rm I}$ infectious neighbors in a time window of duration $\delta t$ is given by
\begin{equation}
(1 - \beta \delta t)^{N_{\rm I}} .
\end{equation}
Therefore, the probability that $v_i$ is infected after a time $\tau = r \delta t$ (i.e., $v_i$ gets infected exactly in the $(r+1)$th time window of length $\delta t$ and not before) is given by
\begin{equation}
p_{\text{I} \to \text{R}} = \left[(1 - \beta \delta t)^{N_{\rm I}} \right]^r \times
\left[1 - (1 - \beta \delta t)^{N_{\rm I}} \right] .
\label{eq:superposition discrete time 1}
\end{equation}
Here, the factor $\left[(1 - \beta \delta t)^{N_{\rm I}} \right]^r$ is the survival probability that an infection does not happen for a time $\tau = r \delta t$. The factor $\left[1 - (1 - \beta \delta t)^{N_{\rm I}} \right]$ is the probability that any of $v_i$'s infectious neighbors infects $v_i$ in the next time window, $t \in [\tau, \tau + \delta t)$.

Using the exponential identity $\lim_{x\to 0} (1+x)^{1/x} = e$ with $x = -\beta \delta t$ as we did in Section~\ref{sub:waiting_time}, we obtain in the continuous-time limit that
\begin{equation}
\lim_{\delta t \to 0}
\left[(1 - \beta \delta t)^{N_{\rm I}} \right]^r 
= \lim_{\delta t \to 0} \left[(1 - \beta \delta t)^{1/(-\beta\delta t)} \right]^{-N_{\rm I}\beta\tau}
= e^{-N_1 \beta \tau} ,
\label{eq:superposition discrete time 2}
\end{equation}
where the first equality is obtained by noting that $r = \tau/\delta t$ and rearranging the terms.
In the same limit of $\delta t \to 0$, we obtain from
Taylor expansion that
\begin{equation}
\left[1 - (1 - \beta \delta t)^{N_{\rm I}} \right]
\approx
1 - (1 - N_{\rm I} \beta \delta t) = N_{\rm I} \beta \delta t .
\label{eq:superposition discrete time 3}
\end{equation}
By combining Eqs.~\eqref{eq:superposition discrete time 1}, \eqref{eq:superposition discrete time 2}, and \eqref{eq:superposition discrete time 3}, we obtain $p_{\text{I}\to\text{R}} \approx N_{\rm I}\beta e^{-N_{\rm I}\beta\tau}\delta t$. Therefore, the probability density with which $v_i$ gets infected at time $\tau$ is given by
\begin{equation}
\psi_{\text{I} \to \text{R}}(\tau) = N_{\rm I} \beta e^{- N_{\rm I} \beta \tau},
\label{eq:p(tau) S->I when N_1 infected}
\end{equation}
i.e., the exponential distribution with rate parameter $N_{\rm I} \beta$. 
By comparing Eqs.~\eqref{eq:psi(tau) infection} and \eqref{eq:p(tau) S->I when N_1 infected}, we see that the effect of having $N_{\rm I}$ infectious neighbors (see Fig.~\ref{fig:local net}(a) for the case of $N_{\rm I}=3$) is the same as having just one infectious neighbor with an infection rate of $N_{\rm I} \beta$.

This is a convenient property of Poisson processes, known as the superposition theorem\index{superposition theorem} (see Box~\ref{box:Poisson_process}, Eq.~\eqref{eq:superposition-Poisson} for the general theorem). To calculate how likely it is that a susceptible node $v_i$ will be infected in time $\tau$, one does not need to examine when the infection would happen or whether the infection happens for each of the infectious individuals contacting $v_i$. We are allowed to agglomerate all those effects into one infectious supernode as if the supernode infects $v_i$ with rate $N_{\rm I} \beta$. We refer to such a superposed Poisson process that induces a particular state transition in the system (in the present case, the transition from the S to I state for $v_i$) as a \emph{reaction channel}\index{reaction channel}, following the nomenclature in chemical reaction systems.

This interpretation remains valid even if $v_i$ is adjacent to other irrelevant individuals. In the network shown in Fig.~\ref{fig:local net}(b), the susceptible node $v_i$ has degree\index{degree} (i.e., number of other nodes that are connected to $i$ by an edge) $k_i = 5$. Three neighbors of $v_i$ are infectious, one is susceptible, and one is recovered. In this case, $v_i$ will be infected at a rate of $3 \beta$, same as in the case of $v_i$ in the network shown in Fig.~\ref{fig:local net}(a).

\begin{figure}
\centering
\includegraphics[width=\textwidth]{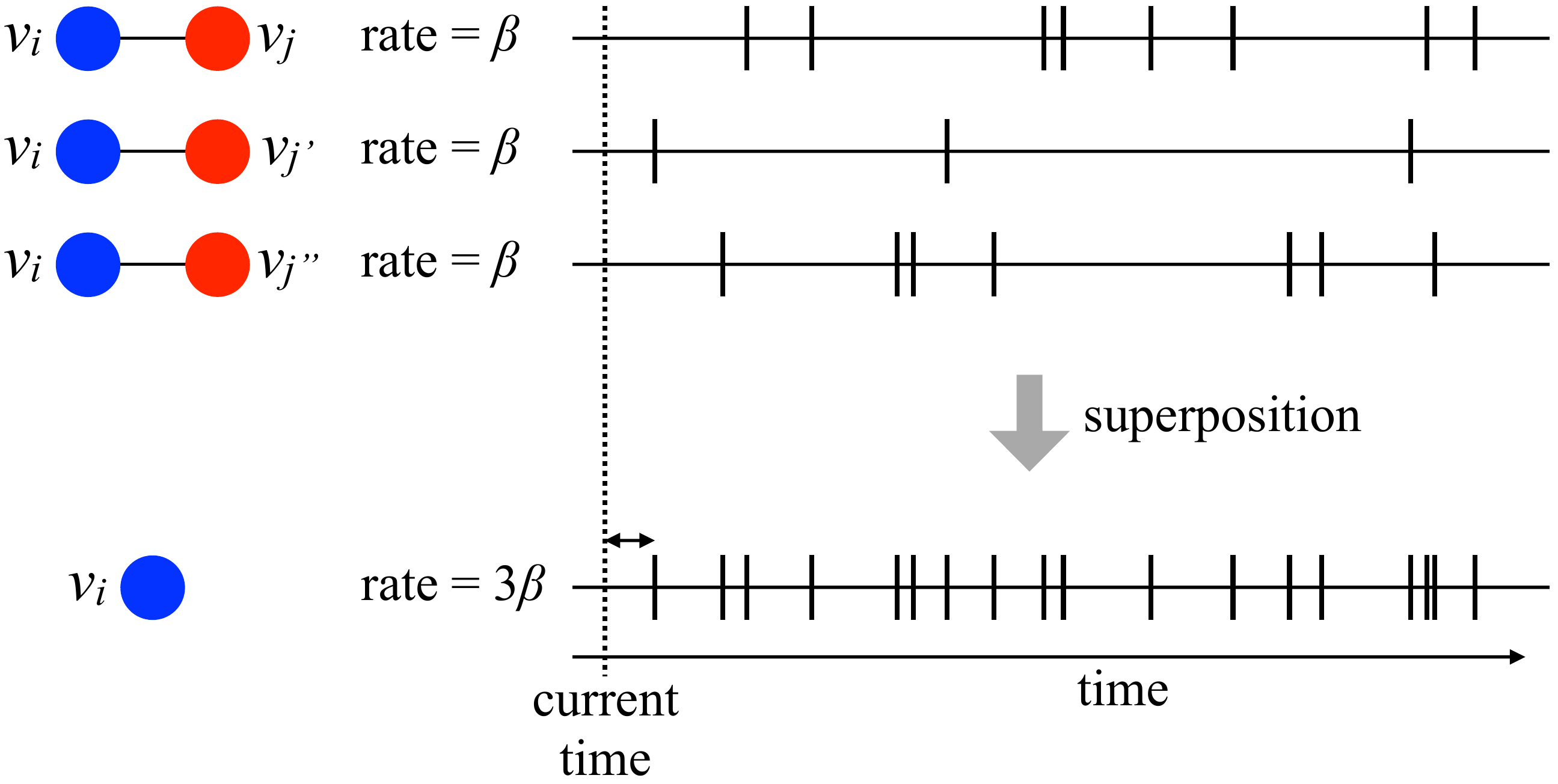}
\caption{Superposition of three Poisson processes. The event sequence in the bottom is the superposition of the three event sequences corresponding to each of the three edges connecting $v_i$ to its neighbors $v_j$, $v_{j'}$, and $v_{j''}$.
The superposed event sequence generates an event whenever one of the three individual processes does.
Note that edges ($v_i$, $v_j$), ($v_i$, $v_{j'}$), and ($v_i$, $v_{j''}$) generally carry different numbers of events in a given time window despite the rate of the processes (i.e., the infection rate, $\beta$) being the same. 
This is due to the stochastic nature of Poisson processes.
}
\label{fig:superposition 3}
\end{figure}

In both cases, we are replacing three instances of the probability density of the time to the next infection event, each given by $\beta e^{- \beta \tau}$, by a single probability density $3\beta e^{-3\beta \tau}$.
Representing the three infectious nodes by one infectious supernode, i.e., one reaction channel, with 3 times the infection rate is equivalent to superposing the three Poisson processes into one. 
Figure~\ref{fig:superposition 3} illustrates this superposition, showing the putative event times generated by each Poisson process as well as those generated by their superposition.
The superposition theorem dictates that the superposition is a Poisson process with a rate of $3\beta$. 
This in particular means that we can draw the waiting time $\tau$ until the first of the events generated by all the three Poisson processes happens (shown by the double-headed arrow in Fig.~\ref{fig:superposition 3}) directly from the exponential distribution $\psi(\tau) = 3\beta e^{-3\beta \tau}$. 
Note that Poisson processes are defined as generating events indefinitely, and for illustrative purposes we show multiple events in Figure~\ref{fig:superposition 3}. 
However, in the SIR model only the first event in the superposed process will take place in practice. For example, once the event changes the state of $v_i$ from S to I, it cannot be infected anymore, and therefore none of the three infection processes can generate any more events.

Let us consider again the snapshot of the SIR dynamics shown in Figure~\ref{fig:6 node SIR}, but this time we consider all the possible infection and recovery events. 
We can represent all the possible events that may occur by four reaction channels (i.e., Poisson processes). One channel represents the infection of the node $v_i$ by any of its neighbors, which happens at a rate $3\beta$. We refer to this reaction channel as the first reaction channel. The three other channels each represent the recovery process of one of the infectious nodes. We refer to these three reaction channels as the second to the fourth reaction channels. 
We can use the same approach as above to obtain the probability density for the waiting time until the first event generated by any of the channels.
However, to completely describe the dynamics, it is not sufficient to know when the next event happens.
We also need to know which channel generates the event. Precisely speaking, we need to know the probability $\Pi_i$ that it is the $i$th reaction channel that generates the event.
Using the definition of conditional probability, we obtain
\begin{equation}
    \Pi_i = \frac{\{\text{probability that an event in the } i\text{th reaction channel occurs}\}} {\{\text{probability\ that\ an event in any reaction channel } j\in\{1,2,3,4\} \text{ occurs}\}} .
    \label{eq:Pi_i def}
\end{equation}
In a discrete-time description, the numerator in Eq.~\eqref{eq:Pi_i def} is simply $\lambda_i\delta t$, where $\lambda_1=3\beta$ and $\lambda_2 = \lambda_3 = \lambda_4 = \mu$ are the rates of the reaction channels.
The denominator is equal to $1 - \prod_{j=1}^4 (1 - \lambda_j\delta t)$, which in the limit of small $\delta t$ can be Taylor expanded to $\sum_{j=1}^4 \lambda_j \delta t = 3(\beta + \mu)\delta t$. 
Thus, the probability that the $i$th reaction channel has generated an event that has taken place is
\begin{equation}
    \Pi_i = \frac{\lambda_i}{\sum_{i=1}^4 \lambda_j} ,
\end{equation}
i.e., $\Pi_i$ is simply proportional to the rate $\lambda_i$.

The same result holds true for general superpositions of Poisson processes (see Box~\ref{box:Poisson_process}).

\subsection{\label{sub:ODE}Ignoring stochasticity --- differential equation approach}

We have introduced the types of models we are interested in and have explored their basic mathematical properties. We now turn our attention to the problem of how we can solve such models in practice.
We consider again the SIR model.
One simple strategy to solve it is to forget about the true stochastic nature of infection and recovery and approximate the processes as being deterministic.
In this approach, we only track the dynamics of the mean numbers of susceptible, infectious, and recovered individuals. Such deterministic dynamics is described by a system of ordinary differential equations (ODEs). The ODE version of the SIR model has a longer history than the stochastic one, dating back to the seminal work by Kermack and McKendrick in the 1920s \citep{Kermack1927ProcRSocLondA}. For the basic SIR model described above, the corresponding ODEs are given by
\begin{align}
\frac{\text{d}\rho_{\text{S}}}{\text{d}t} =& -\beta \rho_{\text{S}} \rho_{\text{I}},
\label{eq:dS/dt}\\
\frac{\text{d}\rho_{\text{I}}}{\text{d}t} =& \ \beta \rho_{\text{S}} \rho_{\text{I}} - \mu \rho_{\text{I}},
\label{eq:dI/dt}\\
\frac{\text{d}\rho_{\text{R}}}{\text{d}t} =& \ \mu \rho_{\text{I}} ,
\label{eq:dR/dt}
\end{align}
where $\rho_{\text{S}} = N_S/N$, $\rho_{\text{I}} = N_I/N$, and $\rho_{\text{R}} = N_R/N$ are the fraction of S, I, and R individuals, respectively. The $\beta \rho_{\text{S}} \rho_{\text{I}}$ terms in Eqs.~\eqref{eq:dS/dt} and \eqref{eq:dI/dt} represent infection events, through which the number of S individuals decreases and the number of I individuals increases by the same amount. 
The $\mu \rho_{\text{I}}$ terms in Eqs.~\eqref{eq:dI/dt} and \eqref{eq:dR/dt} represent recovery events. 

One can solve Eqs.~\eqref{eq:dS/dt}, \eqref{eq:dI/dt}, and \eqref{eq:dR/dt} either analytically, to some extent, or numerically using an ODE solver implemented in various programming languages. 
Suppose that we have coded up Eqs.~\eqref{eq:dS/dt}, \eqref{eq:dI/dt}, and \eqref{eq:dR/dt} into an ODE solver to simulate the infection dynamics (such as time courses of $\rho_{\rm I}$) for various values of $\beta$ and $\mu$. Does the result give us complete understanding of the original stochastic SIR model? The answer is negative \citep{Mollison1994JRStatSocSerA}, at least for the following reasons.

First, the ODE is not a good approximation when $N$ is small. In Eqs.~\eqref{eq:dS/dt}, \eqref{eq:dI/dt}, and \eqref{eq:dR/dt}, the variables are the fraction of individuals in each state. 
For example, $\rho_{\rm I} = N_{\rm I}/N$. 
The ODE description assumes that $\rho_{\rm I}$ can take any real value between 0 and 1 and that $\rho_{\rm I}$ changes continuously as time goes by. 
However, in reality $\rho_{\rm I}$ is quantized, so it can take only the values $0$, $1/N$, $2/N$, $\ldots$, $(N-1)/N$, and $1$, and it changes in steps of $1/N$ (e.g., it changes from $3/N$ to $4/N$ discontinuously). 
This discrete nature does not typically cause serious problems when $N$ is large, in which case $\rho_{\rm I}$ is close to being continuous. 
By contrast, the ODE model is not accurate when $N$ is small due to the quantization effect.
(Note that the ODE approach is problematic in some cases even when $N$ is large, i.e. near critical points as we discuss below.)

Second, even if $N$ is large, the actual dynamic changes in $\rho_{\rm I}$, for example, are not close to what the ODEs describe when the number of infectious individuals is small. For example, if $\rho_{\rm I} = 2/N$, there are two infectious individuals. If one of them recovers, $\rho_{\rm I}$ changes to $1/N$, and this is a 50\% decrease in $\rho_{\rm I}$. The ODE assumes that $\rho_{\rm I}$ changes continuously and is not ready to describe such a change. As another example, suppose that we initially set $\rho_{\rm S} = (N-1)/N$, $\rho_{\rm I} = 1/N$, and $\rho_{\rm R} = 0$. In other words, there is a single infectious seed, and all the other individuals are initially susceptible. In fact, the theory of the ODE version of the SIR model shows that $\rho_{\rm I}$ increases deterministically, at least initially, if $\beta > \mu$, corresponding to the situation in which an outbreak of infection happens. However, in the stochastic SIR model, the only initially infectious individual may recover before it infects anybody even if $\beta > \mu$. When this situation occurs, the dynamics terminates once the initially infectious individual has recovered, and no outbreak is observed.
Although the probability with which this situation occurs decreases as $\beta/\mu$ increases, it is still not negligibly small for many large $\beta/\mu$ values.  This is inconsistent with the prediction of the ODE model. 
It should be noted that another common way to initialize the system is to start with a small fraction of infectious individuals, regardless of $N$.
In this case, if we start the stochastic SIR dynamics in a large well-mixed population and, for example, with 10\% initially infectious individuals, the ODE version is sufficiently accurate at describing the stochastic SIR dynamics.

Third, ODEs are not accurate at describing the counterpart stochastic dynamics when the system is close to a so-called critical point. 
For example, in the SIR model, given the value of the infection rate (i.e., $\mu$), there is a value of the infection rate called the epidemic threshold\index{epidemic threshold}, which we denote by $\beta_{\rm c}$. 
For $\beta < \beta_{\rm c}$, only a small number of the individuals will be infected (i.e., the final epidemic size is of $O(1)$). 
For $\beta > \beta_{\rm c}$, the final epidemic size is large (i.e., $O(N)$) with a positive probability. 
In analogy with statistical physics, $\beta_{\rm c}$ is termed a critical point of the SIR model. 
Near criticality the fluctuations of $\rho_S$, $\rho_I$, and $\rho_R$ are not negligible compared to their mean values, even for large $N$, and the ODE generally fails.

Fourth, ODEs are not accurate when dynamics are mainly driven by stochasticity rather than by the deterministic terms on the right-hand sides of the ODEs. This situation may happen even far from criticality or in a model that does not show critical dynamics. 
The voter model\index{voter model} (see Section~\ref{sub:voter model} for details) is such a case. In its simplest version, the voter model describes the tug-of-war between two equally strong opinions in a population of individuals. Because the two opinions are equally strong, the ODE version of the voter model predicts that the fraction of individuals supporting opinion A (and that of individuals supporting opinion B) does not vary over time, i.e., one obtains ${\rm d}\rho_{\rm A}/{\rm d}t = {\rm d}\rho_{\rm B}/{\rm d}t = 0$, where $\rho_{\rm A}$ and $\rho_{\rm B}$ are the fraction of individuals supporting opinion A and B, respectively. However, in fact, the opinion of the individuals flips here and there in the population due to stochasticity, and it either increases or decreases over time.

To summarize, when stochasticity manifests itself, the approximation of the original stochastic dynamics by an ODE model is not accurate.

\subsection{\label{sub:rejection sampling}Rejection sampling algorithm}

The most intuitive method to simulate the stochastic SIR model, while accounting for the stochastic nature of the model, is probably to discretize time and simulate the dynamics by testing whether each possible event takes place in each step.
This is called the rejection sampling\index{rejection sampling} algorithm. 
Let us consider the stochastic SIR model on a small network composed of $N=6$ nodes as shown in Fig.~\ref{fig:6 node SIR} to explain the procedure.

Assume that the state of the network (i.e., the states of the individual nodes) is as shown in Fig.~\ref{fig:6 node SIR}(a) at time $t$; three nodes are susceptible, two nodes are infectious, and the other node is recovered. In the next time step, which accounts for a time length of $\Delta t$ and corresponds to the time interval $[t, t+\Delta t)$, an infection event may happen in five ways: $v_2$ infects $v_1$, $v_2$ infects $v_3$, $v_2$ infects $v_5$, $v_4$ infects $v_1$, and $v_4$ infects $v_5$.  
Recovery events may happen for $v_2$ and $v_4$. 
Therefore, there are seven possible events in total, some of which may simultaneously happen in the next time step.

With the rejection sampling method, we sequentially (called asynchronous updating) or simultaneously (called synchronous updating) check whether or not each of these events happens in each time step of length $\Delta t$. Note that it is not possible to go to the limit of $\Delta t \to 0$ in rejection sampling. In our example, $v_2$ infects $v_1$ with probability $\beta \Delta t$ in a time step. With probability $1 - \beta \Delta t$, nothing occurs along this edge. 
In practice, to determine whether the event takes place or not, we draw a random number $u$ uniformly from $[0, 1)$. If $u \ge \beta \Delta t$, the algorithm rejects the proposed infection event (thus the name rejection sampling). 
If $u < \beta \Delta t$, we let the infection occur. 
Then, under asynchronous updating, we change the state of $v_1$ from S to I and update the set of possible events accordingly right away, and then proceed to check the occurrence of each of the remaining possible events in turn. 
Under synchronous updating, we first check whether each of the possible state changes takes place and note down the changes that take place. We then implement all the noted changes simultaneously. 
Regardless of whether we use asynchronous or synchronous updating, the infection event occurs with probability $\beta \Delta t$. 

If $v_4$ recovers, which occurs with probability $\mu \Delta t$, and none of the other six possible events occurs in the same time step, the status of the network at time $t + \Delta t$ is as shown in Fig.~\ref{fig:6 node SIR}(b).
Then, in the next time step, $v_1$ may get infected, $v_2$ may recover, $v_3$ may get infected, and $v_5$ may get infected, which occurs with probabilities $\beta \Delta t$, $\mu \Delta t$, $\beta \Delta t$, and $\beta \Delta t$, respectively. In this manner, we carry forward the simulation by discrete steps until no infectious nodes are left.

There are several caveats to this approach. First, the asynchronous and the synchronous updating schemes of the same stochastic dynamics model may lead to systematically different results \citep{Huberman1993PNAS,Cornforth2005PhysicaD,Greil2005PhysRevLett}.

Second, one should set $\Delta t$ such that both $\beta \Delta t < 1$ and $\mu \Delta t < 1$ always hold true. In fact, the discrete-time interpretation of the original model is justified only when $\Delta t$ is small enough to yield $\beta \Delta t \ll 1$ and $\mu \Delta t \ll 1$.

Third, in the case of asynchronous updating, the order of checking the events is arbitrary, but it affects the outcome, particularly if $\Delta t$ is not tiny. For example, we can sequentially check whether each of the five infection events occurs and then whether each of the two recovery events occurs, completing one time step. One can alternatively check the recovery events first and then the infection events. If we do so and $v_4$ recovers in the time step, then it is no longer possible that $v_4$ infects $v_1$ or $v_5$ in the same time step because $v_4$ has recovered. If the infection events were checked before the recovery events, it is possible that $v_4$ infects $v_1$ or $v_5$ before $v_4$ recovers in the same time step. 

Fourth, some of the seven types of event cannot occur simultaneously in a single time step regardless of whether the updating is asynchronous or synchronous, and regardless of the order in which we check the events in the asynchronous updating. For example, if $v_2$ has infected $v_1$, then $v_4$ cannot infect $v_1$ in the same time step (or anytime later) and vice versa. In fact, from the susceptible node $v_1$'s point of view, it does not matter which infectious neighbor, either $v_2$ or $v_4$, infects $v_1$. What is primarily important is whether $v_1$ gets infected or not in the given time step, whereas one wants to know who infected whom in some tasks such as contact tracing. 

A useful method to mitigate the effect of overlapping events of this type 
is to take a node-centric view. The superposition theorem implies that $v_1$ will get infected according to a Poisson process with rate $2\beta$ because it has two infectious neighbors (Section~\ref{sub:superposition}). By exploiting this observation, let us redefine the list of possible events at time $t$. 
The node $v_1$ will get infected with probability $2\beta \Delta t$ (and will not get infected with probability $1- 2\beta \Delta t$). Nodes $v_3$ and $v_5$ will get infected with probabilities $\beta \Delta t$ and $2\beta \Delta t$, respectively. As before, $v_2$ and $v_4$ recover with probability $\mu \Delta t$ each. In this manner, we have reduced the number of possible events from seven to five. We are usually interested in simulating such stochastic processes in much larger networks or populations, where nodes tend to have a degree larger than in the network shown in Fig.~\ref{fig:6 node SIR}. For example, if a node $v_i$ has 50 infected neighbors, implementing the rejection sampling using the probability that $v_i$ gets infected, $50\beta \Delta t$, rather than checking if $v_i$ gets infected with probability $\beta \Delta t$ along each edge that $v_i$ has with an infectious neighbor, will confer a fiftyfold speed up of the algorithm.

Rejection sampling is a widely used method, particularly in research communities where continuous-time stochastic process thinking does not prevail. In a related vein, many people are confused by being told that the infection and recovery rates $\beta$ and $\mu$ can exceed 1. They are accustomed to think in discrete time such that they are not trained to distinguish between the rate and probability. They \textit{are} different; simply put, the rate is for continuous time, and the probability is for discrete time. 
Here we advocate that we should not use the discrete-time versions in general, despite their simplicity and their appeal to our intuition, for the following reasons (see \cite{Gomez2011PhysRevE} and \cite{Fennell2016PhysRevE} for similar arguments).

First, the use of a small $\Delta t$, which is necessary to assure an accurate approximation of the actual continuous-time stochastic process, implies a large computation time. If the duration of time that one run of simulation needs is $T$, one needs $n = T/\Delta t$ discrete time steps, which is large when $\Delta t$ is small. How small should $\Delta t$ be? It is difficult to say. If you run simulations with a choice of a small $\Delta t$ and calculate statistics of your interest or draw a figure for your report or paper, a good practice is to try the same thing after halving $\Delta t$. If the results do not noticeably change, then your original choice of $\Delta t$ is probably small enough for your purpose. Otherwise, you need to make $\Delta t$ smaller. It takes time to carry out such a check just to determine an appropriate  $\Delta t$ value. Many people skip it. The Gillespie algorithms do not rely on a discrete-time approximation and are also typically faster than rejection sampling with a reasonably small $\Delta t$ value.

Second, no matter how small $\Delta t$ is, the results of rejection sampling are only approximate. This is because it is exact only in the limit $\Delta t \to 0$. By contrast, the Gillespie algorithms are always exact.

Proponents of the rejection sampling method may say that they want to define the model (such as the SIR model) in discrete time and run it, rather than to consider the continuous-time version of the model and worry about the choice of $\Delta t$ or the accuracy of the rejection sampling. We recommend against this as well. In the SIR model in discrete time, any infectious individual $v_i$ infects a neighboring susceptible individual $v_j$ with probability $\beta'$, and each infectious individual recovers with probability $\mu'$. Then, there are at least two problems related to this.
First, the order of the events affects dynamics in the case of asynchronous updating. 
Second, and more importantly, we do not know how to change the time resolution of the simulation when we need to. For example, if one simulation step currently corresponds to one hour, one may want to now simulate the same model with some more temporal detail such that one step corresponds to ten minutes. Because the physical time is now one sixth of the original one, should we multiply $\beta'$ and $\mu'$ by $1/6$ and do the same simulations? The answer is no. If the original time step corresponds to $\Delta t = 1$, which is often implicit, then probability that an infectious individual recovers in the continuous-time stochastic SIR model within time $\Delta t (=1)$ is $1-e^{-\mu \Delta t} = 1 - e^{-\mu}$, which we equate with $\mu'$.
Then, if we scale the time $c$ times (e.g., $c=1/6$), the probability that the recovery occurs in a new single time step is $1-e^{-\mu c\Delta t} = 1-e^{-c\mu}$, which is not equal to $c \mu'$. For example, with $\mu = 1$ and $c=1/6$, one obtains $1-e^{-c\mu} \approx 0.154$, whereas $c \mu' \approx 0.105$.

\section{Classic Gillespie algorithms}
\label{sec:classic_Gillespie}

The Gillespie algorithms overcome the two major drawbacks of the rejection sampling algorithm that we discussed near the end of Section~\ref{sub:rejection sampling}; namely, its computational inefficiency and its reliance on a discrete-time approximation of the dynamics.
The Gillespie algorithms are typically faster than rejection sampling, and they are stochastically exact (i.e., they generate exact realizations of the simulated jump processes). In this section, we present the two basic Gillespie algorithms for simulating coupled Poisson processes, largely in their original forms proposed by Daniel Gillespie: the \emph{first reaction method} and the \emph{direct method}. 
The two methods are mathematically equivalent \citep{Gillespie1976JComputPhys}.
Nevertheless, the two algorithms have pros and cons in terms of ease of implementation and computational efficiency. 
Because these two factors depend on the model to be simulated, which algorithm one should select depends on the model as well as personal preference.

We first provide a brief history of the Gillespie algorithms (Section~\ref{sub:history}). We then introduce the first reaction method (Section~\ref{sub:first-reaction}) because it is conceptually the simpler of the two, followed by the direct method (Section~\ref{sub:direct}), which builds on elements of the first reaction method but makes use of the superposition theorem (Box~\ref{box:Poisson_process}) to directly draw the waiting time between events.
We end this section with example implementations in Python of the stochastic SIR dynamics.

\subsection{\label{sub:history}Brief history}

As the name suggests, the Gillespie algorithms are ascribed to American physicist Daniel Thomas Gillespie \citep{Gillespie1976JComputPhys,Gillespie1977JPhysChem}. 
He originally proposed them in 1976 for simulating stochastic chemical reaction systems, and they have seen many applications as well as further algorithmic developments in this field. Nevertheless, the algorithms only rely on general properties of Poisson processes and not on any particular properties of chemical reactions. Therefore, the applicability of the Gillespie algorithms is much wider than to chemical reaction systems. In fact, they have been extensively used in simulations of multiagent systems both in unstructured populations and on networks. The only assumptions are that the system undergoes changes via sequences of discrete events (e.g., somebody infects somebody, somebody changes its internal state from a low-activity state to a high-activity state) and with a rate that stays constant in-between events.
Nevertheless, the latter assumption has been relaxed in recent extensions of the algorithms; we review these in Section~\ref{sec:renewal_and_temporal}.

There were precursors to the Gillespie algorithms.
The American mathematician Joseph Leo Doob developed in his 1942 and 1945 papers the mathematical foundations of continuous-time Markov chains which underlie the Gillespie algorithms \citep{Doob1942TransAmMathSoc,Doob1945TransAmMathSoc}. In the second of the two papers, he effectively proposed the direct method, although the focus of the paper was mathematical theory and he did not propose a computational implementation \cite[pp.~465--466]{Doob1945TransAmMathSoc}.
Due to this, the algorithm is sometimes called the Doob-Gillespie algorithm\index{Doob-Gillespie algorithm}. David George Kendall, who is famous for Kendall's notation in queuing theory\footnote{Not to be confused with another British statistician of the time, Maurice Kendall, famous for Kendall tau rank correlation. Both Kendalls were awarded the honor of the Royal Statistical Society, the Guy Medal in Gold.}, 
implemented an equivalent of the direct method to simulate a stochastic birth-death process\index{birth-death process} on a computer as early as in 1950 \citep{Kendall1950JRStatSocSerB}. In 1953, Maurice Stevenson Bartlett, a British statistician, simulated the SIR model in a well-mixed population\index{well-mixed population} (i.e., every pair of individuals is directly connected to each other) using the direct method
\citep{Bartlett1953JRStatSocSerCApplStat}.

Independently of Gillespie, Alfred B. Bortz and colleagues also proposed the same algorithm as the direct method to simulate stochastic dynamics of Ising spin systems in statistical physics in 1975 \citep{Bortz1975JComputPhys}. Therefore, the direct method is also called the \emph{Bortz-Kalos-Lebowitz algorithm}\index{Bortz-Kalos-Lebowitz algorithm} (or the \emph{$n$-fold way}\index{$n$-fold way} following the naming in their paper, and also \emph{rejection-free kinetic Monte Carlo}\index{rejection-free kinetic Monte Carlo} and the \emph{residence-time algorithm}\index{residence-time algorithm}).
An even earlier paper published in 1966 in the same field proposed almost the same algorithm, with the only difference being that the waiting time between events was assumed to take a deterministic value rather than being stochastic as in the Gillespie algorithms \citep{Young1966ProcPhysSoc}.

\subsection{\label{sub:first-reaction}First reaction method}

\begin{figure}
\centering
\includegraphics[width=\textwidth]{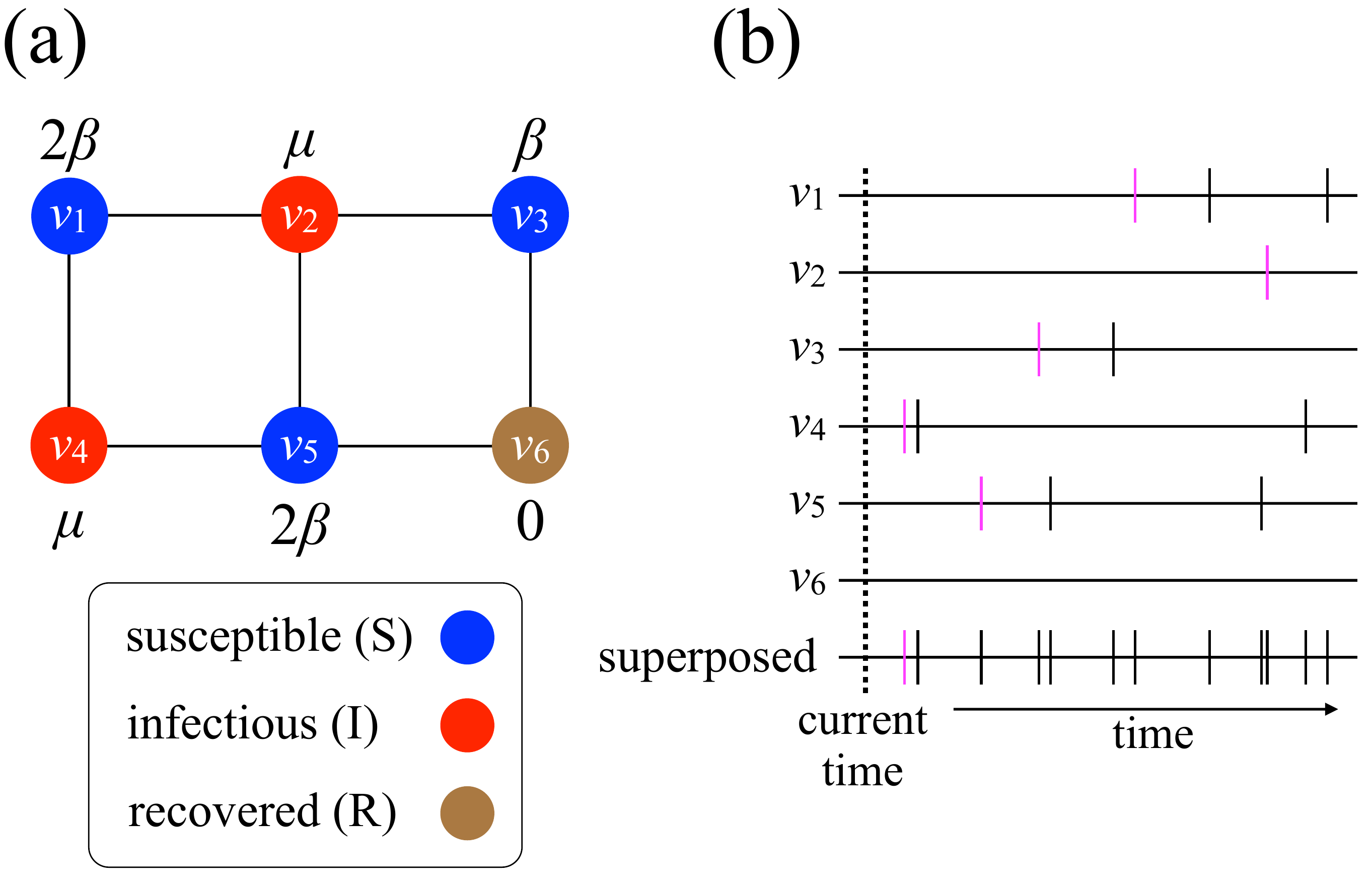}
\caption{Determination of the time to the next event in the SIR model using the Gillespie algorithms. 
(a) The current state of the system. The rate with which each node changes its state is shown next to the node. 
(b) The putative events generated by the Poisson process corresponding to each node with the first event of each process shown in magenta. There is no event on the timeline of $v_6$. This is because $v_6$ is in the recovered state and no longer undergoes any state change; therefore the event rate of the Poisson process associated with $v_6$ is equal to 0.
The first reaction method generates a putative waiting time until the first event for each process and selects the smallest one.
The direct method directly generates the waiting time until the first event in the superposed process (bottom of panel (b)).
}
\label{fig:6 node SIR Gillespie}
\end{figure}

To introduce the first reaction method, we consider our earlier example of the SIR model on a 6-node network (see Fig.~\ref{fig:6 node SIR}(a)).
Here two types of events may happen next: either an susceptible node becomes infected (S $\to$ I), or an infectious node recovers (I $\to$ R). 
The rates at which each node experiences a state transition are shown in Fig.~\ref{fig:6 node SIR Gillespie}(a), which replicates Fig.~\ref{fig:6 node SIR}(a). For example, $v_1$ is twice as likely to be infected next as $v_3$ is because $v_1$ has two infectious neighbors whereas $v_3$ has one infectious neighbor. 
Each event obeys a separate Poisson process. 
Therefore, let us first generate hypothetical event sequences according to each Poisson process with their respective rates (see Fig.~\ref{fig:6 node SIR Gillespie}(b)). In fact, we need to use at most only the first event in each sequence (shown in magenta in Fig.~\ref{fig:6 node SIR Gillespie}(b)). For example, in the event sequence for $v_1$, the first event may be used, in which case $v_1$ will be infected. Once $v_1$ is infected, the subsequent events on the same event sequence will be discarded because $v_1$ will never be infected again. If $v_1$ is infected, it will undergo another type of event, which is recovery. However, we cannot reuse the second or any subsequent events in the same sequence as the recovery event because the recovery occurs with rate $\mu$, which is different from the rate $2 \beta$ with which we have generated the event sequence for $v_1$. A lesson we learn from this example is that we should not prepare many possible event times beforehand because most of them would be discarded. 

Given this reasoning, Gillespie
proposed the first reaction method in one of his two original papers on the Gillespie algorithms \citep{Gillespie1976JComputPhys}. 
The idea of the first reaction method is to generate only the first putative event time (shown in magenta in Fig.~\ref{fig:6 node SIR Gillespie}(b)) for each node. 
These putative times correspond to when each node would experience a next event if nothing else happens in the rest of the system.
We do not generate an event time for $v_6$ because this node is recovered, so it will never undergo any event. 
We then figure out which event occurs first. 
(In the example in Fig.~\ref{fig:6 node SIR Gillespie}(a), it is node $v_4$ that will change its state, and the state change is from I to R.)

To generate a putative waiting time for each node $v_1$, $\ldots$, $v_5$ in practice, we use a general technique called \emph{inverse sampling}\index{inverse sampling}, which proceeds as follows. For example, the time to the first event for $v_1$ obeys the exponential distribution $\psi_1(\tau) \equiv 2\beta e^{-2\beta \tau}$. The probability that the time to the next event is larger than $\tau$, called the survival probability\index{survival probability} (also called the survival function and the complementary cumulative distribution function), is given by
\begin{equation}
\Psi_1(\tau) \equiv
\int_{\tau}^{\infty} \psi_1(\tau') \text{d}\tau'
=
\int_{\tau}^{\infty} 2\beta e^{-2\beta \tau'} \text{d}\tau' = e^{-2\beta \tau} .
\label{eq:survival probability}
\end{equation}
By definition, $\Psi_1(\tau)$ is a probability, so $0 <  \Psi_1(\tau) \le 1$. We have excluded $\Psi_1(\tau)=0$ because it happens only in the limit $\tau \to \infty$. This exclusion does not cause any problem in the following development.

Then, we draw a number $u$ from an unbiased random number generator that generates random numbers uniformly in the interval $(0,1]$. The generated number $u$ is called a uniform $(0,1]$ random variate. Any practical programming language has a function to generate uniform (pseudo) random variates. 
We do however advise against using a programming language's standard pseudo-random number generator, which is typically of poor quality. You should instead use one from a scientific programming library or code it yourself. We will discuss some good practices for pseudo-random number generation in Section~\ref{sub:PRNG}.
Given $u$, we then generate $\tau$ from the implicit equation 
\begin{equation}
u = \Psi_1(\tau) .
\label{eq:u=P(tau)}
\end{equation}
For example, if we draw $u=0.3$, we can find the unique $\tau$ value that satisfies Eq.~\eqref{eq:u=P(tau)}. This method for generating random variates obeying a given distribution is called inverse (transform) sampling \citep{VonNeumann1951ApplMathSer}.
One can use this method for a general probability density function, $\psi(\tau)$, as long as one can calculate its survival function, $\Psi(\tau) = \int_{\tau}^{\infty} \psi(\tau') \text{d}\tau'$, like in Eq.~\eqref{eq:survival probability}. In the present case, by combining Eqs.~\eqref{eq:survival probability} and \eqref{eq:u=P(tau)}, we obtain
\begin{equation}
u = e^{-2\beta \tau} ,
\end{equation}
which leads to
\begin{equation}
\tau = - \frac{\ln u}{2\beta} .
\label{eq:tau from u}
\end{equation}
Note that $\tau > 0$ because $\ln u < 0$. Equation~\eqref{eq:tau from u} is reasonable in the sense that a large event rate $2\beta$ will yield a small waiting time $\tau$ on average.

In the same manner, one can generate the putative times to the next event for $v_1$, $\ldots$, $v_5$, denoted by $\tau^{\rm put}_1$, $\ldots$, $\tau^{\rm put}_5$, using five uniform random variates. If the realized $\tau^{\rm put}_1$, $\ldots$, $\tau^{\rm put}_5$ values are as shown in Fig.~\ref{fig:6 node SIR Gillespie}(b), we conclude that node $v_4$ recovers next. Then, we change the state of $v_4$ from I to R and advance the clock by time $\tau = \tau^{\rm put}_4$. Once $v_4$ recovers, the configuration of the six nodes will be the one given in Fig.~\ref{fig:6 node SIR}(b). 
We then repeat the same procedure to find the next event given the updated set of processes corresponding to the nodes' states after the event (i.e., we now have three infection processes with rate $\beta$ and one recovery process with rate $\mu$), and so on.

Note that in our example, the event rate changed for $v_1$, $v_4$, and $v_5$, while it remained unchanged for $v_2$ and $v_3$. 
Therefore, we do not need to generate entirely new putative waiting times for $v_2$ and $v_3$. We just have to update $\tau^{\rm put}_2$ and $\tau^{\rm put}_3$ as $\tau^{\rm put}_2 \to \tau^{\rm put}_2 - \tau$ and $\tau^{\rm put}_3 \to \tau^{\rm put}_3 - \tau$,
respectively, to account for the time $\tau$ that has elapsed.

In this manner, we can reuse $\tau^{\rm put}_2$ and $\tau^{\rm put}_3$ (by subtracting $\tau$) and avoid having to generate new pseudo-random numbers for redrawing $\tau^{\rm put}_2$ and $\tau^{\rm put}_3$. 
The effect of this frugality becomes important for larger systems where an event generally affects only a small fraction of the processes.
The so-called \emph{next reaction method} \citep{GibsonBruck2000JPhysChemA} exploits this idea to improve the computational efficiency of the first reaction method. 
Although the classic first reaction method did not make use of this trick, we include it here because it is simple to implement.

Let us go back to our example. For $v_4$, we no longer need to generate the time to the next event because $v_4$ is now in the R state. For $v_1$ and $v_5$, we need to discard $\tau^{\rm put}_1$ and $\tau^{\rm put}_5$ because they were generated under the assumption that the event rate was $2\beta$. Now, we need to redraw the time to the next event for the two nodes according to the new distribution $\psi_1(\tau) = \psi_5(\tau) = \beta e^{-\beta \tau}$. For example, we reset $\tau^{\rm put}_1 = - \ln u' / \beta$, where $u'$ is a new uniform $(0,1]$ random variate. Although the time $\tau$ has passed to transit from the status of the network shown in Fig.~\ref{fig:6 node SIR}(a) to that shown in Fig.~\ref{fig:6 node SIR}(b), we do not need to take into account the elapsed time (i.e., $\tau$) when generating the new $\tau^{\rm put}_1$ and $\tau^{\rm put}_5$ values. 
This is due to the memoryless property of Poisson processes (see Box~\ref{box:Poisson_process}), i.e., what happened in the past, such as how much time has passed to realize the state transition of $v_4$, is irrelevant. 

The first reaction method in its general form is given in Box~\ref{box:first-reaction_classic}.

\begin{figure}
\begin{boxedtext}{} 
\captionof{floatbox}{\textbf{Gillespie's first reaction method.}}
\label{box:first-reaction_classic}


\begin{enumerate}[start=0]
    \item Initialization: 
      \begin{enumerate}
        \item Define the initial state of system, and set $t = 0$.
        \item Calculate the rate $\lambda_j$ for each reaction channel $j \in \{1, \ldots, M\}$.
        \item Draw $M$ random variates $u_j$ from a uniform distribution on $(0,1]$.
        \item Generate a putative waiting time $\tau^{\rm put}_j = -\ln u_j /\lambda_j$ for each reaction channel. 
      \end{enumerate}
    \item Select the reaction channel $i$ with the smallest $\tau^{\rm put}_i$, and set $\tau = \tau^{\rm put}_i$.
    \item Perform the event on reaction channel $i$.
    \item Advance the time according to $t \to t + \tau$. 
    \item Update $\lambda_i$ and all other $\lambda_j$ that are affected by the event produced.
    \item Update putative waiting times:
      \begin{enumerate}
          \item Draw new waiting times for reaction channel $i$ and for the other reaction channels $j$ whose $\lambda_j$ has changed, according to $\tau^{\rm put}_j = -\ln u_j /\lambda_j$ with $u_j$ being newly drawn from a uniform distribution on $(0,1]$.
          \item Update the waiting times for all reaction channels $j$ that have not been affected by the last event according to $\tau^{\rm put}_j \to \tau^{\rm put}_j - \tau$.
      \end{enumerate}
    \item Return to Step 1.
\end{enumerate}

\end{boxedtext}
\end{figure}

\subsection{\label{sub:direct}Direct method}

The direct method exploits the superposition theorem to directly generate the waiting times between successive events in the full system of coupled Poisson processes. For expository purposes, we hypothetically generate an event sequence on each node with the respective rate, although only at most the first event in each sequence will be used. 
We then superpose the nodal event sequences into one Poisson process (see Fig.~\ref{fig:6 node SIR Gillespie}(b)). Owing to the superposition theorem (Box~\ref{box:Poisson_process}, Eq.~\eqref{eq:superposition-Poisson}), the superposed event sequence is itself a Poisson process with a rate $\Lambda$ that is equal to the sum of the individual rates, i.e., $\Lambda = 2\beta + \mu +\beta + \mu + 2 \beta + 0 = 5\beta + 2\mu$. 

Therefore, we can generate the time to the next event in the entire population using the inverse sampling method, which we introduced in Section~\ref{sub:first-reaction}, 
according to $\tau = -\ln u / \Lambda$, where $u$ is a uniform random variate in the interval $(0,1]$.
We now know the time to the next event but not which node (or more generally, which one individual Poisson process) is responsible for the next event. This is because the superposition lacks information about the individual constituent event sequences. 

\begin{figure}
\centering
\includegraphics[width=\textwidth]{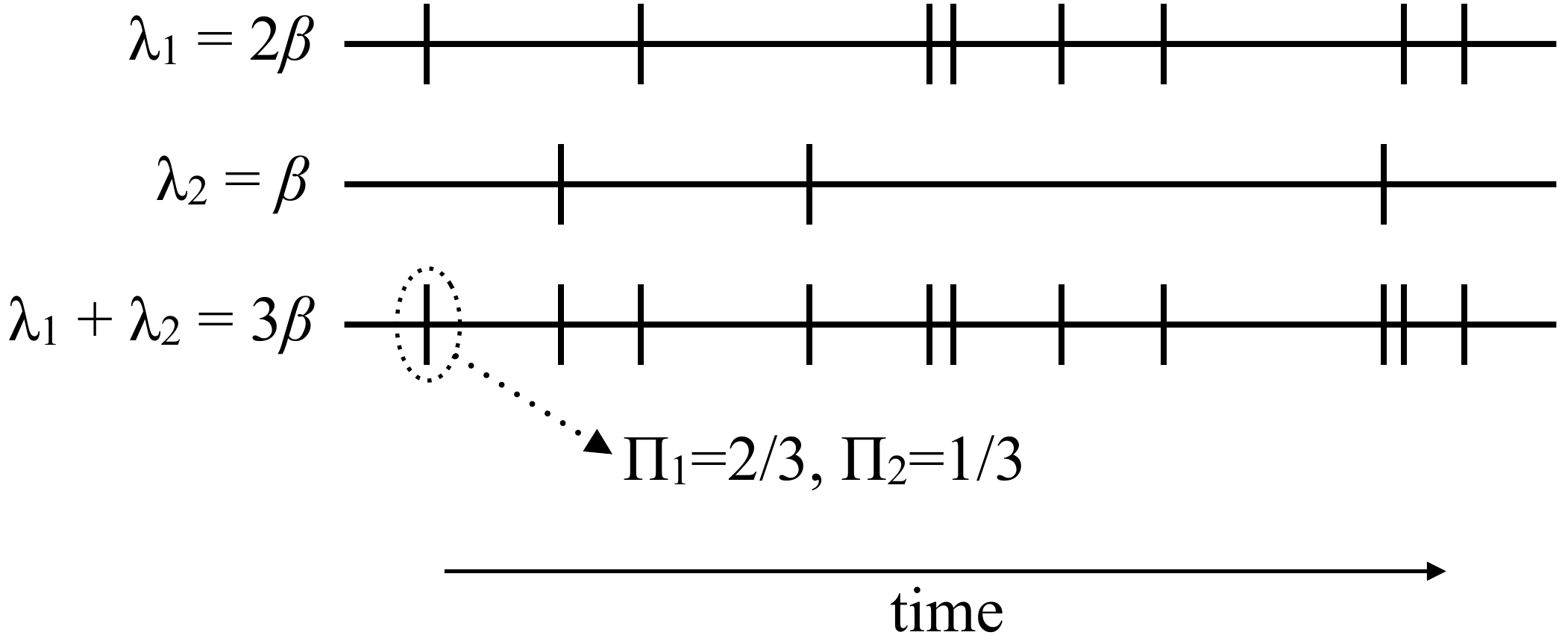}
\caption{Superposition of Poisson processes and how to determine which component Poisson processes contribute to an event in the superposed event sequence. We consider $M=2$ Poisson processes, one with rate $\lambda_1 = 2 \beta$ and the other with $\lambda_2 = \beta$. The superposed Poisson process has rate $\lambda_1 + \lambda_2 = 3 \beta$. The next event in the superposed event sequence (shown in the dotted circle) belongs to process 1 with probability $\Pi_1 = \lambda_1/(\lambda_1 + \lambda_2) = 2/3$ and process 2 with probability $\Pi_2 = \lambda_2/(\lambda_1 + \lambda_2) = 1/3$.}
\label{fig:superposition of 2}
\end{figure}

We thus need to determine which node produces the event. 
The mathematical properties of Poisson processes guarantee that the probability that a given node produces the event is proportional to its event rate (see Box~\ref{box:Poisson_process}, Eq.~\eqref{eq:Pi_superposition_Poisson}). 
For example, in Fig.~\ref{fig:superposition of 2}, where the event rates of the two nodes $v'_1$ and $v'_2$ are $2\beta$ and $\beta$, respectively, each event in the superposed event sequence comes from $v'_1$ and $v'_2$ with probability $\Pi_1 = 2/3$ and $\Pi_2 = 1/3$, respectively. Therefore, the probability that the first event is generated by $v'_1$ is $2/3$. The probability it is generated by $v'_2$ is $1/3$. This is natural because the sequence for $v'_1$ has on average twice as many events as that for $v'_2$. 
In our example (see Fig.~\ref{fig:6 node SIR Gillespie}), the $v_i$ (with $i=1, \ldots, 5$) that produces the next event is drawn with probability $\Pi_i$, where $\Pi_1 = \Pi_5 = 2\beta / (5\beta + 2\mu)$, $\Pi_2 = \Pi_4 = \mu / (5\beta + 2\mu)$, and 
$\Pi_3 = \beta / (5\beta + 2\mu)$. 
We will explain computational methods for doing this later.

Assume that $v_4$ generates the next event and transitions from state I to state R.
Then, the new event rate for each node is as shown in Fig.~\ref{fig:6 node SIR}(b). 
We advance the clock by $\tau$ and go to the next step. 
Again, to determine the time to the following event, regardless of which node produces the event, we only need to consider the sum of the event rates of the six nodes, which is now given by $\Lambda = \beta + \mu + \beta + 0 + \beta + 0 = 3\beta + \mu$. Then, the time to the next event, which we again denote by $\tau$, is given by $\tau = - \ln u / (3\beta + \mu)$, where $u$ is a new uniform random variate. 
We draw another random number to determine which of the four eligible nodes, $v_1$, $v_2$, $v_3$, or $v_5$, produces the event and changes its state. 
(Note that $v_4$ and $v_6$ are recovered, so they cannot undergo a further state change.)
The four remaining nodes are each selected with probability $\Pi_1 = \Pi_3 = \Pi_5 = \beta/(3\beta + \mu)$ and $\Pi_2 = \mu/(3\beta + \mu)$. 

In this manner, we draw $\tau$, determine which node produces the event, implement the state change, advance the clock by $\tau$, and repeat. This is Gillespie's direct method.

\begin{figure}
\centering
\includegraphics[width=0.35\textwidth]{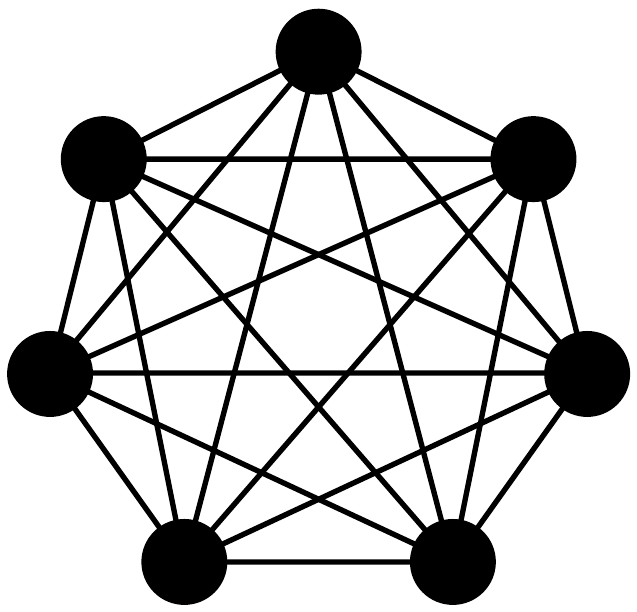}
\caption{Complete graph with $N=7$ nodes.}
\label{fig:complete}
\end{figure}

Let us take a look at another example, which is the SIR model in a population composed of $N$ individuals, in which everybody is directly connected to anybody else 
(called a well-mixed population\index{well-mixed population}; equivalent to a complete graph\index{complete graph}; see Fig.~\ref{fig:complete} for an example).
In contrast to the general network case, each individual in a well-mixed population is indistinguishable from the others. 
In the well-mixed population, it is not the case that, e.g., an individual has two neighbors while another has three neighbors (such as in Fig.~\ref{fig:6 node SIR}); they all have $N-1$ neighbors.
Recall that recovered individuals do not change their state again. 
Therefore, it suffices to consider $N_{\rm S} + N_{\rm I}$ event sequences and their superposition to determine the time to the next event and which individual will produce the next event.

However, in the well-mixed population, we can make the procedure more efficient.
Because everybody is alike, we do not need to keep track of the state of each individual. 

In the case of a network, we generally need to distinguish between different susceptible individuals. 
For example, in Fig.~\ref{fig:6 node SIR}(a), the susceptible nodes $v_1$ and $v_5$ are different because they have different sets of neighbors. Furthermore, $v_1$ has degree 2, while $v_5$ has degree 3.
Therefore, if $v_1$ gets infected and changes its state, it is not equivalent to $v_5$ getting infected. So, we must keep track of the state of each individual in a general network.
By contrast, in the well-mixed population, such a distinction is irrelevant. Everybody is adjacent to all the other $N-1$ nodes, and the number of infectious neighbors is the same for any susceptible individual, i.e., it is equal to $N_{\rm I}$. It does not matter which particular node gets infected in the next event. The only thing that matters for the SIR model in the well-mixed population is the number of susceptible, infectious, and recovered individuals, which is $N_{\rm S}$, $N_{\rm I}$, and $N_{\rm R}$, respectively.

It thus suffices to monitor these numbers.
If an infection event happens, then $N_{\rm S}$ decreases by one, and $N_{\rm I}$ increases by one. 
If an infectious individual recovers, then $N_{\rm I}$ decreases by one, and $N_{\rm R}$ increases by one. Because the number of individuals is preserved over time, it holds true that
\begin{equation}
N_{\rm S} + N_{\rm I} + N_{\rm R} = N
\end{equation}
at any time.
Because anybody is connected to everybody else, any susceptible individual has $N_{\rm I}$ infectious neighbors, so it gets infected at rate $\beta N_{\rm I}$. 
Because recovery occurs independently of a node's neighbors, every infectious individual recovers at the same rate $\mu$. 

Based on this reasoning, we can aggregate the event sequences of the $N_{\rm S}$ susceptible individuals into one even before considering which method we should use to simulate the SIR dynamics. (Therefore, this logic also works for rejection sampling.) 
Each event sequence corresponding to a single susceptible individual has the associated event rate $N_{\rm I}\beta$.
The superposed event sequence is a realization of a single Poisson process with rate $N_{\rm S} \times N_{\rm I}\beta = N_{\rm S} N_{\rm I} \beta$. If an event from this Poisson process occurs, one arbitrary susceptible individual gets infected. Likewise, we do not need to differentiate between the $N_{\rm I}$ infectious individuals. So, we superpose the $N_{\rm I}$ event sequences, each of which has the associated event rate $\mu$, into an event sequence, which is a realization of a Poisson process with rate $N_{\rm I} \times \mu$. If an event from this Poisson process occurs, then an arbitrary infectious individual recovers.

In summary, in a well-mixed population we only need to consider two coupled Poisson processes, one corresponding to contracting infection at a rate $\beta N_{\rm S} N_{\rm I}$, and the other corresponding to recovery at a rate $N_{\rm I} \mu$. In a single step of the direct method, we first determine the waiting time to the next event, $\tau$. We set $\tau = - \ln u / (\beta N_{\rm S} N_{\rm I} + \mu N_{\rm I})$, where $u$ is a uniformly random $(0,1]$ variate. Next, we determine which type of event happens, either the infection of a susceptible node (with probability $\Pi_{\text{S} \to \text{I}}$) or the recovery of an infectious node (with probability $\Pi_{\text{I} \to \text{R}}$). 
We obtain
\begin{align}
\Pi_{\text{S} \to \text{I}} =& \frac{\beta N_{\rm S} N_{\rm I}}
{\beta N_{\rm S} N_{\rm I} + \mu N_{\rm I}}
= \frac{\beta N_{\rm S}}{\beta N_{\rm S} + \mu},\\
\Pi_{\text{I} \to \text{R}} =& \frac{\mu N_{\rm I}}
{\beta N_{\rm S} N_{\rm I} + \mu N_{\rm I}}
= \frac{\mu}{\beta N_{\rm S} + \mu}.
\end{align}
If an infection event occurs, we decrease $N_{\rm S}$ by 1 and increase $N_{\rm I}$ by 1. If a recovery event occurs, we decrease $N_{\rm I}$ by 1 and increase $N_{\rm R}$ by 1. In either case, we advance the clock by $\tau$ and go to the next step. We repeat the loop until $N_{\rm I}$ hits 0.

\begin{figure}
\centering
\includegraphics[width=\textwidth]{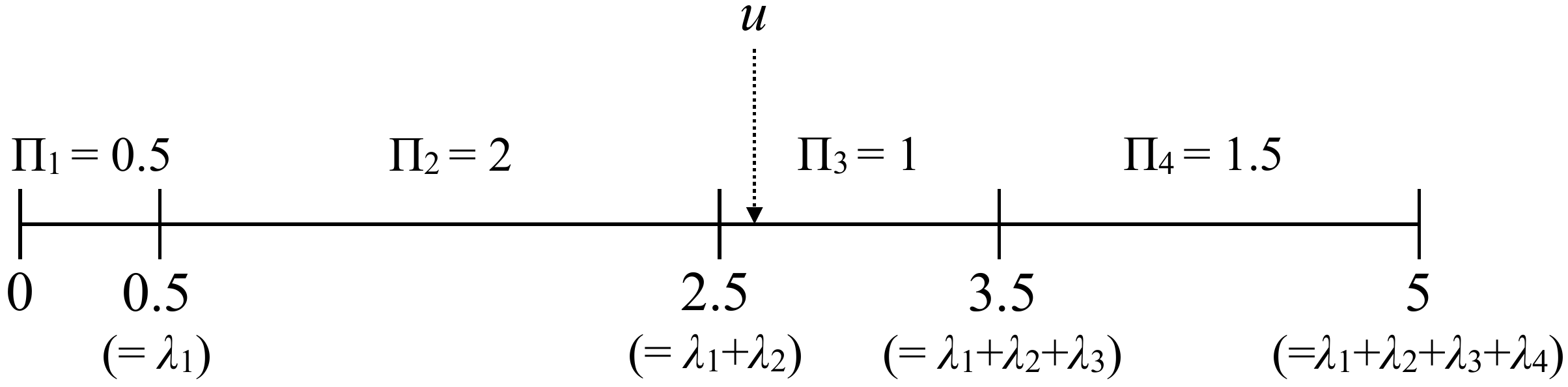}
\caption{Linear search method for computing which Poisson process produces the next event for the entire population.
We consider $M=4$ possible events, with respective rates $\lambda_1 = 0.5$, $\lambda_2 = 2$, $\lambda_3 = 1$, and $\lambda_4 = 1.5$. Suppose that we draw a uniform random variate ranging between 0 and $\sum_{i=1}^M \lambda_i = 5$ whose value is $u = 2.625$. 
We first check whether $u$ falls inside the first interval; in practice we check if $u \leq \lambda_1 = 0.5$.
Because this is not the case, we then check iteratively if it lies in each following interval.
Because $2.5 < u \leq 3.5$, we find that $u$ falls in the third interval from the left (dotted arrow). 
The iteration over $i$ as described in Step~2 in Box~\ref{box:Gillespie_classic} thus stops in the third interval, and the method will select $i=3$.
}
\label{fig:draw Pi_i}
\end{figure}

In general applications of the direct method, we consider a set of $M$ independent Poisson processes with rates $\lambda_i$ ($1\le i\le M$). 
The superposition of the $M$ Poisson processes is a single Poisson process with rate $\Lambda = \sum_{i=1}^M \lambda_i$ by the superposition theorem. 
Therefore, the time to the next event in the entire population, $\tau$, follows the exponential distribution given by 
\begin{equation}
  \psi(\tau) = \Lambda e^{- \Lambda \tau} .
  \label{eq:phi(tau) Gillespie}
\end{equation}
After time $\tau$, the $i$th process produces the next event with probability
\begin{equation}
\Pi_i = \frac{\lambda_i}{\Lambda} .
\label{eq:Pi(i) Gillespie}
\end{equation}
By drawing a random number obeying the categorical distribution over the $M$ possibilities given by $\{ \Pi_1, \ldots, \Pi_M \}$, we can then determine which Poisson process $i$ generates one event. Gillespie's original implementation does this by iterating over the list of $\Pi_i$ values (see Fig.~\ref{fig:draw Pi_i}).

We summarize the steps of the direct method in Box~\ref{box:Gillespie_classic}. 
Similarly to the first reaction method, the direct method is easy to implement, but it is not very fast in its original form (see Fig.~\ref{fig:draw Pi_i}) when $M$ is large.
For this reason more efficient algorithms have been proposed. We review them in Section~\ref{sec:computational_complexity}.

Although we have assumed in our example that a reaction channel (i.e., a Poisson process in the present case) is attached to each node/individual, this does not always have to be the case. 
As we have seen, in the case of the well-mixed population, we only need to track two reaction channels, i.e., the number of susceptible individuals $N_{\rm S}$ and the number of infectious individuals $N_{\rm I}$. 
In a more complicated setting where the network structure changes in addition to the nodes' states, some reaction channels are assigned to nodes, and other reaction channels may be assigned to the state of the edges, which may switch between on (i.e., edge available) and off (edge unavailable, or only weakly available) \citep{Volz2007ProcRSocB,Clementi2008PODC,Kiss2012ProcRSocA,Ogura2016IeeeTransNetwSciEng,Zhang2017EurPhysJB,Fonsecadosreis2020PhysRevE}. In all of the above cases, the key assumptions are that all types of reaction channels that trigger events are Poisson processes and that their event rate may only change in response to events generated by Poisson processes occurring anywhere in the population/network.

\begin{figure}[t]
\begin{boxedtext}{} 
\captionof{floatbox}{\textbf{Gillespie's direct method.}}
\label{box:Gillespie_classic}

\begin{enumerate}[start=0]
    \item Initialization:
      \begin{enumerate}
        \item Define the initial state of the system, and set $t = 0$.
        \item Calculate the rate $\lambda_j$ for each reaction channel $j$.
        \item Calculate the total rate $\Lambda = \sum_{j=1}^M \lambda_j$.
      \end{enumerate}
    \item Draw a random variate $u_1$ from a uniform distribution on $(0,1]$, and generate the waiting time by $\tau = -\ln u_1 /\Lambda$.
    \item Draw $u_2$ from a uniform distribution on $(0,\Lambda]$. Select the event $i$ to occur by iterating over $i=1,2,\ldots,M$ until we find the $i$ for which $\sum_{j=1}^{i-1} \lambda_j < u_2  \le \sum_{j=1}^i \lambda_j$. 
    \item Perform the event on reaction channel $i$.
    \item Advance the time according to $t \to t + \tau$. 
    \item Update $\lambda_i$ as well as all other $\lambda_j$ that are affected by the produced event.
    \item Return to Step 1.
\end{enumerate}
\end{boxedtext}
\end{figure}

\subsection{\label{sub:code}Codes}

Here we present Python codes for the two classic Gillespie algorithms for simulating the SIR model. Then, we compare their output and runtimes to each other and to the rejection sampling algorithm presented in Section~\ref{sub:rejection sampling}.
We first show codes for the SIR model in a well-mixed population (Sec.~\ref{sub:code-mean-field}).
We then present implementations for the SIR model on a network (Sec.~\ref{sub:code-network}), which requires additional bookkeeping to track the dependencies between nodes and the varying number of reaction channels.

All our example codes rely on the \numpy{} library in Python for vectorized computation and for generating pseudo-random numbers. The following code imports the \numpy{} library and initializes the pseudo-random number generator (see Sec.~\ref{sub:PRNG} for a discussion of how to generate random numbers on a computer).

\begin{lstlisting}[language=Python]
import numpy as np
from numpy.random import Generator, PCG64

seed = 42                     # Set seed for PRNG state 
rg = Generator(PCG64(seed))  # Initialize random number generator 
\end{lstlisting}

The following codes as well as those for the rejection sampling algorithm and those for producing figures are found in Jupyter notebooks at \url{github.com/naokimas/gillespie-tutorial}.

\subsubsection{SIR model in well-mixed populations}
\label{sub:code-mean-field}

The major part of the codes for simulating the SIR model is identical for the first reaction and direct methods and is simply related to updating and saving the system's state. We thus show only the code needed for generating the waiting time and the next event in each iteration.

The following code snippet implements the first reaction method for the SIR model in a well-mixed population:

\begin{lstlisting}[language=Python]
def draw_next_event_first_reaction(lambda_inf, lambda_rec):
    '''Input: total infection and recovery rates, lambda_inf=S*I*beta_k and lambda_rec=I*mu, respectively.
    Output: selected reaction channel, i_selected, and waiting time until the event, tau.'''

    # Draw a uniform random variate from (0,1] for each waiting time:
    u = 1. - rg.random(2)

    # Draw waiting times:
    waiting_times = - np.log(u) / np.array([lambda_inf, lambda_rec])

    # Select reaction with minimal tau:
    tau        = np.min(waiting_times)
    i_selected = np.argmin(waiting_times)
        
    return(i_selected, tau)
\end{lstlisting}

The following snippet implements the direct method:
\begin{lstlisting}[language=Python]
def draw_next_event_direct(a_inf, a_rec):
    '''Input: total infection and recovery rates, lambda_inf=S*I*beta_k and lambda_rec=I*mu, respectively.
    Output: selected reaction channel, i_selected, and the waiting time until the event, tau.'''

    # Calculate cumulative rate:
    Lambda = lambda_inf + lambda_rec  

    # Draw two uniform random variates from (0,1]:
    u1, u2 = 1. - rg.random(2)

    # Draw waiting time:
    tau = - np.log(u1) / Lambda 

    # Select reaction and update state:
    if u2 * Lambda < lambda_inf: # S->I reaction
        i_selected = 0
    else:                  # I->R reaction
        i_selected = 1

    return(i_selected, tau)
\end{lstlisting}

Finally, the following code snippet, which is common to the two methods, implements the state update after the next event has been selected and recalculates the values of the infection and recovery rates:

\begin{lstlisting}[language=Python]
# Update state:
if i_selected == 0: # S->I reaction
    S -= 1; I += 1
else:               # I->R reaction
    I -= 1 ;R += 1

# Update infection and recovery rates: 
lambda_inf = S * I * beta_k # Infection rate
lambda_rec = I * mu         # Recovery rate
\end{lstlisting}


        
        

        


        
\begin{figure}
\centering
\includegraphics[width=\textwidth]{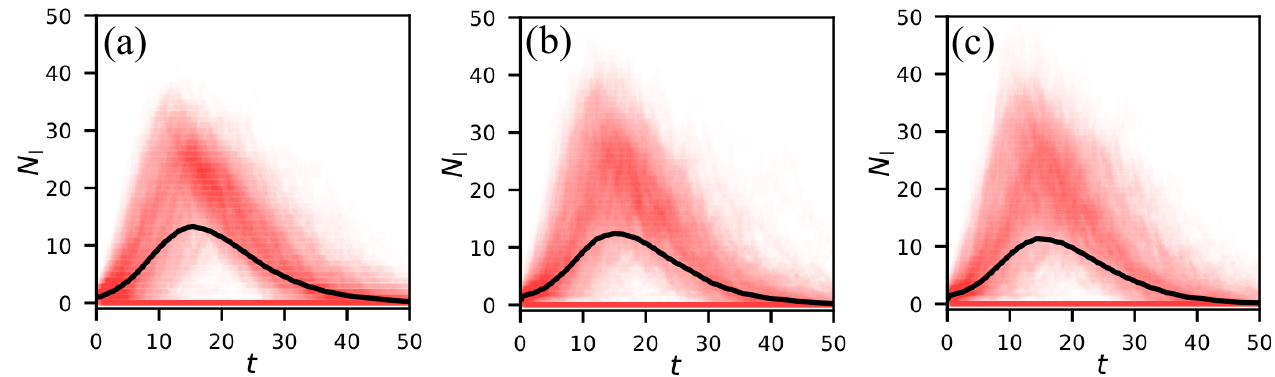}
\caption{Evolution of the number of infectious individuals $N_{\text{I}}$ over time of the SIR model in a well-mixed population simulated using (a) the rejection sampling method, (b) the first reaction method, and (c) the direct method.
The population is composed of $N=100$ individuals. 
The infection rate is $\beta = 0.5$. 
The recovery rate is $\mu=0.2$. 
We carried out 1\,000 simulations with each method. 
Each panel shows the number of infectious individuals over time. 
The overlapping thin red lines show the result for each of the 1\,000 simulations. The thick black lines show the average over the 1\,000 simulations. 
Note that a substantial portion of the simulations do not lead to any secondary infections; the red lines drop rapidly to zero, appearing as a straight red lines at $N_{\rm I} = 0$. Otherwise, $N_{\rm I}$ increases first and then decays towards zero.
The average behavior does not capture this bimodal nature of the stochastic dynamics. 
}
\label{fig:SIR-network_rejection-gillespie}
\end{figure}

We compare simulation results for the SIR model in a well-mixed population with $N=100$ individuals among the rejection sampling, Gillespie's first reaction method, and Gillespie's direct method in Fig.~\ref{fig:SIR-network_rejection-gillespie}.
The time step used for rejection sampling shown in Fig.~\ref{fig:SIR-network_rejection-gillespie}(a) is $\Delta t = 0.1$ 
With this time step, rejection sampling leads to an undershoot of the peak number of infectious individuals; compare Fig.~\ref{fig:SIR-network_rejection-gillespie}(a) to Figs.~\ref{fig:SIR-network_rejection-gillespie}(b) and \ref{fig:SIR-network_rejection-gillespie}(c).
We also note that the average curves, shown by the solid lines, fail to capture the large variation and bimodal nature of the stochastic SIR dynamics in all the panels. 
Finally, we note that the average runtimes of the three different algorithms to generate one simulation are of the order of 300 ms for the rejection sampling algorithm and of the order of 10 ms for both the Gillespie algorithms.

\subsubsection{SIR model on a network}
\label{sub:code-network}

To simulate the SIR model on a network, we rely on the \networkx{} library in addition to \numpy{} to store and update information about nodes in the network as well as their event rates. We import \networkx{} as follows: 
\begin{lstlisting}[language=Python]
import networkx as nx
\end{lstlisting}

The following code implements the generation of a single event of the SIR model on a network \verb|G| using the first reaction method. 
Here, \verb|G| stores the connections between nodes as well as each node's state (S, I, or R), event rate, and putative waiting time. 

\begin{lstlisting}[language=Python]
def draw_next_event_first_reaction(G):
    '''Input: the network G.
    Output: selected reaction channel, i_selected, and the waiting time until the event, tau.'''
     
    # Get waiting times for active reaction channels from G:
    node_indices  = list(nx.get_node_attributes(G, 'tau'))
    waiting_times = list(nx.get_node_attributes(G, 'tau').values()) 
        
    # Select reaction with minimal waiting time:
    tau = np.min(waiting_times)
    
    i_selected = np.where(waiting_times == tau)[0][0]

    return(i_selected, tau)
\end{lstlisting}

The following code implements the generation of a single event of the SIR model on a network \verb|G| using the direct method.
Here, \verb|G| stores the connections between nodes as well as each node's state (S, I, or R) and its event rate. 
\begin{lstlisting}[language=Python]
def draw_next_event_direct(Lambda, G):
    '''Input: the network, G, and the total event 
    rate, Lambda.
    Output: selected reaction channel, i_selected, and the waiting time until the event, tau.'''
    
    # Draw two uniform random variates from (0,1]:
    u1, u2 = rg.random(2)

    # Draw waiting time:
    tau = - np.log(1. - u1) / Lambda

    # Select reaction by linear search:
    target_sum = u2 * Lambda
    sum_i = 0

    for i,attributes in G.nodes(data=True): 
        sum_i += attributes['lambda']

        if sum_i >= target_sum:
            break
    
    return(i, tau)
\end{lstlisting}
Lines 13--20 implement the selection of the reaction channel that generates the next event using the simple algorithm illustrated in Fig.~\ref{fig:draw Pi_i}.

\section{Computational complexity and efficient implementations}
\label{sec:computational_complexity}

In this section we investigate the computational efficiency of the Gillespie algorithms. We also review improvements that have been developed to make the algorithms more efficient when simulating systems with a large number of reaction channels.

A typical way to quantify the efficiency of stochastic algorithms, and the one we shall be concerned with here, is their \emph{expected time complexity}\index{expected time complexity}. 
In the context of event-based simulations, it measures how an algorithm's average runtime depends on the number of reaction channels, $M$.
While it is generally impossible to exactly calculate the expected runtime of an algorithm for all different use cases, we are often able to show how the algorithm's average runtime scales with $M$ for large $M$. 
We indicate the complexity of an algorithm using \emph{big-O notation}\index{big-O notation}\footnote{In the computer science literature, the big-O notation is often informally used to denote the expected time complexity of an algorithm. This differs from the formal definition of the big-O notation, which is pertinent to the worst-case complexity~\citep{knuth1976big}. To keep things simple and to keep our notation consistent with the literature, we also adopt the big-O notation to denote the average time complexity.}, where $\O(f(M))$ means that the algorithm's expected runtime is proportional to $f(M)$ for large $M$.
For example, an algorithm with expected runtime $T_1(M) = 7 M + 10$ and another with expected runtime $T_2(M) = 0.5M + \log M$ both have linear time complexity, i.e., $T_i(M) = \O(M)$ for $i=1, 2$.

As we shall see in Section~\ref{sub:complexity-classic_Gillespie}, the classic implementations of the Gillespie algorithms presented in Section~\ref{sec:classic_Gillespie} have $\O(M)$ time complexity for each simulation step of the algorithms. While a linear time complexity may not seem computationally expensive, it is typical that the number of events taking place per time unit also scales linearly with the size of the system, i.e., $N$ in most of our examples. 
Thus, the number of computations per simulated unit of time then scales as $\O(NM)$.
This means that the overall time complexity of running a single simulation using the classic Gillespie algorithms is $\O(NMT)$, where $T$ is the typical duration of a single simulation. 
An $\O(NMT)$ computation time may be prohibitively expensive for large $N$ and $M$. Because $M$ scales linearly with $N$ at least, which occurs for sparse networks (i.e., networks with relatively few edges), we have at least $\O(N^2T)$ time complexity in this scenario.   

To make the Gillespie algorithms more efficient for large systems, researchers have come up with many algorithmic improvements to lower the computational complexity of both the bookkeeping and simulation steps of the algorithms. 
While these improved algorithms are more complex than the simple Gillespie algorithms presented in Section~\ref{sec:classic_Gillespie}, many of these techniques deserve the effort to learn because they usually shorten the computation time immensely without sacrificing the exactness of the simulations. With these techniques, we may be able to simulate a large system that we could not simulate otherwise. However, we should note that, for systems with a small number of reaction channels, these methods will not confer a significant speedup and may even be slightly slower because they introduce some additional overhead. In this section we review several general methods that we believe to be the most important ones to be aware of for researchers looking to simulate social systems, although many more exist (see~\cite{Marchetti2017Springer} for a recent review).

We detail in Section~\ref{sub:complexity-classic_Gillespie} the computational complexity of each step of the original first reaction and direct methods.
We next discuss in Section~\ref{sub:regrouping} a simple way to improve the computational efficiency of the direct method by grouping similar processes together to reduce the number of reaction channels. 
We then review algorithmic improvements that decrease the expected complexity of both the direct (Section~\ref{sub:binary_tree}) and first reaction (Section \ref{sub:binary_heap}) methods to $\O(\log M)$ time. 
More recent methods further decrease the expected runtime of the direct method, which we review in Sections \ref{sub:composition-rejection}, \ref{sub:recycling}, and \ref{sub:network-considerations}. 
Other methods have been developed that sacrifice the exactness of the Gillespie algorithms to some extent for additional speed gains.
Such methods are not the main focus of this paper, but we briefly review one such method, the \emph{tau-leaping method}, in Section~\ref{sub:other_extensions}.
We provide a short note on how to generate pseudo-random numbers needed for stochastic numerical simulations in Section~\ref{sub:PRNG}. We end this section with example codes and simulation results (Section~\ref{sub:codes-improved}).

\subsection{Average complexity of the classic Gillespie algorithms}
\label{sub:complexity-classic_Gillespie}

In this subsection we will analyze the runtime complexity of each step in Gillespie's two algorithms.
Knowing which parts of the algorithms are the most computationally expensive will also tell us which parts of the algorithms we should focus on to make them more efficient.

We first analyze the complexity of the steps of the first reaction method in the order they appear in Box~\ref{box:first-reaction_classic}:
\begin{itemize}
    \item \textbf{Step 1:} To find the smallest putative waiting time, we go through the entire list $\{\tau^{\rm put}_1, \ldots, \tau^{\rm put}_M\}$. Because there are $M$ elements in the list, this step requires $\O(M)$ time.
    \item \textbf{Steps 2--4:} Updating the system's state and event rates following an event requires a number of operations that is proportional to the number of reaction channels that are affected by the event\footnote{Note that the original implementation of the first reaction method updates all $\lambda_i$ at each time step, which is an $\O(M)$-time operation. However, we only need to update the rates that are affected by the event $i$, which is an $\O(1)$-time operation if the average number of affected reaction channels is of constant order, i.e.,  $\O(1)$.}. 
    In the case of a network, the number is often proportional to the average node degree. 
    Typically the average node degree is relatively small and does not grow much with the size of the system, which we conventionally identify with the number of nodes. Such a network is referred to as a sparse network. So, this step takes $\O(1)$ time. (However, for dense or heterogeneous networks, the number of reaction channels affected by an event may scale with $M$, in which case this step may take $\O(M)$ time -- see Section~\ref{sub:network-considerations}.)
    \item \textbf{Step 5:} To update the putative waiting times,  we only have to generate new random variates for the reaction channel that generated the event and for those that changed their event rate due to the event.
    However, all the other waiting times still need to be updated in Step 5(b) in Box~\ref{box:first-reaction_classic}. Therefore, this step also has $\O(M)$ time complexity.   
\end{itemize}
We observe that, although we have avoided some costly parts of the original implementation of the first reaction method, which we introduced in Section~\ref{sub:first-reaction}, the algorithm still has linear time complexity. 
Thus, for systems with large $M$, the first reaction method may be slow.

Let us similarly analyze the time complexity per iteration of the direct method step-by-step in the order they appear in  Box~\ref{box:Gillespie_classic}.
\begin{itemize}
    \item \textbf{Step 1:} Generating the waiting time requires generating a random variate and transforming it. This is a constant-time operation, $\O(1)$, because the runtime does not depend on $M$.
    \item \textbf{Step 2:} To find the reaction channel $i$ which generates the event, we have to iterate through half of the list of the event rates on average. This step thus takes $\O(M)$
     time.
    \item \textbf{Steps 3--5:} The steps for updating the system's state and event rates are identical to Steps 2--4 for the first reaction method. Therefore, these steps typically have $\O(1)$ time complexity.   
\end{itemize}
Our implementation of the first reaction method has two steps of linear time complexity, whereas the direct method only has a single step of linear time complexity. Therefore, the direct method may be slightly faster than the first reaction method. 
However, the former's overall scaling with $M$ is still linear. So, the direct method may be slow for large $M$, just like the first reaction method.

\subsection{\label{sub:regrouping}Grouping reaction channels}

A simple way to reduce the effective number of reaction channels in the direct method, and thus accelerate the sampling of the next event, is available when the rates $\lambda_i$ only take a small number of different values.
The strategy is to group $i$'s that have the same $\lambda_i$ value and then apply a rounding operation, which is fast for a computer, to determine a unique value of $i$ to be selected \citep{Schulze2002PhysRevE}. 

\begin{figure}
\centering
\includegraphics[width=\textwidth]{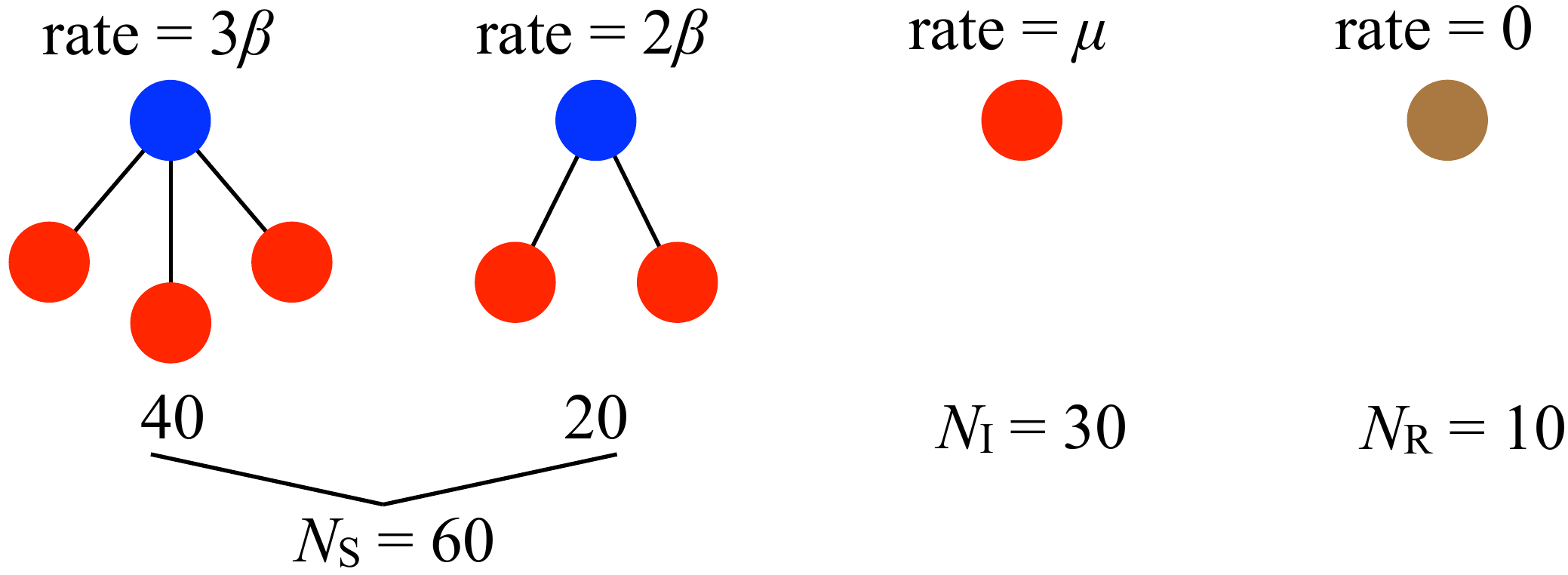}
\caption{Diagram of the state of each node undergoing the SIR dynamics. Out of the $N=100$ nodes, 40 nodes are susceptible and have three infectious neighbors (and other susceptible or recovered neighbors), 20 nodes are susceptible and have two infectious neighbors, 30 nodes are infectious, and 10 nodes are recovered.}
\label{fig:rounding}
\end{figure}

This method works as follows. Consider an idealized situation in the SIR dynamics on a network in which $N=100$, $N_{\rm S} = 60$, $N_{\rm I} = 30$, and $N_{\rm R} = 10$. Let us further assume that, at the present moment in time, 40 out of the 60 susceptible individuals are adjacent to three infectious nodes, and the other 20 are adjacent to two infectious nodes (Fig.~\ref{fig:rounding}). We can safely ignore the $N_{\rm R} = 10$ recovered individuals because they do not generate an event. 
For each susceptible individual with three and two infectious neighbors, the rate to get infected is $3\beta$ and $2\beta$, respectively.
For each infectious individual, the recovery rate is $\mu$. 
Therefore, the total (i.e., cumulative) event rate is equal to $\Lambda = 40 \times 3\beta + 20 \times 2 \beta + 30 \times \mu = 160\beta + 30\mu$. If the population is well-mixed and one just wants to track the numbers of the susceptible, infected, and recovered individuals, one needs to prepare only two reaction channels, one for infection (i.e., $N_{\rm S}$ decreases by 1, and $N_{\rm I}$ increases by 1) with rate $160\beta$, and the other for recovery (i.e., $N_{\rm I}$ decreases by 1, and $N_{\rm R}$ increases by 1) with rate $30\mu$. However, the population that we are considering is not well-mixed because the number of infectious neighbors that a susceptible individual has at each moment in time depends explicitly on the states of its neighbors in the network.
We thus need to keep track of each individual's state individually to be able to simulate the dynamics.

Each susceptible individual will generate the next event with probability
\begin{equation}
\Pi_i = \frac{3\beta}{160\beta+30\mu}
\end{equation}
or
\begin{equation}
\Pi_i = \frac{2\beta}{160\beta+30\mu},
\end{equation}
depending on whether it has three or two infectious neighbors, respectively, while each infectious individual will generate the event with probability
\begin{equation}
\Pi_i = \frac{\mu}{160\beta+30\mu}.
\end{equation}
We group together 
the 40 susceptible individuals with three infectious neighbors, which altogether have a total rate of $\lambda'_1 \equiv 40 \times 3\beta = 120\beta$.
Likewise, we group together the 20 susceptible individuals with two infectious neighbors, whose total rate is
$\lambda'_2 \equiv 20 \times 2\beta
= 40\beta$. 
The group of infectious individuals finally have a total rate of $\lambda'_3 \equiv 30 \times \mu = 30\mu$.
Then, we determine which group is responsible for the next event. Because there are only three groups, this is computationally easy. In other words, 
we draw $u_2$ from a uniform distribution on $[0, 160\beta + 30\mu)$, and if $u_2 < \lambda'_1$, then it is group 1; if $\lambda'_1 \le u_2 < \lambda'_1 + \lambda'_2$, then it is group 2; otherwise, it is group 3. 

We can next easily determine which individual in the selected group experiences the event. 
If group 1 is selected, we need to select one from the 40 susceptible nodes. Because their event rate is the same (i.e., $=3\beta$), they all have the same probability to be selected. 
Therefore, one can select the $i$th individual (with $i=1, \ldots, 40$) according to
\begin{equation}
i = \left\lfloor \frac{u_2}{\lambda'_1} \times 40 \right\rfloor + 1,
\label{eq:sample i group 1}
\end{equation}
where $\lfloor \cdot \rfloor$ denotes rounding down to the nearest integer. Note that, because group 1 was selected, we have conditioned on $0 \le u_2 < \lambda'_1$, and $u_2/\lambda'_1$ is thus a uniform random variate on $[0, 1)$. 
Therefore, $u_2/\lambda'_1 \times 40$ is uniformly randomly distributed on $[0, 40)$. By rounding down this number, we can sample each integer from 0 to 39 with equal probability, i.e., $1/40$. The term
$+1$ on the right-hand side of Eq.~\eqref{eq:sample i group 1} lifts the sampled integer by one to guarantee that $i$ is integer between 1 and 40.
Likewise, if group 2 has been selected, we set
\begin{equation}
i = \left\lfloor \frac{u_2 - \lambda'_1}{\lambda'_2} \times 20 \right\rfloor + 41.
\label{eq:sample i group 2}
\end{equation}
Because $\lambda'_1 \le u_2 < \lambda'_1 + \lambda'_2$ when group 2 is selected, $(u_2 - \lambda'_1)/\lambda'_2$ is a uniform random variate on $[0, 1)$, and $(u_2 - \lambda'_1)/\lambda'_2 \times 20$ is uniformly distributed on $[0, 20)$. 
Therefore, $\left\lfloor \frac{u_2 - \lambda'_1}{\lambda'_2} \times 20 \right\rfloor$ yields an integer between 0 and 19 with equal probability (i.e., $=1/20$), and
Eq.~\eqref{eq:sample i group 2} yields an integer between 41 and 60, each with probability $1/20$.
Finally, if group 3 has been selected, we set
\begin{equation}
i = \left\lfloor \frac{u_2 - (\lambda'_1 + \lambda'_2)}{\lambda'_3} \times 30 \right\rfloor + 61
\label{eq:sample i group 3}
\end{equation}
such that $i$ is an integer between 61 and 90, each with probability $1/30$.

Although we have considered an idealized scenario, the assumption that we can find groups of individuals sharing the event rate value is not unrealistic. In the homogeneous SIR model (i.e.\ where all individuals have the same susceptibility, infectiousness, and recovery rate), all the infectious individuals share the same event rate $\mu$. Furthermore, we may be able to group the susceptible nodes according to their number of infectious neighbors and other factors. 
As events occur, the grouping will generally change and must thus be updated after the event. 
For example, if an infected node recovers, then the group of infected nodes, whose total event rate (for recovery) was $\mu N_{\rm I}$, loses one member such that the total event rate is updated to $\mu (N_{\rm I}-1)$.

\subsection{\label{sub:binary_tree}Logarithmic-time event selection in the direct method using a binary tree}

We now move on to a general method for speeding up the direct method. 
To speed up the implementation, we use a binary tree data structure to store the $\lambda_i$ values. 
This allows us to select the reaction channel $i$ that will produce the next event (Step 2 in Box~\ref{box:Gillespie_classic}) in $\O(\log M)$ operations instead of $\O(M)$ operations\footnote{Note that each update of a $\lambda_j$ value in the binary tree also takes 
 $\O(\log M)$ time, which is slower than the original algorithm's $\O(1)$ runtime. However, the overall runtime of the new algorithm is logarithmic, compared to the linear time complexity of the original one, which will make a huge difference for large systems.} 
\citep{GibsonBruck2000JPhysChemA}. 
(See also \cite{Blue1995PhysRevE} for earlier studies and \cite{Wong1980SiamJComput} for the general case of sampling from urns with a general categorical probability distribution $\{\Pi_1, \ldots, \Pi_M\}$.) 
Because the other steps of the direct method typically have constant time complexity, improving the time complexity of the event selection step will speed up the entire algorithm.

\begin{figure}
\centering
\includegraphics[width=\textwidth]{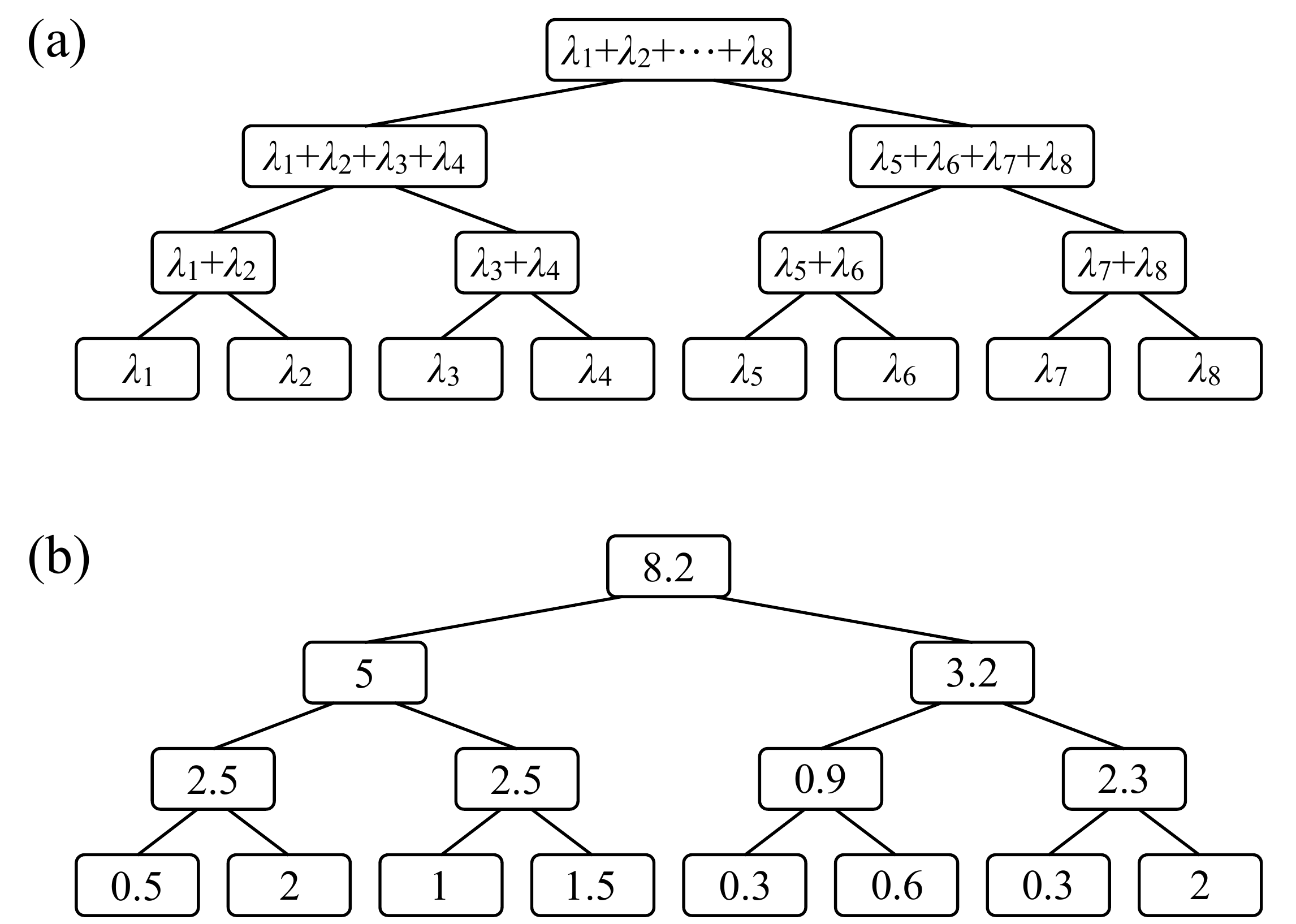}
\caption{Binary tree for drawing $i$ from a discrete distribution $\{\lambda_1, \ldots, \lambda_M\}$. We assume $M=8$. (a) General case. (b) An example. The value in each non-leaf node of the tree is equal to the sum of its two child nodes' values. There are $\log_2 M + 1 = 4$ hierarchical levels.}
\label{fig:binary tree}
\end{figure}

The main idea is to store the $\lambda_i$ values in the leaves of a binary tree and let each parent node store the sum of the values in its two child nodes (see Fig.~\ref{fig:binary tree}(a)). 
By repeating this procedure for all the internal nodes of the tree, we reach the single root node on the top of the tree, to which the value $\Lambda = \sum_{i=1}^M \lambda_i$ is assigned. 
For simplicity, we assume that $M$ is a power of 2 in Fig.~\ref{fig:binary tree}(a) such that the tree is a perfect binary tree, i.e., a binary tree where every level is completely filled. 
In fact, $M$ varies in the course of a single simulation in general, but if $M$ is not a power of 2, we can simply pad leaves of the binary tree with zeros to get a perfect tree. For example, if $M=6$, we pad the two rightmost leaves in Fig.~\ref{fig:binary tree}(a) with $\lambda_7 = \lambda_8 = 0$. Then, these two reaction channels are never selected for event generation. If a next event changes the number of reaction channels from $M=6$ to $M=7$, then we fill $\lambda_7$ by a designated positive value as well as possibly have to renew the values of some of $\lambda_1$, $\ldots$, $\lambda_6$.

To determine which event occurs, we first draw a random variate $u_2$ from a uniform distribution on $(0,\Lambda]$. 
We then start from the root node in the binary tree and look at the node's left child. If $u_2$ is smaller than or equal to the value stored in the left child, we move to the left child and repeat the procedure. 
Otherwise, we subtract the value in the left child from $u_2$, move to the right child, and repeat. 
For example, if $u_2 = 5.5$ and the binary tree is as given in Fig.~\ref{fig:binary tree}(b), we move to the right child of the root node because $u_2 > 5$. 
Then, we update $u_2$ by subtracting the value in the left child node: $u_2 \to 5.5 - 5 = 0.5$. 
Note that the new $u_2$ value is a realization of a random variate uniformly distributed on $(0, 3.2]$. 
Because the new $u_2 < 0.9$, we next move to the left child of the node with value $3.2$. We repeat this procedure until we reach a leaf. 
This leaf's index is the selected value of $i$. 
In the current example, we eventually reach the leaf node $i=6$.
There are $\log_2 M$ levels in the binary tree if we do not count the root node of the tree. Therefore, determining a value of $i$ given $u_2$ requires $O(\log M)$ time.

Once we have carried out the event associated with the selected $i$ value, we need to update the $\lambda_j$ values that are affected by the event (typically including $\lambda_i$). 
In the binary tree we can complete the updating locally for each $j$, i.e., by only changing the affected leaf and its parent nodes in the tree. 
For example, if $\lambda_2$ changes due to an event generated by the $i$th reaction channel, then, first of all, we replace $\lambda_2$ by the new value. Then, we need to replace the internal node of the binary tree just above $\lambda_2$ by a new value owing to the change in the value of $\lambda_2$. For example, suppose that the new value of $\lambda_2$ is $2.4$, which is $0.4$ larger than the previous $\lambda_2$ value ($=2$ as shown in Fig.~\ref{fig:binary tree}(b)). Then, we need to increase the value of the parent of $\lambda_2$ by $0.4$ so that we replace $2.5$ by $2.9$.
We repeat this procedure up through the hierarchical levels of the tree and update the values of the relevant internal nodes and finally that of the root node.
Therefore, we need to update only $\log_2 M + 1$ values per $\lambda_j$ value that changes. 
This is a small number compared to the total number of nodes in the binary tree, which is $2M-1$. 
Even if we need to update $\lambda_j$ for several $j$'s, the total number of nodes in the binary tree to be updated is typically still small compared to $2M-1$. 
(However, if we need to update a large fraction of the $\lambda_j$, the number of updates may become comparable to or even surpass $2M-1$.) 
The steps for implementing the direct method with binary tree search and updating of the tree are shown in Box.~\ref{box:direct_binary_tree}

\begin{figure}
\begin{boxedtext}{} 
\captionof{floatbox}{\textbf{Gillespie's direct method with binary tree search and updating of the tree.}}
\label{box:direct_binary_tree}


\begin{enumerate}
   \item[0.] {Initialization:}
      \begin{enumerate}
        \item Define the initial state of the system, and set $t = 0$.
        \item Calculate the rate $\lambda_j$ for each reaction channel $j\in\{1, \ldots, M\}$.
        \item {Initialize the binary tree:}
        \begin{enumerate} 
            \item Store each $\lambda_j$ in a leaf of a perfect binary tree with $2^{\lceil\log_2 M\rceil}$ leafs, where $\lceil\log_2 M\rceil$ denotes the smallest integer larger than $\log_2 M$. 
            \item Fill the remaining leaf nodes with zeros.
            \item Move up through the remaining $\lceil \log_2 M \rceil$ levels of the tree, setting the value of each node equal to the sum of the values of its two child nodes. 
            \item The value in the root node is equal to the total rate $\Lambda = \sum_{j=1}^M \lambda_j$.
        \end{enumerate}
      \end{enumerate}
    \item Draw a random variate $u_1$ from a uniform distribution on $(0,1]$, and generate the waiting time by $\tau = -\ln u_1 /\Lambda$.
    \item {Binary tree search:}
        \begin{enumerate}
            \item Draw $u_2$ from a uniform distribution on $\left(0, \Lambda\right]$.
            \item Start from the root node.
            \item If $u_2 \leq a_l$, where $a_l$ is the value in the left child of the current node, then go to the left child. Otherwise, set $u_2 \to u_2 - a_l$ and go to the right child.
            \item Repeat Step (c) until a leaf node is reached. The index $i$ of the leaf node gives the reaction channel that produces the next event.
        \end{enumerate}
    \item Perform the event on reaction channel $i$.
    
    \item Advance the time by setting $t \to t + \tau$.
    \item Update $\lambda_i$ as well as all other $\lambda_j$s that are affected by the produced event.
    \item {Update the binary tree:}
        \begin{enumerate}
            \item For a reaction channel whose rate $\lambda_j$ changes, set $\Delta \lambda_j = \lambda_j^{(\rm new)} - \lambda_j^{(\rm old)}$, where $\lambda_j^{(\rm new)}$ and $\lambda_j^{(\rm old)}$ are the new and old event rates,  respectively.
            \item Increase the value of the $j$th leaf and all its parents including the root node of the tree by $\Delta \lambda_j$.
            \item Repeat Step (b) for all reaction channels $j$ to be updated.
        \end{enumerate}
    \item Return to Step 1.
\end{enumerate}

\end{boxedtext}
\end{figure}

\subsection{\label{sub:binary_heap}Next reaction method: logarithmic-time version of the first reaction method}

It is also possible to make the runtime of the first reaction method scale logarithmically with the number of reaction channels, i.e., to make it have a $\O(\log M)$ time complexity. 
The improvements, collectively referred to as the \emph{next reaction method}\index{next reaction method}, were proposed in \cite{GibsonBruck2000JPhysChemA}.
Because both finding the smallest waiting time (Step 1 in Box~\ref{box:first-reaction_classic}) and updating the waiting times (Step 5 in Box~\ref{box:first-reaction_classic}) have $\O(M)$ time complexity, we need to decrease the complexity of both steps to reduce the time complexity of the entire algorithm.
The next reaction method implements three distinct improvements of the steps of the first reaction method. 
We describe each in turn.

\subsubsection{\label{sub:switch-to-absolute-time}Switch to absolute time}

One can make Step 5 of the first reaction method (see Box~\ref{box:first-reaction_classic}) more efficient simply by switching from storing the putative waiting times $\tau_j^{\rm put}$ for each reaction channel $j$ to storing the putative absolute times of the next event, denoted by $t^{\rm put}_j$. 
Analogous to the original first reaction method, the event with the smallest time $t = \min\{t^{\rm put}_1, \ldots, t^{\rm put}_M\}$ is selected to happen next, and the current time is set to $t$. 
Following the event, we only need to draw new waiting times for the reaction channel that generated the event, $i$, as well as for other reaction channels that are affected by the event. For these reaction channels, we reset the putative absolute time of the next event by adding the new waiting time drawn to the current time $t$. 
Thus, for each reaction channel $j$ that must be updated, we set $t^{\rm put}_j \to t + \tau^{\rm put}_j$, where $\tau^{\rm put}_j$ is the new putative waiting time drawn.  
There is no need to update the putative absolute times of the next event for the other reaction channels because the absolute times of the next event for these reaction channels do not depend on the current time.

In comparison, the putative waiting time until reaction channel $j$ will generate an event, i.e., $\tau^{\rm put}_j$, does change for all the reaction channels following an event. This is because the waiting times are measured relative to the current time and thus change whenever the time advances. Therefore, the putative waiting times for all reaction channels need to be updated after each event in the original first reaction method. 
Because only a small number of reaction channels are affected by each event on average (except for systems that are densely connected or are in a critical phase), the use of the absolute time in place of the waiting time reduces the complexity of this step from $\O(M)$ to $\O(1)$. 

\subsubsection{\label{sub:reuse-putative}Reuse putative times to generate fewer random numbers}

Gibson and Bruck also developed a procedure for generating new putative waiting times $\tau^{\rm put}_j$ for the reaction channels that are different from $i$ and are affected by the event. Their idea is to reuse the old putative event time for each affected reaction channel. 
With this new procedure, one does not have to generate a new random variate to determine the new  putative waiting time $\tau^{\rm put}_j$ for each of these reaction channels. 
Let us denote by $M'$ the number of reaction channels affected by the event in the $i$th reaction channel besides $i$ itself. Then, this new procedure brings the number of random variates generated per reaction down from $M' + 1$ for the first reaction method to one for the next reaction method.
While the introduction of this procedure does not change the computational complexity of the step, which remains $\O(1)$, the reduction in the actual computation time may be considerable when pseudo-random number generation is much slower than arithmetic operations. 
However, this is generally less of a concern for newer pseudo-random number generators than it was earlier.

The procedure takes advantage of the memoryless property of Poisson processes (see Box~\ref{box:Poisson_process}). 
Suppose that the rate for reaction channel $j$ has changed from $\lambda_j^{({\rm old})}$ to $\lambda_j^{({\rm new})}$ owing to the event that has occurred in reaction channel $i$. 
The memoryless property means that if the reaction channel's rate had remained unchanged, i.e., if $\lambda_j^{({\rm new})} = \lambda_j^{({\rm old})}$, then the waiting time until the next event for reaction channel $j$, i.e., $\tau^{\rm put}_j = t^{\rm put}_j - t$, would follow the same exponential distribution as that of the original waiting time.
Furthermore, any rescaling of an exponentially distributed random variable, $\tau^{'} = a \tau$, is also an exponentially distributed variable, with a rescaled rate $\lambda^{'} = \lambda/a$. 
Thus, we define the new waiting time
\begin{equation}
\tau^{{\rm put}({\rm new})} \equiv t_j^{{\rm put}({\rm new})} - t,
\label{eq:def_new_waiting_time}
\end{equation}
which is related to the old waiting time by the rescaling
\begin{equation}
    \tau^{{\rm put}({\rm new})}_j = 
    \frac{\lambda_j^{({\rm old})} \tau_j^{{\rm put}({\rm old})}}{ \lambda_j^{({\rm new})}} . 
    \label{eq:reused_waiting_time}
\end{equation}
This $\tau^{({\rm new})}$ follows the desired distribution of the new waiting time, i.e.,
\begin{equation}
    p\left( \tau^{{\rm put}({\rm new})} \right) = \lambda_j^{({\rm new})}\,e^{-\lambda_j^{({\rm new})}\,\tau_j^{{\rm put}({\rm new})}} .
\end{equation}
The combination of Eqs.~\eqref{eq:def_new_waiting_time}, \eqref{eq:reused_waiting_time}, and the definition $\tau_j^{{\rm put}(\text{old})} = t_j^{\text{put}(\text{old})} - t$ implies that, for the reaction channels $j$ whose rates have changed, we can generate the new putative time of the next event according to
\begin{equation}
    t_j^{\text{put}({\rm new})} = \frac{\lambda_j^{({\rm old})} \left(t_j^{\text{put}({\rm old})} - t\right)} { \lambda_j^{({\rm new})}} + t .
\end{equation}

\subsubsection{Use an indexed priority queue for selecting the next event}

\begin{figure}
    \centering
    \includegraphics[width=0.8\textwidth]{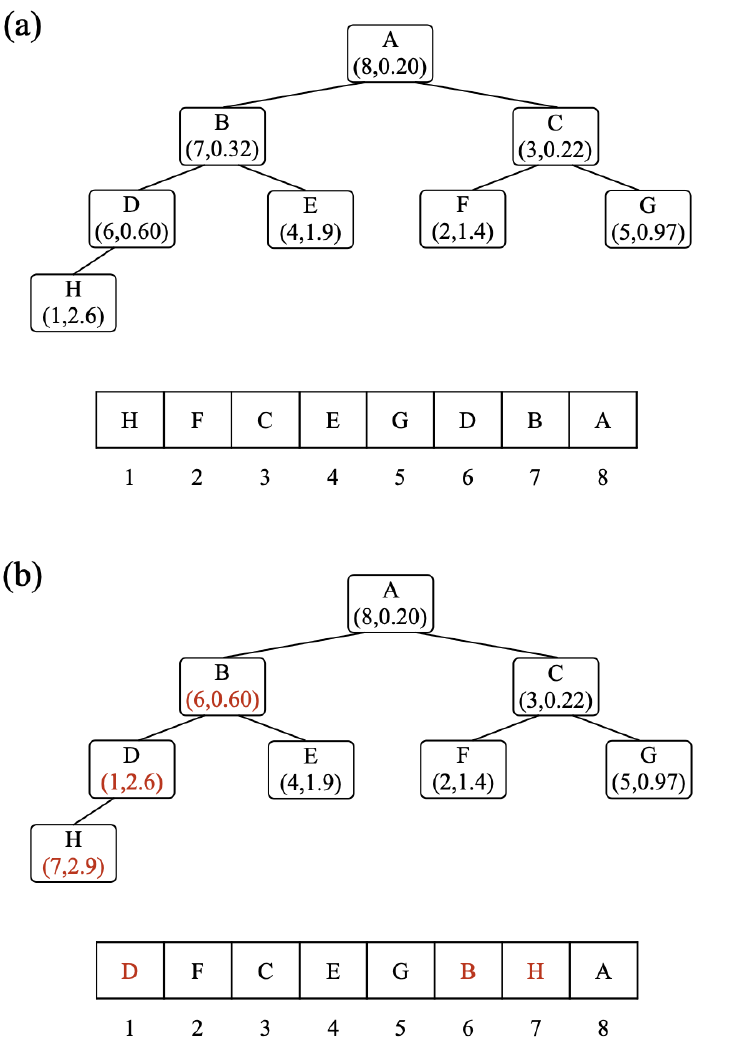}
    \caption{Indexed priority queue for storing putative reaction times in the next reaction method.
    (a) Example of an indexed priority queue. 
    The indexed priority queue consists of a binary heap (top) and an index array (bottom).
    The binary heap contains tuples $(i, t^{\rm put}_i)$, where $i$ is the reaction channel number and $t^{\rm put}_i$ the putative time when it would generate its next event. The nodes in the binary heap are ordered vertically by the value of $t^{\rm put}_i$ they store. The index array points to the node in the binary heap that corresponds to each reaction channel.
    (b) Configuration of the indexed priority queue after the value of $t^{\rm put}_7$ has been updated to 2.9 and the values stored in the nodes have been rearranged to satisfy the vertical ordering of the $t^{\rm put}_i$ values.
    All entries of the binary heap and the index array that the updating has affected are marked in red.
     }
    \label{fig:priority_queue}
\end{figure}

Increasing the efficiency of Step 1 of the first reaction method, which finds the reaction channel with the smallest waiting time, is more involved than the first two improvements described in Sections~\ref{sub:switch-to-absolute-time} and \ref{sub:reuse-putative}. 
It relies on a data structure similar to the binary tree discussed in Section~\ref{sub:binary_tree}.
Gibson and Bruck dubbed this structure an \emph{indexed priority queue}\index{indexed priority queue}, which is a binary heap\index{binary heap}, i.e., a type of binary tree that is optimized for implementing a priority queue, coupled to an index array (see Fig.~\ref{fig:priority_queue}). The binary heap stores the putative times of the next event, $t^{\rm put}_i$, for all reaction channels, ordered from the smallest to largest, and provides lookup of the smallest of them, i.e., $\min \left\{t^{\rm put}_1, \ldots, t^{\rm put}_M \right\}$, in $\O(1)$ time. The index array contains pointers to each reaction channel's position in the binary heap to provide fast updating of the $t_i$ values, i.e., in $\O(\log M)$ time.

The binary heap is a complete binary tree that stores a pair $(i, t^{\rm put}_i)$ in each node and is ordered such that each node has a $t^{\rm put}_i$ value that is smaller than that of both its children and larger than that of its parent, as shown in Fig.~\ref{fig:priority_queue}(a). Therefore, the heap stores the smallest $t^{\rm put}_i$ value in the root node. This implies that finding the smallest putative event time requires only a single operation, i.e., it has $\O(1)$ time complexity.

Because the nodes in the binary heap are not ordered by their reaction channel number $i$, the index array  (see Fig.~\ref{fig:priority_queue}) stores for each reaction channel $i$ a pointer to the position of the node in the binary heap that corresponds to $i$. Specifically, the $i$th entry of the index array points to the node in the binary heap that contains $(i, t^{\rm put}_i)$. For example, in Fig.~\ref{fig:priority_queue}(a), the reaction channel 2 is located at node F in the tree.
The index array removes the need to search through the binary heap to locate a given reaction channel and the corresponding event time. 
The index array thus enables us to find the nodes that need to be updated after an event in $\O(1)$ time.

After updating the waiting time in a given node of the binary heap, we may need to update the ordering of the nodes in the binary heap to respect the order of putative event times across the hierarchical levels (i.e., descending order as one goes from any leaf node towards the root node). 
We perform this reordering  by ``bubbling'' the values up and down: 
Starting at the node whose value has changed, corresponding to reaction channel $j$, say, we repeat either Step (1) or (2) below, depending on the value of $t^{\rm put}_j$, until one of the stopping conditions is satisfied.
\begin{itemize}
  \item[(1)] If the new $t^{\rm put}_j$ value is smaller than the $t^{\rm put}$ value stored in its parent node, swap $(j, t^{\rm put}_j)$ with the value in the parent node, and also swap the two nodes' pointers in the index array. We repeat this procedure for the parent node. 
  \item[(2)] Otherwise, i.e., if the new $t^{\rm put}_j$ value is larger than or equal to the $t^{\rm put}$ value stored in its parent node, compare $t^{\rm put}_j$ to the $t^{\rm put}$ values in its two child nodes. If $t^{\rm put}_j$ is larger than the minimum of the two, swap $(j, t^{\rm put}_j)$ with the values in the child node that attains the minimum, and also swap the two nodes' pointers in the index array. We repeat this procedure for the child node that attained the minimum of $t^{\rm put}$ before the swapping.
  \item \textbf{Stopping conditions:} We repeat the procedure until the new $t^{\rm put}_j$ value is larger than its parent's and smaller than both of its children's. Alternatively, if the value $(j, t^{\rm put}_j)$ has bubbled up to the root node or down to a leaf, we also terminate the procedure.
\end{itemize}
This procedure allows us to update the binary heap in $\O(\log M)$ time for each reaction channel whose event rate has changed following an event. 
This is the most costly part of the algorithm. So, the next reaction method improves the overall runtime of the first reaction method from $\O(M)$ to $\O(\log M)$.
Box~\ref{box:bubbling} shows an implementation of the bubbling algorithm.

\begin{figure}
\begin{boxedtext}{} 
\captionof{floatbox}{\textbf{Bubbling algorithm.}} 
\label{box:bubbling}


\verb|bubbling|(node $n$):
\begin{enumerate}
    \item[] 
      \begin{itemize}
          \item If the $t^{\rm put}$ value in the node $n$ is smaller than the $t^{\rm put}$ value in $n$'s parent node, $\verb|parent|(n)$, then
            \begin{enumerate}
                \item swap $n$ and $\verb|parent|(n)$, and update the index array correspondingly;
                \item run  \verb|bubbling|(\verb|parent|($n$)).
            \end{enumerate}
          \item Else if the $t^{\rm put}$ value in $n$ is larger than the smaller $t^{\rm put}$ value of its two children, then
            \begin{enumerate}
                \item swap $n$ and the corresponding child node, $\verb|min_child|(n)$, and update the index array correspondingly;
                \item run \verb|bubbling|(\verb|min_child|($n$)).
            \end{enumerate}
         \item Else, stop the bubbling algorithm.
      \end{itemize}
\end{enumerate}

\end{boxedtext}
\end{figure}

To illustrate the bubbling procedure we turn to the example shown in Fig.~\ref{fig:priority_queue}(a). Suppose that the putative event time for reaction channel 7 changes from $t^{\rm put}_7 = 0.32$ to $t^{\rm put}_7 = 2.9$ following an event.
We first update the value in node B of the binary heap.
We then compare the value of $t^{\rm put}_7$ to the value in the parent node, node A, in Step (1). 
Because $t^{\rm put}_7$ is larger than the value stored in node A, we then compare $t^{\rm put}_7$ to the values stored in node B's two child nodes in Step (2). 
Because $t^{\rm put}_7$ is larger than the values in both the child nodes, we swap the values with the node containing the smallest of the two, which is node D, containing $t^{\rm put}_6 = 0.60$. 
We also update the index array by swapping the pointers of reaction channels 6 and 7.
We then repeat the procedure for node D, which now
contains $(7, t^{\rm put}_7)$. 
Because we just swapped the content of nodes B and D, we know that $t^{\rm put}_7$ is larger than the value stored in node D's parent node (i.e., node B). So, we compare $t^{\rm put}_7$ to the value stored in the only child node of node D, i.e., node H.
We find that $t^{\rm put}_7$ is larger than the value stored in node H (i.e., $t^{\rm put}_1 = 2.6$). Therefore, we swap the content of the two nodes, and we update the index array by swapping the pointers to channels 1 and 7.
Because $t^{\rm put}_7$ is now stored in a leaf node, a stopping condition is satisfied, and we stop the procedure.
Figure~\ref{fig:priority_queue}(b) shows the indexed priority queue after being updated. 

One can also use the bubbling operation to initialize the indexed priority queue by successively adding nodes corresponding to each reaction channel. Because we initially need to add $M$ values of $t^{\rm put}_i$ (with $i=1, \ldots, M$), the initialization using bubbling takes $\O(M\log M)$ time. 
More efficient methods to initialize the priority queue exist \citep{Chen2012Lncs}. 
However, one runs the initialization only once during a simulation. Therefore, using an efficient initialization method would typically not much contribute to the algorithm's runtime as a whole.

The binary search tree for the direct method and the binary heap for the next reaction method are both binary tree data structure and accelerate search. However, their aims are different. The binary tree for the direct method enables us to efficiently draw $i$ with probability $\Pi_i$, and the tree holds and updates all the $\lambda_i$ values ($i=1, \ldots, M$). 
The binary heap used in the next reaction method enables us to efficiently determine the $i$ that has the smallest $t^{\rm put}_i$ value, and the tree holds and updates all the $t^{\rm put}_i$ values ($i=1, \ldots, M$). In both these structures, updating the values stored in the node (i.e., $\lambda_j$ for the direct method or $t^{\rm put}_j$ for the next reaction method) following an event is \emph{less} efficient than for the linear arrays used in their original implementation. Specifically, updating a value in the tree structures takes $\O(\log M)$ time as opposed to $\O(1)$ time for the linear array. 
However, the improved direct and the next reaction methods still have an overall $\O(\log M)$ time complexity. In contrast, the original direct and first reaction methods have an overall $O(M)$ time complexity due to the linear search, which costs $O(M)$ time. For large systems, the time saved by getting rid of the linear search is larger than the added overhead.

Finally, note that the binary heap has only $M$ nodes, which contrasts with the binary tree used for the direct method (Section~\ref{sub:binary_tree}), which has $2M-1$ nodes in the ideal case of $M$ being a power of 2. However, because the nodes in the binary heap are ordered according to their $t^{\rm put}_i$ value and not their index, the binary heap also needs to store the reaction channel's index $i$ in each node.
In addition to the binary heap, the indexed priority queue also needs to store the index array, representing an additional $M$ values.
Thus, the indexed priority queue stores $3M$ values, $M$ of which are floating-point numbers and $2M$ are integers. In contrast, the binary tree stores $2M-1$ floating-point numbers. Therefore, the memory footprints of the indexed priority queue and the binary tree are similar. 

The steps for implementing the next reaction method are shown in Box~\ref{box:next-reaction}. 

\begin{figure}
\begin{boxedtext}{} 
\captionof{floatbox}{\textbf{Next reaction method.}}
\label{box:next-reaction}

\begin{enumerate}[start=0]
    \item {Initialization:} 
      \begin{enumerate}
        \item Define the initial state of the system, and set $t = 0$.
        \item Calculate the rate $\lambda_j$ for each reaction channel $j \in \{1, \ldots, M\}$.
        \item Draw $M$ random variates $u_j$ from a uniform distribution on $(0,1]$.
        \item Generate a putative event time $t^{\rm put}_j = -\ln u_j /\lambda_j$ for each $j$.
        \item {Initialize the indexed priority queue:}\\
              Sequentially for each reaction channel $j$, add a node containing the values $(j, t^{\rm put}_j)$ to the binary heap and its position to the index array, by performing the bubbling algorithm (see Box~\ref{box:bubbling}).
      \end{enumerate}
    \item Select the reaction channel $i$ corresponding to the root node in the heap (which has the smallest $t^{\rm put}_i$).
    \item Perform the event on reaction channel $i$.
    \item Advance the time $t \to t^{\rm put}_i$
    \item Update $\lambda_i$ and all other $\lambda_j$ that are affected by the event produced.
    \item {Update the indexed priority queue:}
      \begin{enumerate}
          \item Draw a new putative waiting time for reaction channel $i$ according to $\tau^{\rm put}_i = -\ln u /\lambda_i$ with $u$ being drawn from uniform distribution on $(0,1]$, and set $t^{\rm put}_i = t + \tau^{\rm put}_i$.
          \item Generate new $t^{\rm put}_j$ values for other reaction channels $j$ whose $\lambda_j$ has changed, according to $t_j^{\text{put}({\rm new})} \to \lambda_j^{({\rm old})} \left(t_j^{\text{put}({\rm old})} - t\right) \big/ \lambda_j^{({\rm new})} + t$.
          \item For each reaction channel $j$ whose $t^{\rm put}_j$ value has changed (including $i$), look up in the index array the node $n$ that stores $(j, t^{\rm put}_j)$ in the binary heap; update $t^{\rm put}_j$ in the node $n$, and run the bubbling algorithm (see Box~\ref{box:bubbling}) to reorder the heap and update the index array. 
      \end{enumerate}
    \item Return to Step 1.
\end{enumerate}

\end{boxedtext}
\end{figure}

\subsection{\label{sub:composition-rejection}Composition and rejection algorithm to draw the next event in the direct method}

Let us discuss a third  method to draw an event $i$ with probability $\Pi_i$ from a categorical distribution $\{\Pi_1, \ldots, \Pi_M\}$ 
in the direct method.
The idea is to use the so-called composition and rejection (CR) algorithm\index{composition and rejection algorithm} \citep{Schulze2008JComputPhys,Slepoy2008JChemPhys}. This is a general method to sample a random variate that obeys a given distribution \citep{VonNeumann1951ApplMathSer}, which typically has a constant time complexity, i.e., $\O(1)$, and thus can be fast even for large systems. 

\begin{figure}
\centering
\includegraphics[width=.5\textwidth]{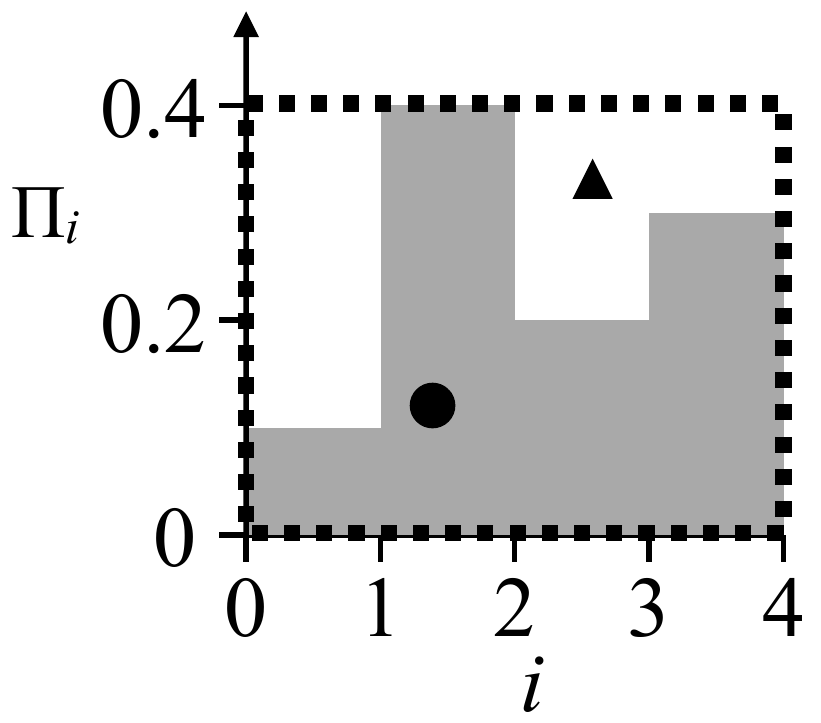}
\caption{Bar graph for the composition and rejection algorithm. The height of each bar represents $\Pi_i$. Like in Fig.~\ref{fig:draw Pi_i}, we assume $N=4$, $\Pi_1 = 0.1$, $\Pi_2 = 0.4$, $\Pi_3 = 0.2$, and $\Pi_4 = 0.3$. The two points uniformly randomly sampled from the dotted rectangle are shown by a circle and triangle.}
\label{fig:draw Pi_i CR}
\end{figure}

The idea is to first represent the categorical distribution $\{\Pi_1, \ldots, \Pi_M\}$ as a bar graph. 
The bar graph corresponding to the distribution given by Fig.~\ref{fig:draw Pi_i} is shown in Fig.~\ref{fig:draw Pi_i CR}. 
The total area of the bar graph is equal to 1. 
We consider a rectangle that bounds the entirety of the bar graph, shown by the dotted lines in Fig.~\ref{fig:draw Pi_i CR}. 
Then, we draw two random variates, denoted by $u_3$ and $u_4$, from the uniform distribution on $(0,1]$ and consider the point ($u_3 M$, $u_4 \Pi_{\max}$), where $\Pi_{\max} = \max \{ \Pi_1, \ldots, \Pi_M \}$. 
By construction, the point drawn is distributed uniformly (i.e., without bias)  in the rectangle. 
If the point happens to be inside the area of the bar graph, it is in fact a uniformly random draw from the bar graph. 
Thus, the probability for the point to land inside the $i$th bar is proportional to $\Pi_i$ in this case.
The composition and rejection algorithm uses this property to draw the event that happens, without having to iterate over a list (or a binary tree) of $\lambda_i$ values, by simply judging which bar the point falls inside. 
In the example shown in Fig.~\ref{fig:draw Pi_i CR}, the point drawn, shown by the filled circle, belongs to $\Pi_2$, so we conclude that reaction channel 2 has produced the event. 
If the point does not fall inside any bar (e.g., the triangle in Fig.~\ref{fig:draw Pi_i CR}), we then reject this point and obtain another point by redrawing two new uniform random variates. 
In practice, we find a putative reaction channel to produce the event by $i = \lfloor u_3 M \rfloor$ and adopt it if $\Pi_i \leq u_4 \Pi_{\max}$; we reject it otherwise.
The steps of the CR algorithm for general cases are shown in Box~\ref{box:CR_algorithm}. 
These steps replace Step~2 of the direct method in Box~\ref{box:Gillespie_classic}.
We note that the meaning of the rejection here is the same as that for the rejection sampling algorithm (see Section~\ref{sub:rejection sampling}) but that the two algorithms are otherwise different. 

If the area of the bar graph, which is always equal to 1, is close to the area of the rectangle, the CR algorithm is efficient. This is because rejection then occurs with a small probability, and we only waste a small fraction of the random variates $u_3$ and $u_4$, whose generation is typically the most costly part of the algorithm. 
In Fig.~\ref{fig:draw Pi_i CR}, the rectangle has area of $4\times 0.4 = 1.6$. Therefore, one rejects $1 - 1/1.6 = 3/8$ of the generated random points on average.
If the area of the box is much larger than one, which will generally happen when the event rates are heterogeneous and a few rates are much larger than the majority, then the CR algorithm is not efficient.

In \cite{Schulze2008JComputPhys} and \cite{Slepoy2008JChemPhys}, the authors went further to improve the CR algorithm to reduce the rejection probability. 
The idea is to organize the individual bars such that bars of similar heights are grouped together into a small number of groups and then draw a rectangle to bound each group of bars.
The probability for a group to generate the next event is proportional to the sum of the areas of the individual bars in the group. 
Because the number of groups is small, one can efficiently select the group that generates an event using a simple linear search.
In many cases the number of groups does not depend on $M$, and this step thus has constant time complexity, $\O(1)$.
We then apply the original CR algorithm, given in Box~\ref{box:CR_algorithm}, inside this group to select the individual reaction channel that generates the next event.
This step is necessarily efficient since the reaction channels were grouped to have similar rates, so the area of the box corresponding to the group is not much larger than one.
This implementation of the CR algorithm conserves its $\O(1)$ time complexity. It makes the rejection step more efficient at the cost of requiring an additional random variate and having to iterate through the list of groups to select the one that generates the event.

In both the original and improved CR algorithms, the time to determine $i$ does not depend on $M$, so it has $\O(1)$ time complexity in terms of $M$. 
In practice, the efficiency of the algorithm depends on the probability of rejection and on the complexity of re-grouping the bars in the case of the improved CR algorithm.

\begin{figure}[t]
\begin{boxedtext}{}
\captionof{floatbox}{\textbf{Composition and rejection algorithm.}}
\label{box:CR_algorithm}
\begin{enumerate}
    \item Generate two uniform random variates $u_3, u_4 \in (0, 1]$.
    \item Set $i = \lfloor u_3 M \rfloor$.
    \item If $u_4 \Pi_{\max} \leq \Pi_i$, we conclude that the $i$th reaction channel produces the event. Otherwise, return to Step 1.
\end{enumerate}
\end{boxedtext}
\end{figure}

\subsection{\label{sub:recycling}Recycling pseudo-random numbers in the direct method}

Each iteration of the direct method requires the generation of two uniform random variates, $u_1$ and $u_2$, one for generating the waiting time and another for selecting the reaction channel that produces an event. 
Generating a pseudo-random number is generally more costly than simple arithmetic operations.
However, as we mentioned in Section~\ref{sub:binary_heap}, recent pseudo-random number generators have substantially reduced the computational cost of generating random numbers. 
Nevertheless, there may be situation where it is preferable to generate as few random numbers as possible.
\cite{Yates2013JChemPhys} proposes a method to recycle a single pseudo-random number to generate both $u_1$ and $u_2$ (see also \cite{Masuda2018SiamRev}). 
The method works as follows.

First, we generate a uniform random variate on $(0, \Lambda]$  denoted by $u_2$, where $\Lambda = \sum_{j=1}^M \lambda_j$. Second, we determine the reaction channel $i$ that produces the event, which satisfies $\sum_{j=1}^{i-1} \lambda_j < u_2 \le \sum_{j=1}^i \lambda_j$ (see Step 2 in Box~\ref{box:Gillespie_classic}). 
These steps are the same as those of the direct method. 
Now, we exploit the fact that
$u_2 - \sum_{j=1}^{i-1} \lambda_j$ is uniformly distributed on $(0, \lambda_i]$ given that the reaction channel $i$ has been selected. This is because $u_2$ is uniformly distributed between $\sum_{j=1}^{i-1} \lambda_j$ and $\sum_{j=1}^{i} \lambda_j$ (which is equal to $\lambda_i + \sum_{j=1}^{i-1} \lambda_j$). Therefore, we set
\begin{equation}
u_1 = \frac{u_2 - \sum_{j=1}^{i-1} \lambda_j}{\lambda_i},
\end{equation}
which is a uniform random variate on $(0, 1]$. Then, we generate the waiting time by $\tau = -\ln u_1 /\Lambda$ (see Step 1 in Box~\ref{box:Gillespie_classic}).

There are two remarks. First, we need to determine $i$ first and then determine $\tau$ with this method. In contrast, one can first determine either $i$ or $\tau$ as one likes in the original direct method. Second, by generating two pseudo-random numbers from a one pseudo-random number, one is trading speed for accuracy. The variable $u_1$ has a smaller number of significant digits (i.e., less accuracy) compared to when one generates $u_1$ directly using a pseudo-random number generator as in the original direct method. However, this omission probably does not cause serious problems in typical cases as long as $M$ is not extremely large because the $u_1$ generated by the recycling direct method and the original direct method differ only slightly in the numerical value.

\subsection{\label{sub:network-considerations}Network considerations}

In networks, where different nodes may have different degrees and the number of infectious neighbors may be different even for same-degree nodes, it may be difficult to book-keep, select, and update the $\lambda_j$ values efficiently.
So, we need careful consideration of these steps when simulating stochastic processes in networks
\citep{KissMillerSimon2017book,Stonge2019ComputPhysComm}.
A particular problem that arises in dense networks or for dynamic processes in heterogeneous networks that are close to a critical point is that the average number of reaction channels that are affected by each event can become extremely large. Concretely, it may be proportional to the number of reaction channels, $M$.
In this case updating the event rates $\lambda_j$ (Step 4 in Boxes~\ref{box:first-reaction_classic} and \ref{box:next-reaction} and Step 5 in Boxes~\ref{box:Gillespie_classic} and \ref{box:direct_binary_tree}) no longer has an $\O(1)$ time complexity but $\O(M)$.
Because none of the methods discussed above improves this step, they then cannot improve the time complexity of the classic Gillespie algorithms in this situation.

Carefully designed event-based simulations, which are similar in spirit  to the first reaction method, can significantly accelerate exact simulations of coupled jump processes, even for dense or heterogeneous networks and including the case of coupled non-Poissonian renewal processes (see also Section~\ref{sub:event-data-driven_simulation}). An important assumption underlying this approach is that, once a node is infected, it will recover at a time that is drawn according to the distribution of recovery times regardless of what is going to happen elsewhere in the network. In this way, one can generate and store the recovery time of this node in a priority queue to be retrieved when the time comes.
Codes for the SIR model and the susceptible-infectious-susceptible (SIS) model\index{susceptible-infectious-susceptible (SIS) model} (i.e., individuals may get reinfected after a recovery) as well as for generating animation and snapshot figures are available at \cite{KissMillerSimon2017book-code}. The corresponding pseudocode and explanation are available in Appendix A of \cite{KissMillerSimon2017book}.

Another idea that can be used for speeding up simulations in this case is that of phantom processes\index{phantom process}, which is to assign a positive event rate to types of events that actually cannot occur. 
For example, an infectious node attempts to infect an already infectious or recovered node. If such an event is selected in a single step of the Gillespie algorithm, nothing actually occurs, and so the event is wasted. However, by designing such phantom processes carefully, one can simplify the updating of the list of all possible events upon the occurrence of events, leading to overall saving of computation time \citep{Cota2017ComputPhysCommun}.

St-Onge and colleagues have advanced related simulation methods in three main aspects \citep{Stonge2019ComputPhysComm}. First, they noted the fact that, in the case of the SIR and SIS models, any event, i.e., either the infection or recovery event, involves an infected individual. Therefore, we can reorganize the set of possible events such that they are grouped according to the individual infectious nodes. In other words, an infectious node either recovers with rate $\mu$ or infects one of its susceptible neighbors with rate $\beta$. Therefore, the total event rate associated to an infectious node $v_i$ is equal to $k_{i,\text{S}}\beta + \mu$, where $k_{i,\text{S}}$ is the number of susceptible neighbors of $v_i$. In this manner, one only has to monitor $N_{\rm I}$ reaction channels during the process. 
Their second idea is to use the CR algorithm (Section~\ref{sub:composition-rejection}).
We have
\begin{equation}
\Pi_i = \frac{k_{i,\text{S}}\beta + \mu}
{\sum_{j; v_j \text{ is infected}} \left( k_{j,\text{S}}\beta + \mu \right)}
\end{equation}
for infectious nodes $v_i$. Because infected nodes tend to be large-degree nodes \citep{Pastorsatorras2001PhysRevLett,Barrat2008book,Pastorsatorras2015RevModPhys}, some $\Pi_i$'s tend to be much larger than other $\Pi_i$'s. To accelerate the sampling of the event $i$ that occurs in each step in this situation, they employed the improved CR algorithm of \cite{Slepoy2008JChemPhys}. 
Third, they employed phantom processes, corresponding to infections of already infectious nodes.

Their algorithm has a time complexity of $\O(\log\log M)$ and is thus efficient in many cases. Their code, whose computational part is implemented in C++ for efficiency and whose interface is in Python, is available on Github \citep{Stonge2019ComputPhysComm-code}.

\subsection{Tau-leaping method}
\label{sub:other_extensions}

There are various other algorithms that are related to the Gillespie algorithms and introduce some approximations to speed up the simulations. 
We briefly review just one such method here,  
the \emph{tau-leaping method}\index{tau-leaping method}, which Gillespie proposed in 2001. 
The method works by discretizing time into intervals of some chosen length, $\Delta t$.
In a given interval, the method draws a random variate to determine how many events have happened for each process $i$, denoted by $\overline{n}_i$, and then updates the state of the system \citep{Gillespie2001JChemPhys}. 
For example, in a chemical reaction system, $\overline{n}_i$ is the increment in the number of molecules of the $i$th species.
Under the assumption that each $\lambda_i$ stays constant between $[t, t + \Delta t]$, where $t$ is the current time, $\overline{n}_i$ obeys a Poisson distribution with mean $\lambda_i \Delta t$. In other words, the number is equal to $\overline{n}_i$ with probability $(\lambda_i \Delta t)^{\overline{n}_i} e^{-\lambda_i \Delta t}/\overline{n}_i!$. 
It is desirable to make $\Delta t$ large enough to reduce the computation time as much as possible. On the other hand, $\Delta t$ should be small enough to guarantee that $\lambda_i$ stays approximately constant in each time window of length $\Delta t$ in order to assure the accuracy of the simulation. 
In simulations of social dynamics, the tap-leaping methods are probably not relevant in most cases because a single event on the $i$th reaction channel typically produces the state change of, e.g., the $i$th individual. Then, one needs to renew the $\lambda_i$ value. If this is the case, we cannot use the same $\lambda_i$ to produce multiple events on the $i$th reaction channel.

For other methods to accelerate the Gillespie algorithm or related algorithms, we refer to the review paper by \cite{Goutsias2013PhysRep}.

\subsection{\label{sub:PRNG}Pseudo-random number generation}

All stochastic simulation algorithms including the Gillespie algorithms rely on the generation of random numbers.
We here give a brief and practical introduction to computer generation of random numbers for application to Monte Carlo simulations.
We will not address the technical workings of random number generators here. An authoritative introduction to the subject is found in Chapter 7 of~\cite{Press2007numerical}.

A pseudo-random number generator (PRNG)\index{pseudo-random number generator (PRNG)} is a deterministic computer algorithm that generates a sequence of approximately random numbers.
Such numbers are not truly random. So, we refer to them as pseudo-random (i.e., seemingly random) to distinguish them from numbers generated by a truly random physical process. However, the numbers need not be truly random in most applications. 
They just need to be random enough. In the context of Monte Carlo simulations, a working criterion for what constitutes a good PRNG is that simulation results based on it are indistinguishable from those obtained with a truly random source~\citep{Jones2010good}\footnote{An entirely equivalent definition is that any two (good) PRNGs should lead to statistically the same results of the simulations~\citep{Press2007numerical}.}. 
Not all PRNGs satisfy this requirement. In fact, many standard PRNG algorithms still in use today have been shown to have serious flaws. To make this volume self-contained, we here explain some simple good practices to be followed and pitfalls to be avoided to ensure that the pseudo-random numbers that our simulations rely on are of sufficient quality.
For a more detailed, yet not too technical, introduction to good and bad practices in pseudo-random number generation, we recommend~\cite{Jones2010good}.

What constitutes a good PRNG depends on the application. For example, good PRNGs for Monte Carlo simulations are generally not random enough for cryptography applications. Conversely, while cryptographically secure PRNGs produce high-quality random sequences, they are generally much slower and are thus not optimal for Monte Carlo simulations. As this example suggests, the choice of PRNG involves a trade-off between the statistical quality of the generated sequence and the speed of generation.
Nevertheless, many PRNGs now exist that are both fast and produce sequences that are sufficiently random for any Monte Carlo simulation. Ensuring that you use a good PRNG essentially boils down to checking two simple points: (1) do not use your programming language's standard PRNG, and (2) properly seed the PRNG.

First, the most important rule to follow when choosing a PRNG is to never use your programming language's standard PRNG! To ensure backwards compatibility, the standard random number generators in many programming languages are based on historical algorithms that do not produce good pseudo-random number sequences. (A non-exhaustive list of languages with bad PRNGs is found in~\cite{Jones2010good}.) Extensive test suites such as TestU01~\citep{l2007testu01} and Dieharder~\citep{brown2021dieharder} have been developed for testing the statistical quality of PRNGs. 

Several fast and high-quality PRNGs are implemented in standard scientific computing libraries. So, using a good PRNG is as simple as importing it from one of these libraries\footnote{Many sources advise to implement the PNRG for oneself in one's code. While this is instructive, we do not believe that it is necessary for starting your work with Monte Carlo simulations.}. 
For example, we obtained all simulations performed in Python that are shown in this tutorial using the 64-bit Permuted Congruential Generator (PCG64)\index{PCG64}~\citep{ONeill2014pcg}, which is available in the standard \numpy{} library.
For C++ code (see Section~\ref{sub:codes-improved}), we used the Mersenne Twister\index{Mersenne Twister} 19937 algorithm with improved initialization\footnote{The Mersenne Twister 19937 algorithm (MT19937) has long been the reference for Monte Carlo simulations. (It has now been surpassed by the last generation of PRNGs, both in terms of speed and memory requirement, and in terms of its performance on statistical randomness tests.) 
MT19937 passes almost all tests in the test suites (e.g., Dieharder and TestU01), but it fails a few of the more rigorous tests of randomness which recent PRNGs, such as PCG64, pass. 
However, these are very rigorous tests, and this non-randomness is unlikely to pose any problem in Monte Carlo simulations~\citep{Jones2010good}.
While MT19937 is a bit slower than PCG64, the speed difference is generally too small to matter much in practice. (Tests show that PCG64 is around twice as fast as MT19937 when generating 64-bit random numbers~\citep{ONeill2014pcg}.)
Finally, MT19937 has a much larger memory footprint than other PRNGs; it keeps a 20\,032-bit internal state compared to PCG64's 128-bit state, for example.
This could pose a problem when running massively parallel simulations, e.g., on a graphics processing unit (GPU) with a relatively small amount of memory. However, it will not be a problem when running a single or a few parallel simulations on a desktop computer.}, 
i.e., \verb|mt19937ar|~\citep{Matsumoto1998mersenne, mt19937ar}, which is available as part of the Boost and Libstdc++ libraries.

Second, we should properly seed the PRNG. PRNGs rely on an internal state, which they use to generate the next output in the random number sequence. The internal state is updated at each step of the algorithm. 
At the first use of the PRNG, one must initialize, or \emph{seed}, the internal state.
Proper initialization is crucial for the performance of a PRNG. In particular, earlier PRNGs suffered from high sensitivity to the seed value.
For example, the original implementation of the Mersenne Twister algorithm is hard to seed due to its slow mixing time. 
This means that, if the bitstring corresponding to the initial state is not random enough (e.g., if it contains mostly zeros), up to the first one million generated numbers can be non-random. 
Recently proposed PRNGs generally do not show the same pathologies. So, as long as one uses a recently proposed, good PRNG, seeding the generator is not a problem except when one runs simulations in parallel. The initial seeding problems of the Mersenne Twister have been fixed in the \verb|mt19937ar| version in 2002~\citep{mt19937ar}. As another example, the PCG64 PRNG is easy to seed.

We need to be more careful on how to seed each instance of the PRNG when running simulations in parallel. 
An often used method to select the seed for a PRNG is to generate it automatically based on the system clock. However, this is a bad idea for launching many (e.g., thousands or more) simulations in parallel because many simulations will then tend to be initialized with the same or nearly the same seed. Any simulations launched with the same seed will produce exactly the same results, and those launched with close seeds may produce correlated results depending on the mixing properties of the PRNG. In both cases, this is wasteful. 
What is worse, if we are not aware that the simulation results are correlated, we will overestimate the precision of the obtained results. The safest way to seed the PRNG for parallel simulation is to use a \emph{jump ahead} operation that allows us to advance the internal state of the PRNG by an arbitrary amount of steps. With this method, one can intialize the PRNG of individual simulations with states that are sufficiently far from each other such that the pseudo-random number sequences generated by the different simulations are dissimilar. Methods for parallel seeding exist for both the Mersenne Twister~\citep{haramoto2008efficient} and PCG64~\citep{ONeill2014pcg}. 
If one wants to use a PRNG for which no efficient jump ahead method exists, a better source of randomness than the system time should be used. On Unix machines, \verb|/dev/urandom| is a choice. 
Note, however, that all PRNGs have a fixed cycle length, after which it will repeat itself deterministically. 
Therefore, one should use a PRNG with a sufficiently large cycle length (at least $2^{64}$) and a seed with a sufficient number of bits (at least ${64}$) to avoid overlap between the pseudo-random number sequences generated by the PRNG.

\subsection{\label{sub:codes-improved}Codes}

Here we showcase some example simulations of event-based stochastic processes using the Gillespie algorithms. 
In practice, in scientific research in which the Gillespie algorithms are used, we often
need to exactly run coupled jump processes on a large scale. For example, you may need to simulate a system composed of many agents, or you may have to repeat the same set of simulations for various parameter values to investigate the dependence of the results on the parameter values of your model. In such a situation, we often want to implement the Gillespie algorithms in a program language faster than Python. Therefore, we implemented the SIR model and three other dynamics in C/C++, which is typically much faster than in Python. 
We use the Mersenne Twister as the PRNG and only implement the direct method. Our C/C++ codes and the list of edges of the networks used in our demonstrations are available at Github (https://github.com/naokimas/gillespie-tutorial).

\subsubsection{SIR model}

\begin{figure}
\centering
\includegraphics[width=.49\textwidth]{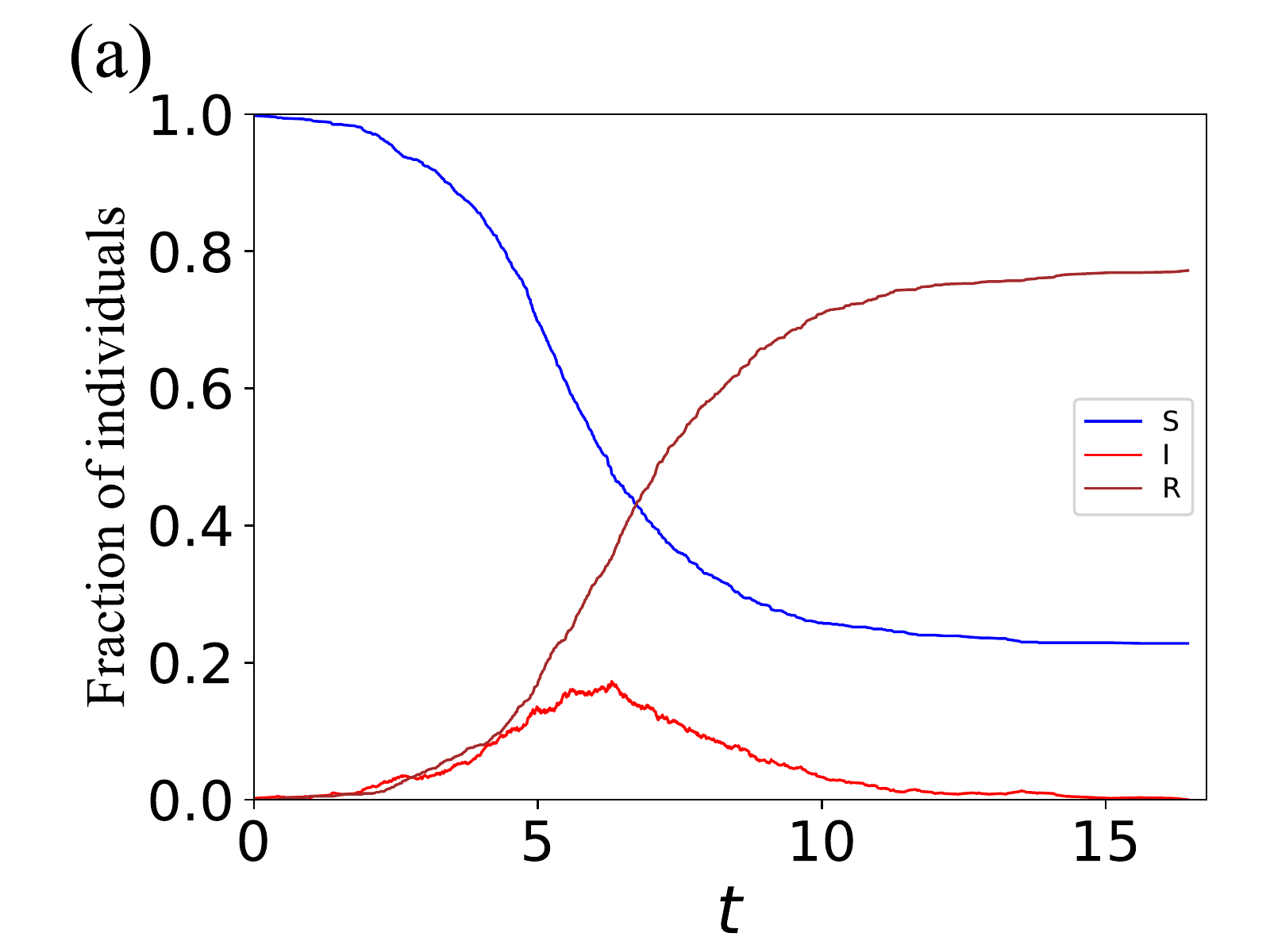}
\includegraphics[width=.49\textwidth]{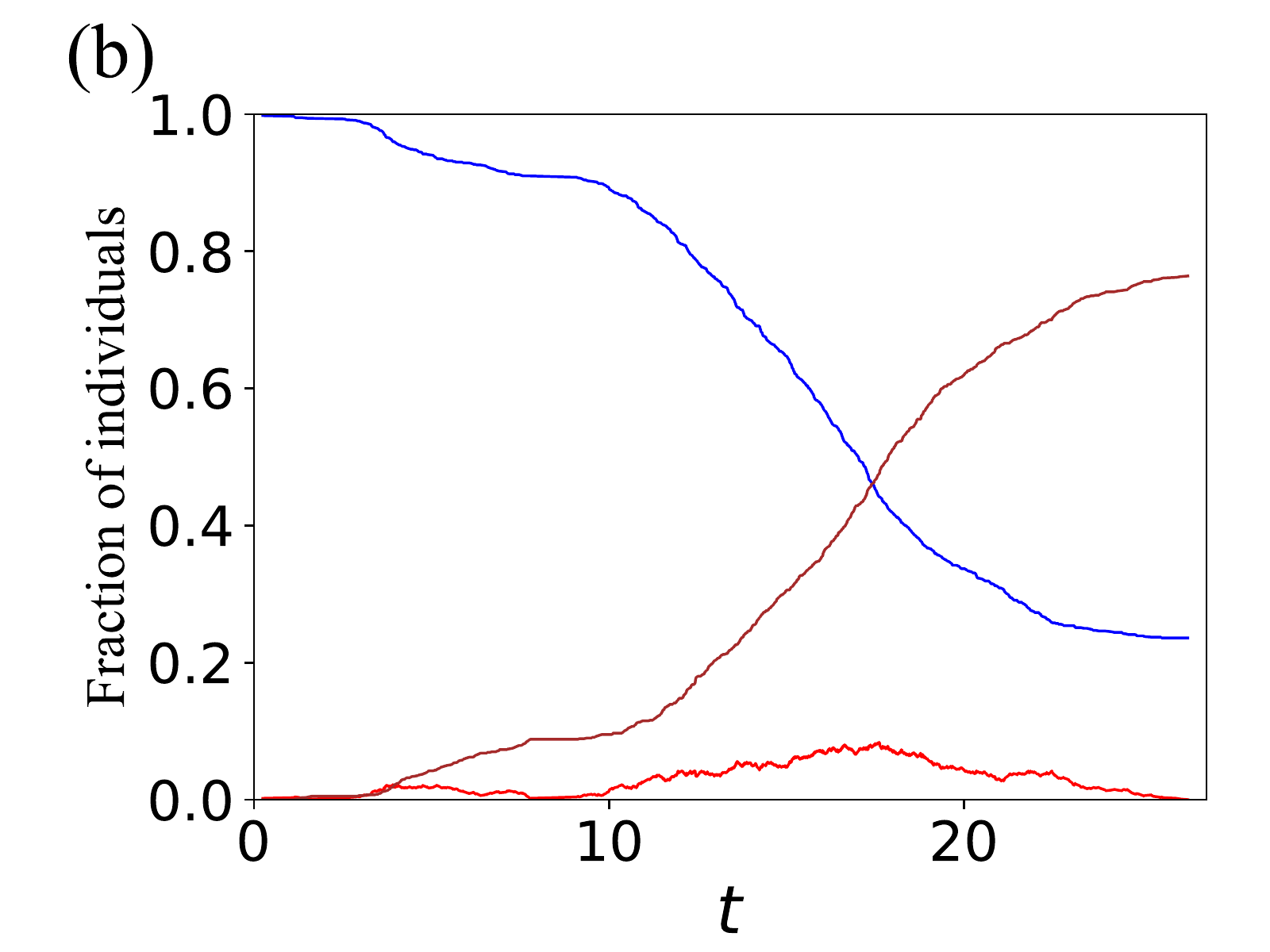}
\caption{Time courses of the fraction of the susceptible, infectious, and recovered nodes obtained from two runs of the SIR model. We set $\beta=0.6$ and $\mu=3$. We used a regular random graph with $N=1000$ nodes and the nodes' degree equal to five. In other words, each node has degree 5, and apart from that, we connect the nodes uniformly at random according to the configuration model \citep{Fosdick2018SiamRev}. 
The network is the same for the two runs shown in (a) and (b).
Each run started from the same initial condition in which a particular node was infectious and the other $N-1$ nodes were susceptible. 
}
\label{fig:sir trajectory}
\end{figure}

We provide codes for simulating the SIR model in well-mixed populations (\verb|sir-wellmixed.cc|), for general networks using Gillespie's original direct method (\verb|sir-net.cc|), and for general networks using the binary search tree (see Section~\ref{sub:binary_tree}) to speed up the selection of the events (\verb|sir-net-binary-tree.cc|).
Time courses of the fractions of the susceptible, infectious, and recovered nodes from two runs of the SIR model with $\beta=0.6$ and $\mu=3$ on a regular random graph with $N=1000$ nodes are shown in Fig.~\ref{fig:sir trajectory}. 
We started both runs from the same initial condition in which just one node, which was the same node in both runs, was infectious and the other $N-1$ nodes were susceptible. The figure illustrates the variability of the results due to stochasticity, which is lacking in the ODE version of the SIR model (Section~\ref{sub:ODE}). 

\subsubsection{Metapopulation model with SIR epidemic dynamics}

\begin{figure}
\centering
\includegraphics[width=0.9\textwidth]{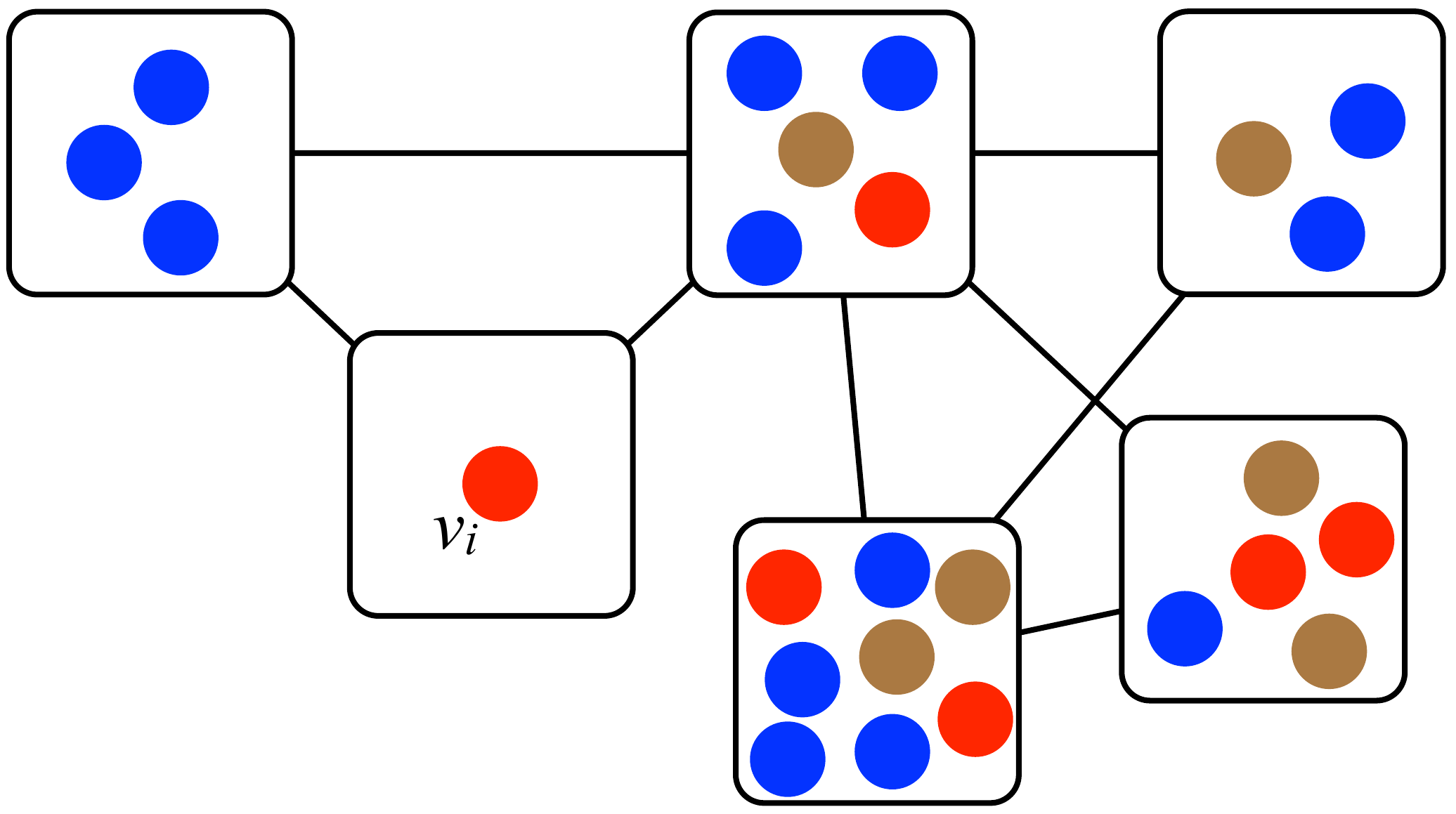}
\caption{A metapopulation model network with $\tilde{N} = 6$ patches and $N=25$ individuals. As in the previous similar figures, the blue, red, and brown circles represent susceptible, infectious, and recovered individuals, respectively.}
\label{fig:metapopulation}
\end{figure}

Another example system that the Gillespie algorithms can be used for is the SIR model in a metapopulation network. 
Mobility may induce different contact patterns at different times.
For example, we typically contact family members in the morning and evening, while we may contact workmates or schoolmates in the day time. 
The metapopulation model\index{metapopulation model} provides a succinct way to model network changes induced by mobility \citep{Anderson1991book,Hanski1998Nature,Diekmann2000book,Hufnagel2004PNAS,Colizza2006PNAS,Colizza2007NatPhys}. 
We consider a network, where a node is a patch, also called a subpopulation, which is a container of individuals, modeling e.g., a home, a workplace, a sports team meeting, or a pub. A network of patches is distinct from a network in which a node is an individual.
In Fig.~\ref{fig:metapopulation}, there are $\tilde{N}=6$ patches connected as a network.
Each individual is in either the S, I, or R state and is assumed to be situated in one patch; there are $N=25$ individuals in 
Fig.~\ref{fig:metapopulation}. An infectious individual infects each susceptible individual in the same patch with rate $\beta$. Crucially, an infectious individual does not infect susceptible individuals in other patches. An infectious individual recovers with rate $\mu$ regardless of who are in the same patch. 

In addition, the individuals move from a patch to another. There are various mobility rules used in the metapopulation model \citep{Masuda2020book}, but a simple one is the so-called continuous-time random walk\index{continuous-time random walk}.
In its simplest variant, each individual moves with constant rate $D$, which is often called the diffusion rate\index{diffusion rate}. In other words, each individual stays in the currently visited patch for a \emph{sojourn time}\index{sojourn time} $\tau$, which follows the exponential distribution, $\psi_{\rm stay}(\tau) = D e^{-D\tau}$, before it moves to a neighboring patch. When the individual moves, it selects each neighboring patch with equal probability. For example, the infectious individual $v_i$ in Fig.~\ref{fig:metapopulation} moves to either of the neighboring patches with probability $1/2$ when it moves. The movement of different individuals are independent of each other, and the moving events occur independently of the infection or recovery events. Because we assumed that the time to the next move of each individual obeys an exponential distribution, we can use the Gillespie algorithms to simulate the SIR plus mobility dynamics as described by the standard metapopulation model.

We provide a code (\verb|sir-metapop.cc|) for simulating the SIR model in the metapopulation model to which one can feed an arbitrary network structure.
Time courses of the numbers of S, I, and R individuals are qualitatively similar to those for the standard SIR model in well-mixed populations and networks.

\subsubsection{\label{sub:voter model}Voter model}

\begin{figure}
\centering
\includegraphics[width=.7\textwidth]{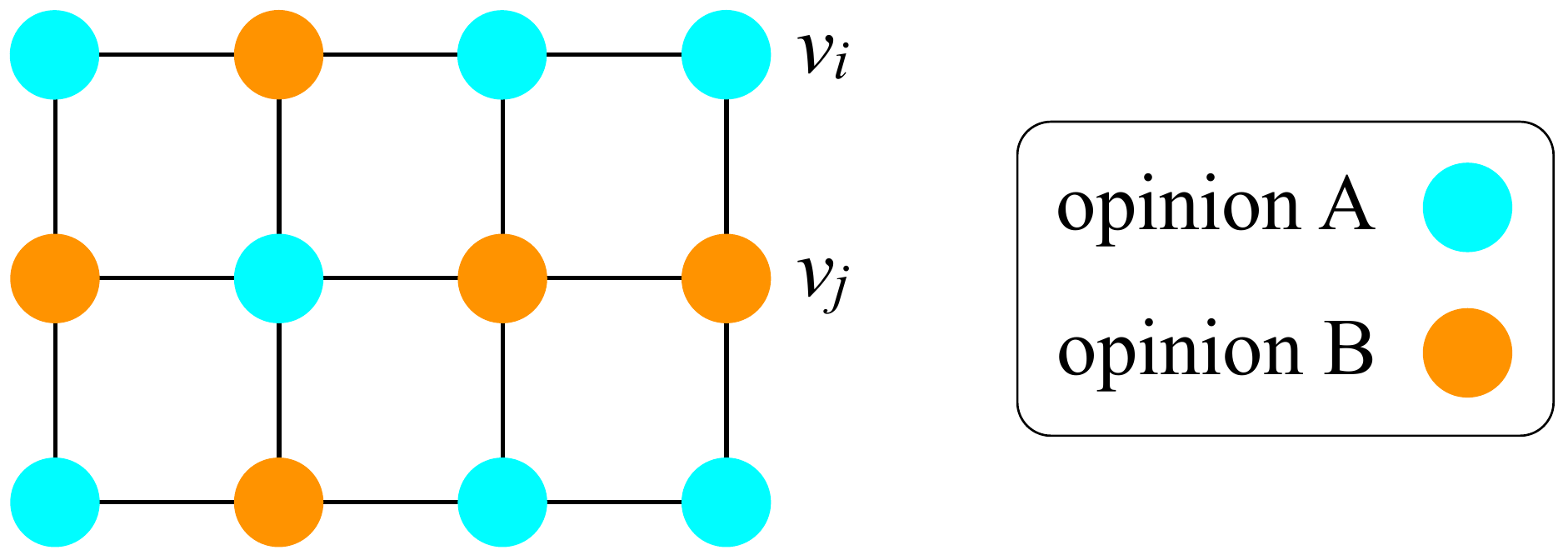}
\caption{Schematic of the voter model on a network.}
\label{fig:voter}
\end{figure}

Another typical example of collective dynamics is the voter model\index{voter model} \citep{Holley1975AnnProb,Liggett1999book,Barrat2008book,Castellano2009RevModPhys,Krapivsky2010book}. Suppose again that the individuals are nodes of a network.
Each individual is a voter and takes either of the two states A and B, referred to as opinions (see Fig.~\ref{fig:voter}). If two individuals adjacent on the network have the opposite opinions, the A individual, denoted by $v_i$, tries to convince the B individual, denoted by $v_j$, into supporting opinion A, in the same manner as an infectious individual infects a susceptible individual in the SIR model. This event occurs with rate $\beta_{\text{B}\to\text{A}}$. At the same time, $v_j$ tries to convince $v_i$, who currently supports opinion A, into supporting opinion B, which occurs with rate $\beta_{\text{A}\to\text{B}}$. Clearly, the two opinions compete with each other. The time before $v_j$ flips its opinion from B to A due to $v_i$ obeys an exponential distribution given by
$\psi_{\text{B}\to\text{A}}(\tau_{\text{B}\to\text{A}}) =  
\beta_{\text{B}\to\text{A}} e^{-\beta_{\text{B}\to\text{A}} \tau_{\text{B}\to\text{A}}}$.
Likewise, the time before $v_i$ flips its opinion from A to B due to $v_j$ obeys an exponential distribution given by $\psi_{\text{A}\to\text{B}}(\tau_{\text{A}\to\text{B}}) = \beta_{\text{A}\to\text{B}} e^{-\beta_{\text{A}\to\text{B}} \tau_{\text{A}\to\text{B}}}$.
If $\tau_{\text{B}\to\text{A}} < \tau_{\text{A}\to\text{B}}$ and nothing else occurs on the network for time $\tau_{\text{B}\to\text{A}}$
from now, $v_j$ flips its opinion from B to A. This implies that $v_j$ loses the chance to convince $v_i$ to take opinion B because $v_j$ itself now supports opinion A.

Such a competition occurs on every edge of the network that connects two nodes with the opposite opinions. The dynamics stops when the unanimity of opinion A or that of opinion B has been reached. Then, there is no opinion conflict in the entire population. Note that opinion B does not emerge if everybody in the network has opinion A, and vice versa. For this and other reasons, the voter model is not a very realistic model of voting or collective opinion formation. However, the model has been extensively studied since its inception in 1970s. The most usual setting is to assume $\beta_{\text{A}\to\text{B}} = \beta_{\text{B}\to\text{A}}$ (i.e., both opinions are as influential as each other) and ask questions such as the time until consensus (i.e., unanimity) and which opinion is likely to win depending on the initial conditions. 
When $\beta_{\text{A}\to\text{B}} \neq \beta_{\text{B}\to\text{A}}$, the model is called the biased voter model\index{biased voter model}, and an additional question to be asked is which opinion is likely to win depending on the imbalance between $\beta_{\text{A}\to\text{B}}$ and $\beta_{\text{B}\to\text{A}}$. Because there are only two types of events, associated with $\beta_{\text{A}\to\text{B}}$ and $\beta_{\text{B}\to\text{A}}$, and they occur with exponentially distributed waiting times, one can simulate the voter models, including biased ones, using the standard Gillespie algorithms.

\begin{figure}
\centering
\includegraphics[width=.49\textwidth]{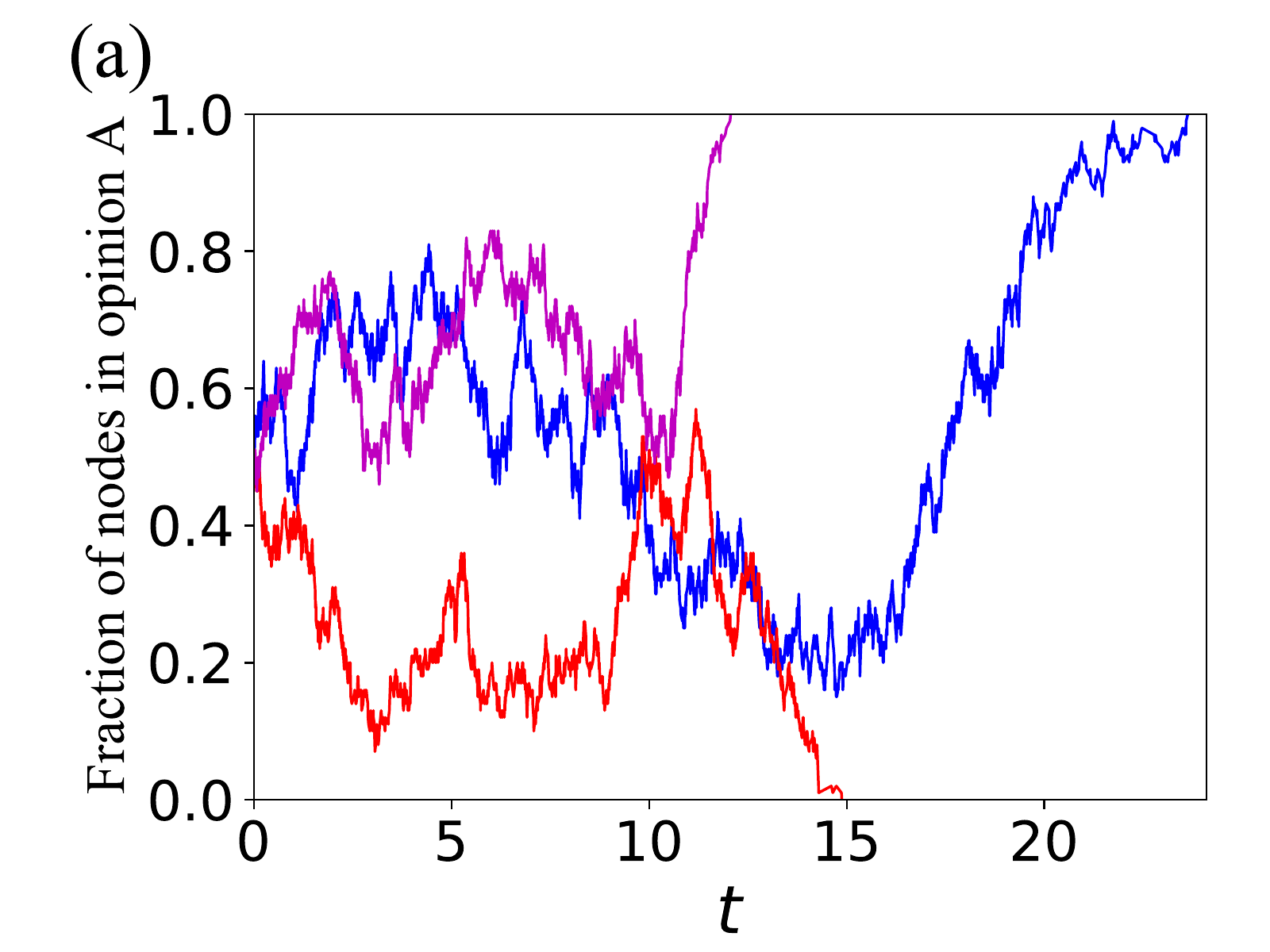}
\includegraphics[width=.49\textwidth]{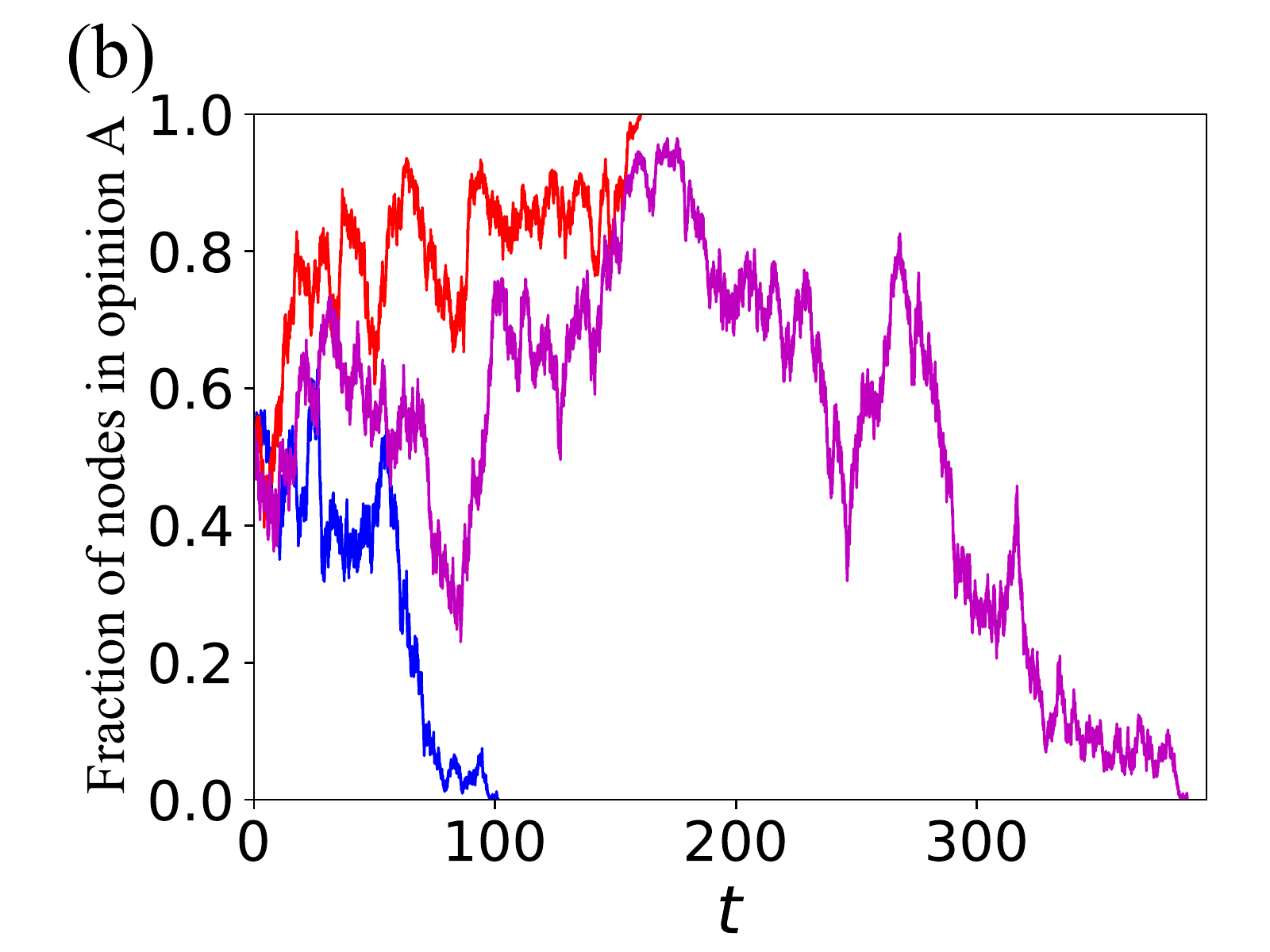}
\caption{Time courses of the fraction of the nodes with opinion A in the voter model in three different simulations. We use a regular random graph with nodes' degree equal to five. We set (a) $N=100$ and (b) $N=1000$. We set $\beta_{\text{A} \to \text{B}} = \beta_{\text{B} \to \text{A}} = 1$. Each run started from the initial condition in which half the nodes were in opinion A and the other half were in opinion B. 
The results for three runs are shown in different colors in each panel.}
\label{fig:voter trajectory}
\end{figure}

We provide codes for simulating the voter model in well-mixed populations (\verb|voter-wellmixed.cc|), for general networks using Gillespie's original direct method (\verb|voter-net.cc|), and for general networks using a binary search tree (\verb|voter-net-binary-tree.cc|). 
Time courses of the fraction of the nodes in opinion A from three runs of the unbiased voter model on a regular random graph with $N=100$ and $N=1000$ nodes are shown in Fig.~\ref{fig:voter trajectory}(a) and Fig.~\ref{fig:voter trajectory}(b), respectively. All the runs for each network started from the same initial condition in which half the nodes are in opinion A and the other half in opinion B. The figure indicates that some runs terminate with the consensus of opinion A and the others with the consensus of opinion B. It takes much longer time before a consensus is reached with $N=1000$ (Fig.~\ref{fig:voter trajectory}(b)) than with $N=100$ (Fig.~\ref{fig:voter trajectory}(a)), which is expected. Results for well-mixed populations (which one can produce with \verb|voter-wellmixed.cc|) are similar to those shown in
Fig.~\ref{fig:voter trajectory}.

\subsubsection{Lotka-Volterra model}

The Lotka-Volterra model\index{Lotka-Volterra model} describes dynamics of the numbers of prey and of predators under predator-prey interaction. It is common to formulate and analyze this dynamics as a system of ODEs, where the dependent variables represent the numbers of the prey and predators, and the independent variable is time. The ODE approach to the Lotka-Volterra model and its variants have been particularly useful in revealing mathematical underpinnings of oscillatory time courses of the numbers of prey and predators \citep{Hofbauer1988book,Murray2002book1}. However, it is indispensable to consider stochastic versions of the Lotka-Volterra models \citep{Parker2009PhysRevE,Dobrinevski2012PhysRevE,Gokhale2013BmcEvolBiol} when the number of prey or of predators is small. (See Section~\ref{sub:ODE} for a general discussion of the problems with ODE models.)

\begin{figure}
\centering
\includegraphics[width=.8\textwidth]{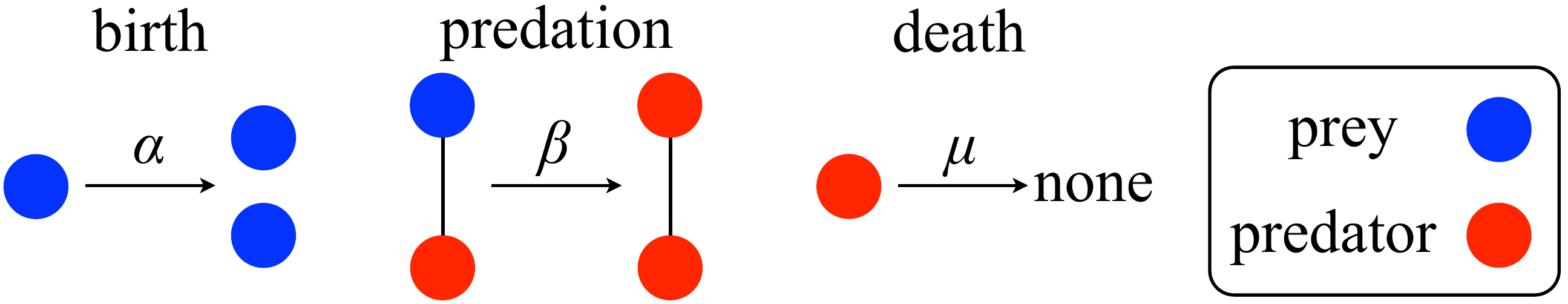}
\caption{Rules of the stochastic Lotla-Volterra model with one prey species and one predator species.}
\label{fig:LV rule}
\end{figure}

Consider a system composed of a single species of prey (which we call rabbits) and a single species of predator (which we call foxes). We denote the number of rabbits and that of foxes by $N_{\rm rab}$ and $N_{\rm fox}$, respectively. The rules of how $N_{\rm rab}$ and $N_{\rm fox}$ change stochastically are shown schematically in Fig.~\ref{fig:LV rule}. A rabbit gives birth to another rabbit with rate $\alpha$. A fox dies with rate $\mu$. A fox consumes a rabbit with rate $\beta$, which by definition results in an increment of $N_{\rm fox}$ by one. This assumption is probably unrealistic because a fox would not give birth to its cub only by consuming one rabbit. (A fox probably has to eat many rabbits to be able to bear a cub.) The model furthermore ignores natural deaths of the rabbits. These omissions are for simplicity. Because the three types of events occur as Poisson processes with their respective rates and we also assume that different types of events occur independently of each other, one can simulate the stochastic Lotka-Volterra dynamics using the Gillespie algorithms.

The extension of the Lotka-Volterra system to the case of many species is straightforward. In this scenario, a species $i$ may act as prey towards some species and as predator towards some other species. A version of the Lotka-Volterra model for more than two species can be described by the birth rate of species $i$, denoted by $\alpha_i$ (with which one individual of species $i$ bears another individual of the same species), the natural death rate of species $i$, denoted by $\mu_i$, and the rate of consumption of individuals of species $j$ by one individual of species $i$, denoted by $\beta_{ij}$ (i.e., an individual of species $i$ consumes an individual of species $j$ with rate $\beta_{ij}$).

\begin{figure}
\centering
\includegraphics[width=.49\textwidth]{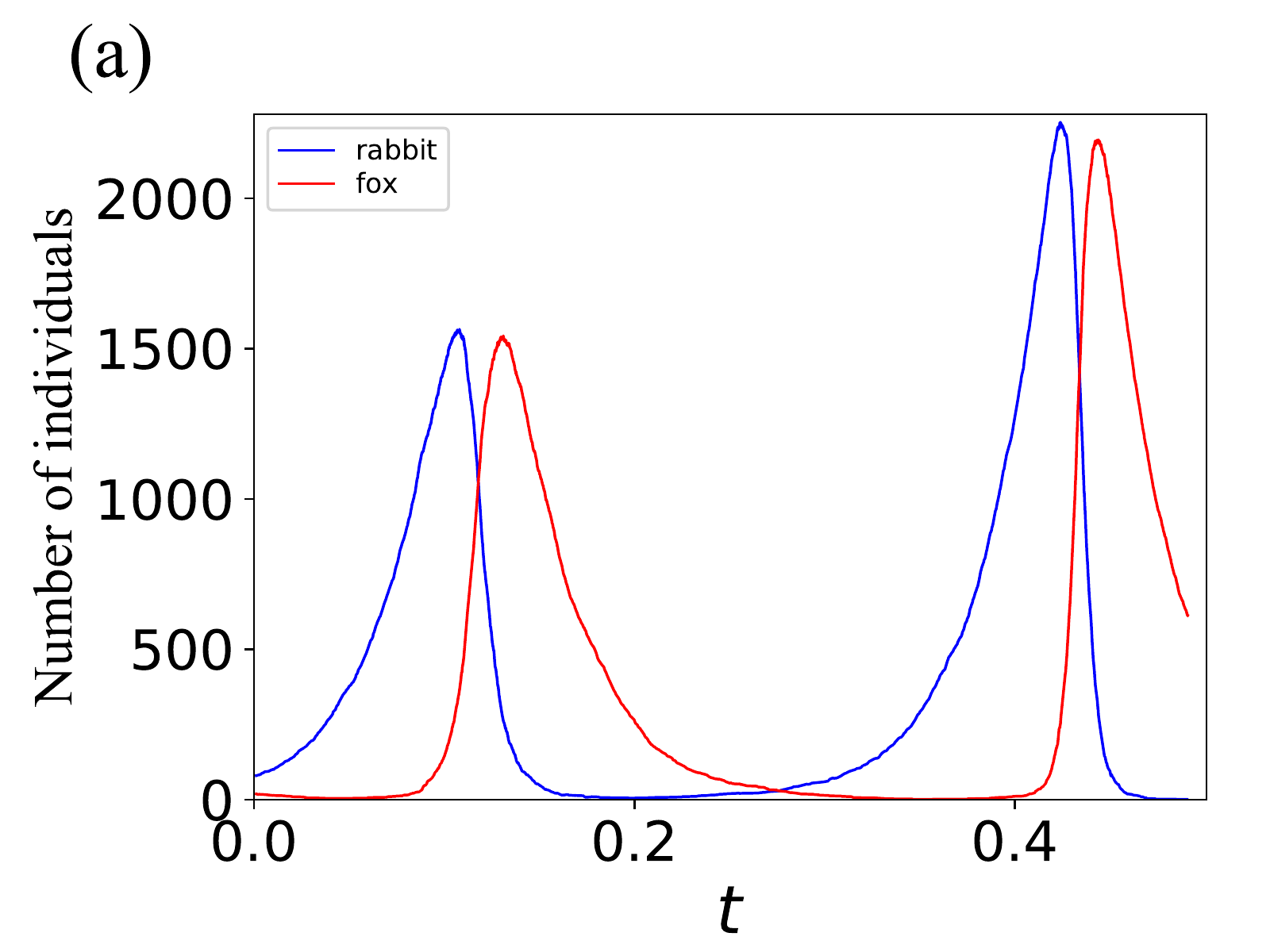}
\includegraphics[width=.49\textwidth]{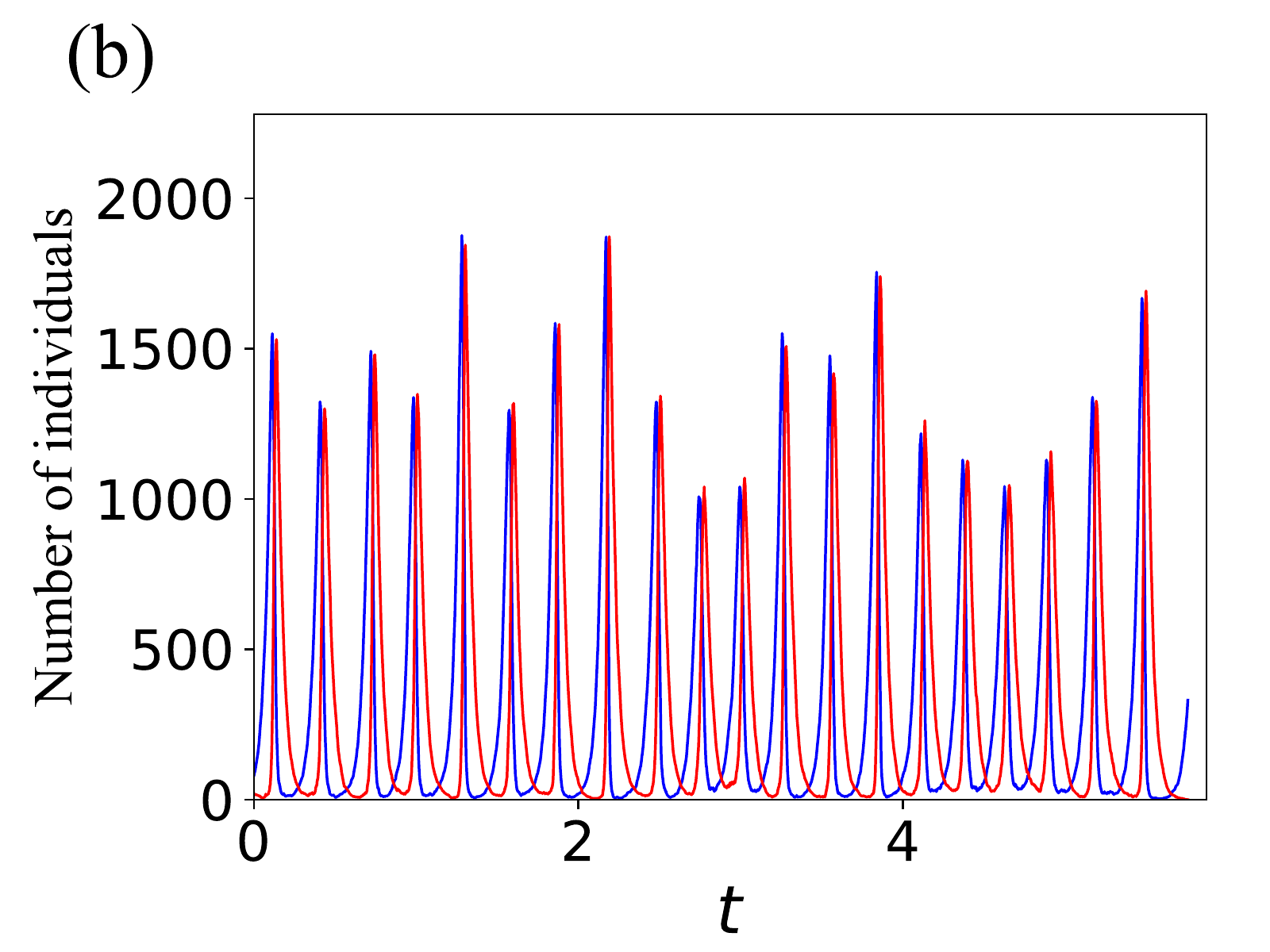}
\caption{Number of rabbits and foxes for two different runs of the stochastic Lotka-Volterra model in a well-mixed population. In both (a) and (b), we set $\alpha=30$, $\beta=0.1$, $\mu = 30$, and there are initially $N_{\text{rab}} = 80$ rabbits and $N_{\text{fox}} = 20$ foxes.}
\label{fig:LV trajectory}
\end{figure}

We provide a code (\verb|lotka-volterra-wellmixed.cc|) for simulating the stochastic single-prey single-predator Lotka-Volterra dynamics in a well-mixed population. Two sample time courses of the number of rabbits and that of foxes are shown in Fig.~\ref{fig:LV trajectory}. In both runs, the initial condition was the same, i.e., $N_{\text{rab}} = 80$ rabbits and $N_{\text{fox}} = 20$ foxes. We see oscillatory behavior of both species with time lags, which is well known to appear in the Lotka-Volterra model. In Fig.~\ref{fig:LV trajectory}(a), the simulation terminated when the rabbits went extinct after two cycles of wax and wane. By contrast, in Fig.~\ref{fig:LV trajectory}(b), the simulation terminated when the foxes went extinct after many cycles of wax and wane. The apparent randomness in the sequence of the height of the peaks in Fig.~\ref{fig:LV trajectory}(b) is due to the stochasticity of the model.

The results shown in Fig.~\ref{fig:LV trajectory} are in stark contrast with those that the ODE version of the Lotka-Volterra model would produce in two aspects.
First, the two time courses from the present stochastic simulations look very different from each other due to the stochasticity of the model. The ODE version will produce the same result every time if the simulation starts from the same initial conditions and one can safely ignore rounding errors. Second, the ODE version does not predict the extinction of one species; $N_{\text{rab}}$ or $N_{\text{fox}}$ can become tiny in the course of the dynamics, but it never hits zero in finite time. By contrast, the stochastic-process version always ends up extinction of either species although it may take long time before the extinction occurs. Once rabbits go extinct, the foxes will necessarily go extinct because there is no prey for the foxes to consume. With our code, a run terminates once rabbits go extinct in this case. On the contrary, if foxes go extinct first, then the number of rabbits will grow indefinitely because the predators are gone. In either case, there is no room for foxes to survive.

\section{Gillespie algorithms for temporal networks and non-Poissonian jump processes\label{sec:renewal_and_temporal}}

Until now we have assumed that all events occur according to Poisson processes and that the interaction network, including the case of the well-mixed population, stays the same over the duration of the simulation. 
However, both of these assumptions are often violated in empirical social systems. In this section, we present algorithms that relax these assumptions and allow us to simulate processes with non-Poissonian dynamics and on networks whose structure evolves over time.

\subsection{\label{sec:temporal-networks}Temporal networks}

\begin{figure}
\centering
\includegraphics[width=\textwidth]{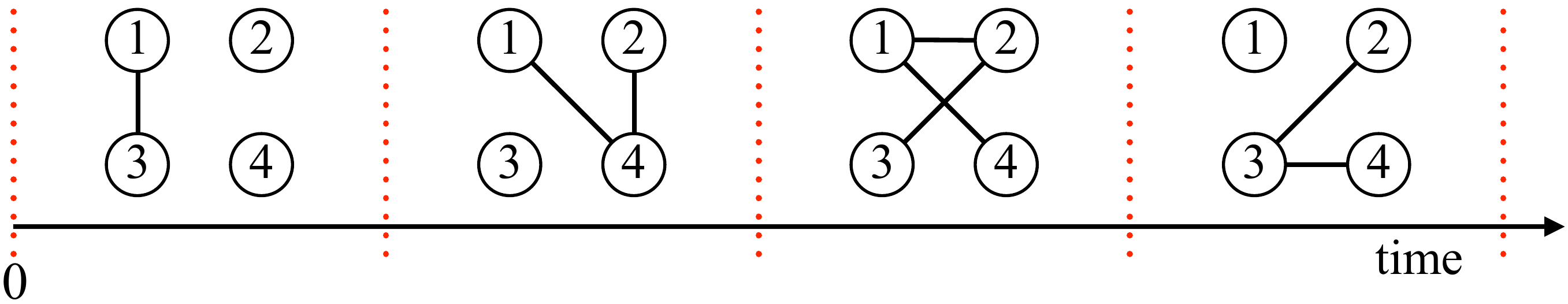}
\caption{Schematic of a ``switching" temporal network with $N=4$ nodes. The network switches from one static graph to another at discrete points in time. 
}
\label{fig:switching}
\end{figure}

In general, interactions between individuals in a social system are not continually active, so the networks they define vary in time (Fig.~\ref{fig:switching}). 
The statistics of both the dynamics of empirical temporal networks and of dynamic processes taking place on them are often strongly non-Poissonian, displaying both non-exponential waiting times and temporal correlations. Both the dynamics of empirical networks and of processes taking place on dynamic networks have been studied under the umbrella term of temporal networks\index{temporal network} \citep{HolmeSaramaki2012PhysRep,HolmeSaramaki2013book,Holme2015EurPhysJB,HolmeSaramaki2019book,Masuda2020book}. 
In the following subsections, we present several recent extensions of the direct method to temporal network scenarios. 
Before that, let us clarify to which situations we want to extend it to.

\begin{figure}
\centering
\includegraphics[width=0.8\textwidth]{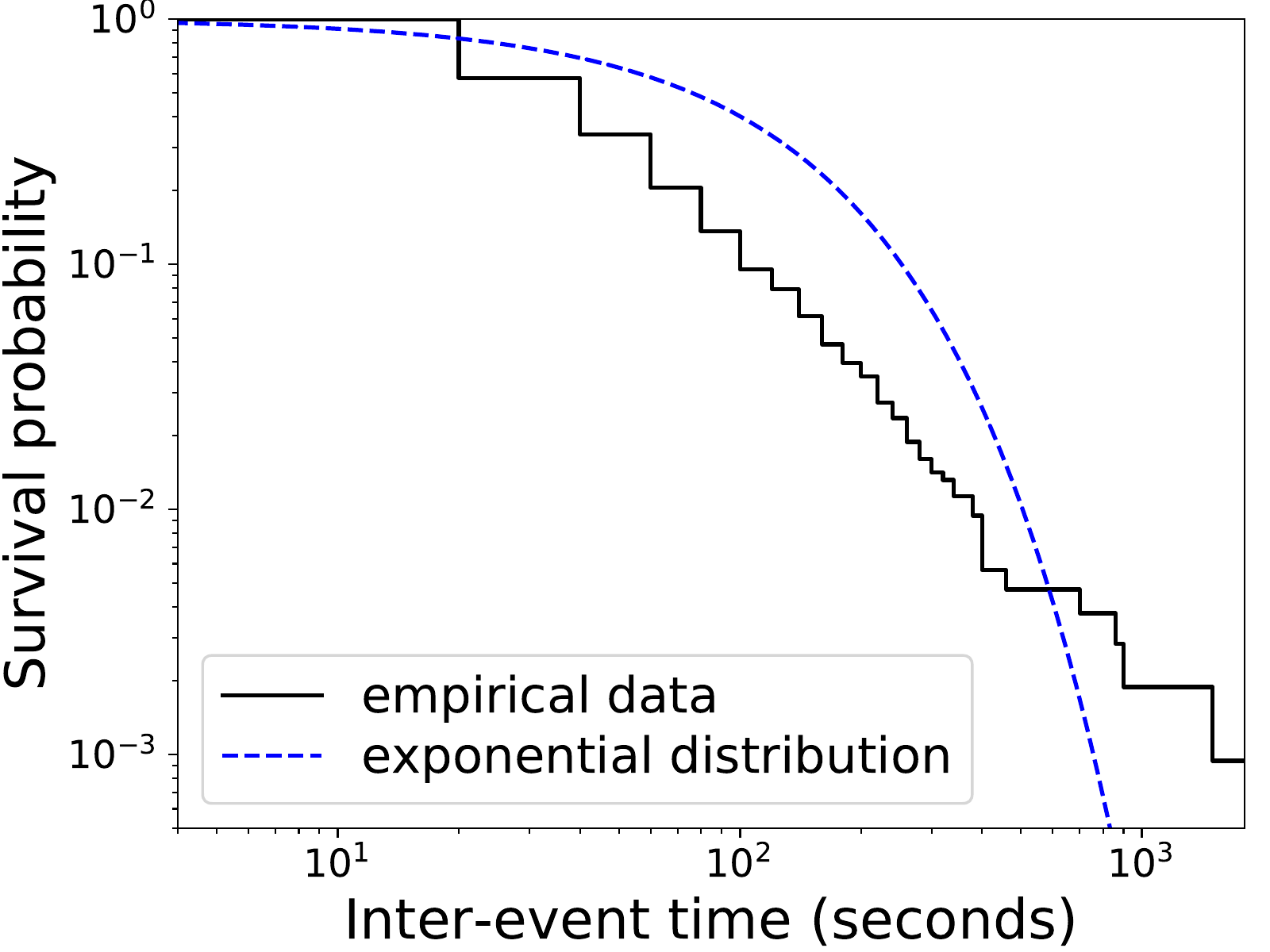}
\caption{Survival probability of inter-event times between face-to-face encounters. 
Solid black line: empirical data; dashed blue line: exponential distribution having the same mean as that of the empirical data. 
The empirical data come from the ``Primary School'' data set from the SocioPatterns project \citep{Isella2011PlosOne}. 
Events are face-to-face proximity relationships between an individual and other individuals in the school. 
We show the survival probability, i.e., $\Psi(\tau) = \int_{\tau}^{\infty} \psi(\tau^{\prime}) \text{d}\tau^{\prime}$, instead of the distribution of inter-event times, $\psi(\tau)$, because $\Psi(\tau)$ is more robust to noise in data.
In other words, the vertical axis represents the fraction of the inter-event times that are larger than the value specified on the horizontal axis. 
We selected the individual with the largest number of events and calculated all the inter-event times for the selected individual.
We omitted the largest inter-event time, which is more than 10 times larger than the second largest one. 
The survival probability of the exponential distribution is given by $\Psi(\tau) = \int_{\tau}^{\infty} \lambda e^{-\lambda\tau^{\prime}}\text{d}\tau^{\prime} = e^{-\lambda\tau}$.}
\label{fig:fat-tailed psi(tau)}
\end{figure}

First, empirical sequences of discrete events tend to strongly deviate from Poisson processes. In a Poisson process, the distribution of inter-event times is an exponential distribution. By contrast, events in empirical human activity data often do not exhibit exponential distributions. 
Figure~\ref{fig:fat-tailed psi(tau)} shows the distribution of inter-event times $\tau$ between face-to-face encounters for an individual in a primary school.
For reference, we also show an exponential distribution whose mean is the same as that of the empirical data. 
The empirical and exponential distributions do not resemble each other. In particular, the empirical distribution is much more skewed than the exponential distribution. It thus has a much larger chance of producing extreme values of $\tau$, both small and large. Typically, the right tail of the distribution (i.e., at large values of $\tau$) is roughly approximated by a power-law distribution $\psi(\tau) \propto \tau^{-\alpha}$, 
where $\propto$ means proportional to, and $\alpha$ is a constant, typically between 1 and 3.
If one replaces the exponential $\psi(\tau)$, which the Gillespie algorithm and the original stochastic multiagent models assume, by a power-law $\psi(\tau)$, the results may considerably change. For example, for given $\beta$ and $\mu$ values, epidemic spreading may be less likely to occur in the SIR model with inter-event times $\tau$ following a power-law distribution than with ones following an exponential distribution with the same mean \citep{Karsai2011PhysRevE,Miritello2011PhysRevE,Kivela2012JStatMech,Masuda2013F1000}. 
Therefore, we are interested in simulating stochastic dynamics where the waiting times obey distributions that may differ from the exponential distribution. 
Such processes are called \emph{renewal processes}\index{renewal process}.
A Poisson process is the simplest example of a renewal process. 
It generates events at a constant rate irrespectively of the history of the events in the past and is thus called memoryless, which is also referred to as the process being \emph{Markovian}\index{Markovian}. 
General renewal processes are not memoryless and often referred to as being non-Markovian, especially in the physics literature.
Note that, even in general renewal processes, each waiting time is independent of the past ones. However, the expected waiting time until the next event depends on the time elapsed since the last event\footnote{For this reason, renewal processes are formally defined as a type of semi-Markov process. We will not delve further into this distinction here. We will simply refer to processes that do not have exponentially distributed inter-event times as non-Poissonian.}. 
In Sections~\ref{sub:nMGA} and \ref{sub:LGA}, we will present two algorithms that simulate stochastic dynamics when $\psi(\tau)$ can be non-exponential distributions.

Second, we may be interested in simulating a dynamic process on an empirically recorded temporal network (e.g., epidemic spread over a mobility network). 
We will here consider a representation of temporal networks in which the network changes discontinuously in discrete time points (Fig.~\ref{fig:switching}), which we call \emph{switching networks}\index{switching (temporal) network}. 
Such a representation is often practical since empirical temporal networks are generally recorded with finite time resolution and thus change only in discrete points in time.
The Gillespie algorithms do not directly apply in this second case either. 
This is because, in switching networks, which events can occur and the rates at which they occur depend on time. In contrast, the classic Gillespie algorithms assume that the event rates stay constant in-between events.
We will present a temporal version of the direct method that can treat switching networks in Section~\ref{sub:TGA}.

Both non-Poissonian statistics of event times and temporally changing event rates can also occur in chemical reaction systems, for which the Gillespie algorithms were originally proposed. Several extensions have been developed in the chemical physics and computational biology literature to deal with these issues. 
Different from social systems, temporally changing event rates are often externally driven in such systems. 
For example, in cellular reaction systems the cell's volume may change over time owing to cell growth. 
Such a volume change leads to changes in molecular concentrations and thus to temporally evolving reaction rates, similar to temporal networks. 
Extensions of the Gillespie algorithms have enabled, for example, simulating chemical reaction systems with time-varying volumes~\citep{Kierzek2002Bioinformatics,LuVolfson2004SystBiol,Carletti2012ComputMathMethodsMed}.
There are also Gillespie algorithms for more generally fluctuating event rates 
\citep{Anderson2007JChemPhys}. 
Another common phenomenon in chemical reaction systems is delays due to, e.g., diffusion-limited reactions. 
Such delays lead to non-exponential waiting times, and several approaches have been developed to deal with this case~\citep{Bratsun2005PNAS,Barrio2006PlosComputBiol,Anderson2007JChemPhys,Cai2007JChemPhys}.
While these issues are similar to those encountered in temporal networks, each has its particularities.
Delays in chemical reaction systems lead to distributions of inter-event times that are less skewed than the exponential distribution. In contrast, typical distributions of inter-event times in social networks are \emph{more} skewed than exponential distributions. 
Another difference is that external dynamics influencing chemical reaction systems are typically much slower than the reaction dynamics, while social network dynamics typically occur on the same scale or faster than the dynamics we want to simulate on networks.
These facts pose specific challenges for the simulation algorithms. In fact, although some extensions of the Gillespie algorithms developed for chemical reaction systems may also be suitable for simulating multiagent systems and temporal networks, algorithms focusing specifically on the temporal network setting have emerged.
We review them in this section.

\subsection{\label{sub:nMGA}Non-Markovian Gillespie algorithm}

The \emph{non-Markovian Gillespie algorithm}\index{non-Markovian Gillespie algorithm} is an extension of the direct method to the case in which inter-event times are not distributed according to exponential distributions \citep{BogunaLafuerza2014PhysRevE}. It relaxes the assumption that the individual jump processes are Poisson processes and enables us to simulate general renewal processes.

We denote by $\psi_i$ the distribution of inter-event times for the $i$th reaction channel (where $i=1, \ldots, M$), which we assume is a renewal process\index{renewal process}. 
If $\psi_i$ is an exponential distribution, the $i$th renewal process is a Poisson process. If all $\psi_i$s are exponential distributions, we can use the original Gillespie algorithms. 
When $\psi_i$ is not an exponential distribution,
we need to know the time $\tilde{t}_i$ since the last event for the process $i$ to be able to generate the time to the next event for that process. 

\begin{figure}
\centering
\includegraphics[width=0.9\textwidth]{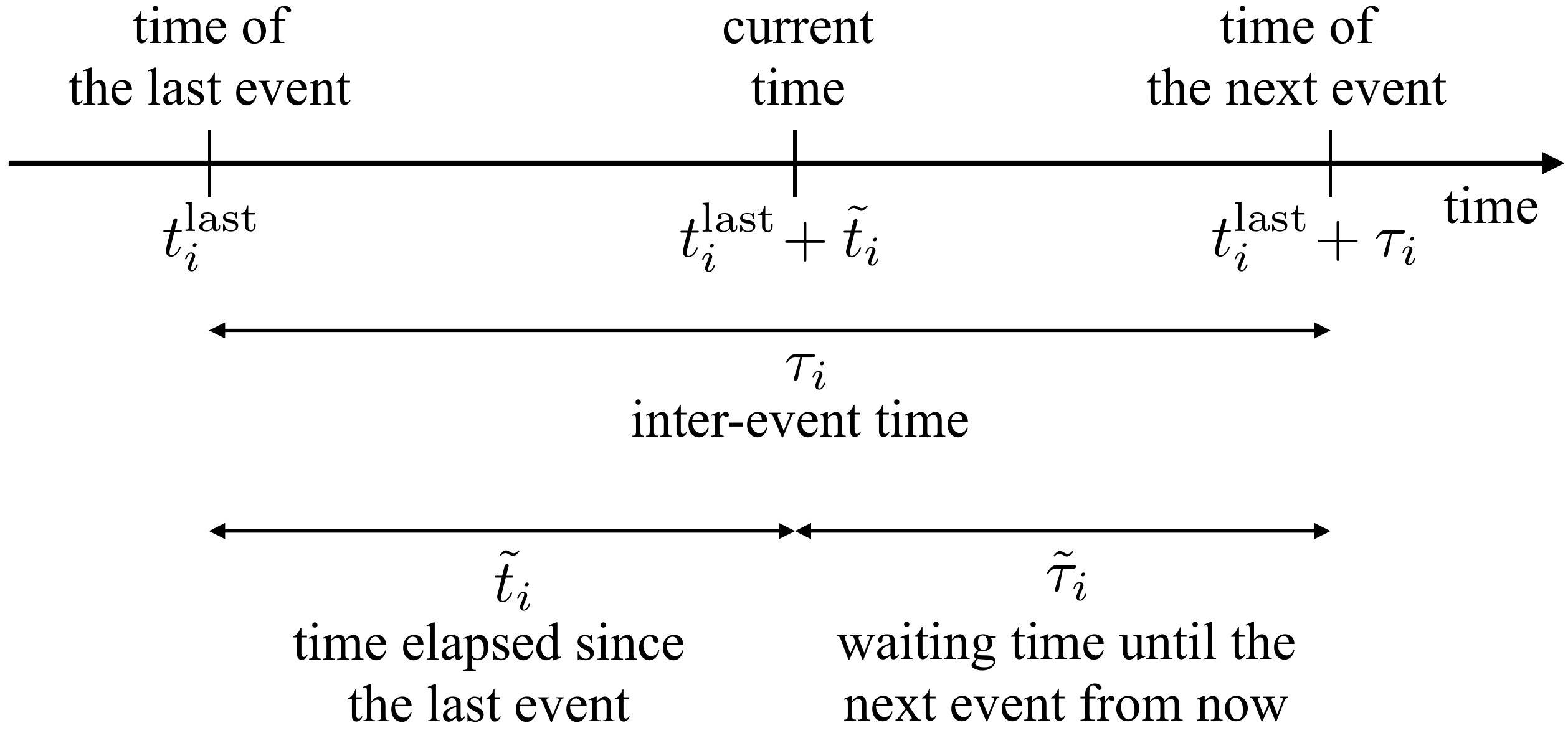}
\vspace{4pt}
\caption{Schematic definition of the different notions of times employed in this section and the relations between them.
}
\label{fig:schem-different-concepts-of-times}
\end{figure}

By definition, the inter-event time $\tau_i$ between two successive events produced by the $i$th process is given by $\psi_i(\tau_i)$. 
We want to know the next event time, $t_i^{\rm last} + \tau_i$, where $t_i^{\rm last}$ is the time of the last event on the $i$th reaction channel. 
The calculation of $t_i^{\rm last} + \tau_i$ is not straightforward to implement using the direct method because knowing the function $\psi_i$ for each reaction channel is not enough on its own to simulate coupled renewal processes. 
In fact, we must be able to calculate not only the waiting time since the last event of the $i$th process, but since an arbitrary time $t$ at which another process may have generated an event. 
Suppose that $i$ has not produced an event for a time $\tilde{t}_i$ after its last event. (The current time is thus $t_i^{\rm last}+\tilde{t}_i$.) 
We denote by $\tilde\tau_i$ the waiting time until the next event starting from time $t_i^{\rm last} + \tilde{t}_i$.
See Fig.~\ref{fig:schem-different-concepts-of-times} for a schematic definition of the different notions of times. The waiting time $\tilde\tau_i$ does not obey $\psi_i(\tau_i)$. 
Instead, $\tilde\tau_i$ obeys the following conditional probability density with which the next event occurs at time $t_i^{\rm last}+\tilde{t}_i + \tilde\tau_i$ given that no event has occurred between $t_i^{\rm last}$ and $t_i^{\rm last}+\tilde{t}_i$: 
\begin{align}
\psi_i^{\rm w}(\tilde\tau_i|\tilde{t}_i) =& 
\frac{(\text{Probability that the next event occurs at } t^{\rm last}_i+\tilde{t}_i+\tilde\tau_i)}
{\text{(Probability of no event between } t_i \text{ and } t^{\rm last}_i+\tilde{t}_i)}\notag\\
=& 
\frac{\psi_i(\tilde{t}_i+\tilde\tau_i)}{\Psi_i(\tilde{t}_i)} .
\label{eq:psi(tau | tildet_i) gen Gillespie}
\end{align}
Here
\begin{equation}
\Psi_i(\tilde{t}_i) = \int_{\tilde{t}_i}^\infty \psi_i(\tau^{\prime}){\rm d}\tau^{\prime}
\label{eq:Psi_i(t_i)}
\end{equation}
is the survival probability, i.e., the probability that the inter-event time is larger than $\tilde{t}_i$. The above argument shows that the waiting time to the next event for each process explicitly depends on $\tilde{t}_i$. Therefore, we need to record when the last event has happened (i.e, $t_i^{\rm last}$, which is $\tilde{t}_i$ before the current time) to generate the waiting time. 

As an example, we consider a power-law distribution of inter-event times given by
\begin{equation}
\psi_i(\tau_i) = \frac{\alpha-1}{(1+\tau_i)^{\alpha}} .
\label{eq:pareto-waiting-time-distribution}
\end{equation}
By substituting Eq.~\eqref{eq:pareto-waiting-time-distribution} into Eq.~\eqref{eq:psi(tau | tildet_i) gen Gillespie}, we find the probability distribution for the waiting time $\tilde{\tau}_i$ until the $i$th renewal process generates its next event given that a time $\tilde{t}_i$ has already elapsed since its last event:
\begin{align}
\psi_i^{\rm w}(\tilde{\tau}_i|\tilde{t}_i)
=& \frac{(\alpha-1)(1+\tilde{t}_i)^{\alpha-1}}
{(1+\tilde{t}_i+\tilde{\tau}_i)^{\alpha}} .
\end{align}
Due to the highly skewed shape of $\psi_i$, the expected waiting time until the next event becomes longer if more time has already elapsed without an event; that is, $\tilde{\tau}_i$ tends to be longer than $\tau_i$. 
One can show this counterintuitive result by comparing the mean values of $\langle\tilde{\tau}_i \rangle$ and  $\langle \tau_i \rangle$. 
The former is equal to
$\langle \tilde{\tau}_i \rangle = \int_0^{\infty} \tau' \psi_i^{\rm w}(\tau'|\tilde{t}_i) \text{d}\tau' = (1+\tilde{t}_i)/(\alpha-2)$. This is larger than the latter, which is given by 
$\langle \tau_i \rangle = \int_{0}^{\infty} \tau' \psi_i (\tau') \text{d}\tau' = 1/(\alpha-2)$.

When $i$ is a Poisson process, $\psi_i$ is an exponential distribution, and such a complication does not occur.
The memoryless property of the exponential distribution yields $\psi_i^{\rm w}(\tilde{\tau}_i|\tilde{t}_i) = \psi_i(\tilde{\tau}_i)$, which we can verify by plugging the exponential distribution into Eq.~\eqref{eq:psi(tau | tildet_i) gen Gillespie}:
\begin{align}
\psi_i^{\rm w}(\tilde{\tau}_i|\tilde{t}_i) 
=& \frac{\lambda e^{-\lambda (\tilde{t}_i + \tilde{\tau}_i)}}
{e^{-\lambda \tilde{t}_i}}\notag\\
=& \lambda e^{-\lambda \tilde{\tau}_i} .
\end{align}
Therefore, $\psi_i^{\rm w}(\tilde{\tau}_i | \tilde{t}_i)$ does not depend on the time elapsed since the last event, $\tilde{t}_i$, and is the same as the original exponential distribution, $\psi_i(\tilde{\tau_i})$. The original Gillespie algorithm fully exploits this property of Poisson processes.

To build a direct Gillespie method for simulating coupled renewal processes, we need to calculate two quantities:
(i) the time until the next event in the entire population, $\tau$, whichever process produces this event; 
(ii) the probability $\Pi_i$ that the next event is produced by the $i$th process.
We denote by $\phi(\tau, i | \{\tilde{t}_j\})$ the probability density for the $i$th process, and not any other process, to generate the next event after a time $\tau$ conditioned on the time elapsed since the last event of all processes in the population, $\{\tilde{t}_j\} \equiv \{ \tilde{t}_1, \ldots, \tilde{t}_M\}$.
It should be noted that we need to condition on each $\tilde{t}_j$.
This is because $\phi(\tau, i | \{\tilde{t}_j\})$ depends not only on the $i$th renewal process generating an event after the waiting time $\tau$ but also on all the other processes not generating any event during this time.
By putting all this together, we obtain
\begin{equation}
\phi(\tau, i | \{\tilde{t}_j\}) = \psi_i^{\rm w}(\tau|\tilde{t}_i) \prod_{j=1; j\neq i}^M \Psi_j(\tau | \tilde{t}_j) ,
\label{eq:phi(tau, i | set of tildet_j) Gillespie}
\end{equation}
where $\Psi_j(\tau | \tilde{t}_j)$ is the conditional survival probability for the waiting time of the $j$th process if it were running in isolation, given that its last event occurred a time $\tilde{t}_j$ ago.

Equation~\eqref{eq:phi(tau, i | set of tildet_j) Gillespie} is composed of two factors. The first factor is the probability density for the $i$th process to generate the next event within a small time window around $\tau$ (i.e., between $\tau$ and $\tau + \text{d}\tau$ from now, where $\text{d}\tau$ is infinitesimally small), corresponding to the probability density $\psi_i^{\rm w}(\tau|\tilde{t}_i)$. The other factor is the probability that none of the other $M-1$ processes generates an event within this time window, corresponding to the product of the survival probabilities $\Psi_j(\tau | \tilde{t}_j)$ over all $j \neq i$.
Using Eq.~\eqref{eq:psi(tau | tildet_i) gen Gillespie}, we obtain $\Psi_j(\tau | \tilde{t}_j)$ as follows:
\begin{equation}
\Psi_j(\tau | \tilde{t}_j) = \int_{\tau}^{\infty} \psi_j^{\rm w}(\tau^{\prime} | \tilde{t}_j) {\rm d}\tau^{\prime}
=
\frac{\Psi_j(\tilde{t}_j+\tau)}{\Psi_j(\tilde{t}_j)} .
\label{eq:Psi_j(tau | tildet_j) Gillespie}
\end{equation}
By substituting Eqs.~\eqref{eq:psi(tau | tildet_i) gen Gillespie} and \eqref{eq:Psi_j(tau | tildet_j) Gillespie} into Eq.~\eqref{eq:phi(tau, i | set of tildet_j) Gillespie}, we obtain
\begin{equation}
\phi(\tau, i | \{\tilde{t}_j\}) = \frac{\psi_i(\tilde{t}_i+\tau)}{\Psi_i(\tilde{t}_i+\tau)} \Phi(\tau | \{\tilde{t}_j\}) ,
\label{eq:phi(tau, i | set of tildet_j) Gillespie 2}
\end{equation}
where
\begin{equation}
\Phi(\tau | \{\tilde{t}_j\}) = \prod_{j=1}^M \frac{\Psi_j(\tilde{t}_j+\tau)}{\Psi_j(\tilde{t}_j)} .
\label{eq:Phi(tau | set of tildet_j) Gillespie}
\end{equation}
We interpret Eq.~\eqref{eq:phi(tau, i | set of tildet_j) Gillespie 2} as follows. 

First, $\Psi_j(\tilde{t}_j)$ is the probability that the $j$th renewal process has not generated any event for a time $\tilde{t}_j$ since its last event. 
The factor $\Psi_j(\tilde{t}_j+\tau)$ is the probability that the same process has not generated any event for time $\tilde{t}_j$ since its last event and it does not generate any event for another time $\tau$. 
Therefore, $\Psi_j(\tilde{t}_j+\tau) / \Psi_j(\tilde{t}_j)$ is the conditional probability that the $j$th process does not generate any event during the next time $\tau$ given that a  time $\tilde{t}_j$ has already elapsed since it generated its last event.
Equation~\eqref{eq:Phi(tau | set of tildet_j) Gillespie} gives the probability that none of the $M$ processes produces an event for time $\tau$. So, it is the survival probability for the entire population. In other words, it is the probability for the next event in the entire population occurs sometime after time $\tau$ from now.

Second, the factor $\psi_i(\tilde{t}_i+\tau) / \Psi_i(\tilde{t}_i+\tau)$ on the right-hand side of Eq.~\eqref{eq:phi(tau, i | set of tildet_j) Gillespie 2} is the probability density function that the $i$th process generates an event at a time $\tilde{t}_i + \tau$ since its last event given that it has not generated any event before this time since the last event. Only this factor creates the dependence of $\phi(\tau, i | \{\tilde{t}_j\})$ on $i$. Given this observation, we define
\begin{equation}
\Pi_i \equiv \frac{\phi(\tau, i | \{ \tilde{t}_j\})}{\sum_{j=1}^M \phi(\tau, j | \{\tilde{t}_j\})}
= \frac{\lambda_i (\tilde{t}_i+\tau)}{\sum_{j=1}^M \lambda_j(\tilde{t}_j+\tau)} ,
\label{eq:Pi(i | tau, set of tildet_j) Gillespie}
\end{equation}
where
\begin{equation}
\lambda_i(t) = \frac{\psi_i(t)}{\Psi_i(t)}
\end{equation}
is the instantaneous rate of the $i$th process.

In the original Gillespie algorithm, we equated the survival probability of the next event time for the entire population to $u$, a random variate obeying a uniform density on $(0, 1]$, to produce $\tau$ using inverse sampling. Similarly, a non-Markovian Gillespie algorithm can use inverse sampling to produce $\tau$ based on Eq.~\eqref{eq:Phi(tau | set of tildet_j) Gillespie}. However, once a uniform random variate $u$ is drawn, solving $\Phi(\tau | \{\tilde{t}_j\}) = u$ is time-consuming because one cannot explicitly solve $\Phi(\tau | \{\tilde{t}_j\}) = u$ for $\tau$ in general, and thus one must solve it by numerical integration to produce each single event.
This restriction does not prevent the algorithm from working but makes it too slow to be of practical use in many cases.

The non-Markovian Gillespie algorithm resolves this issue as follows.
We first rewrite Eq.~\eqref{eq:Phi(tau | set of tildet_j) Gillespie} as
\begin{equation}
\Phi(\tau | \{\tilde{t}_j\}) = \exp \left[ -\sum_{j=1}^M \ln\frac{\Psi_j(\tilde{t}_j)}{\Psi_j(\tilde{t}_j+\tau)}\right].
\label{eq:Phi(tau | set of tildet_j) Gillespie 2}
\end{equation}
When $M$ is large, it is unlikely that no process generates an event
during a long time interval. 
Therefore, the $\tau$ values realized as the solution of
$\Phi(\tau | \{\tilde{t}_j\}) = u$
will generally be small. This is equivalent to the situation in which $\Phi(\tau | \{\tilde{t}_j\})$ is tiny except for $\tau\approx 0$. 
Based on this observation, we approximate Eq.~\eqref{eq:Phi(tau | set of tildet_j) Gillespie 2} by a first-order cumulant expansion around $\tau=0$. 
This is done by the substitution of the following Taylor expansion of $\Psi_j(\tilde{t}_j+\tau)$:
\begin{equation}
\Psi_j(\tilde{t}_j+\tau) = \Psi_j(\tilde{t}_j) - \psi_j(\tilde{t}_j)\tau + O(\tau^2) ,
\label{eq:nMGA first order approximation 1}
\end{equation}
for $j=1, \ldots, M$, into Eq.~\eqref{eq:Phi(tau | set of tildet_j) Gillespie 2}. This substitution leads to the following simplified expression for $\Phi (\tau | \{\tilde{t}_j\})$: 
\begin{align}
\Phi (\tau | \{\tilde{t}_j\}) =& \exp \left[ -\sum_{j=1}^M \ln\frac{\Psi_j(\tilde{t}_j)}{\Psi_j(\tilde{t}_j) - \psi_j(\tilde{t}_j)\tau + O(\tau^2)}\right]\notag\\
=& \exp \left\{ -\sum_{j=1}^M \ln \left[1 + \frac{\psi_j(\tilde{t}_j)}{\Psi_j(\tilde{t}_j)}\tau + O(\tau^2)\right]\right\}\notag\\
=& \exp \left[ -\sum_{j=1}^M \frac{\psi_j(\tilde{t}_j)}{\Psi_j(\tilde{t}_j)}\tau + O(\tau^2) \right]\notag\\
\approx& \exp \left[-\tau M \overline{\lambda}(\{\tilde{t}_j\})\right],
\label{eq:nMGA first order approximation 2}
\end{align}
where 
\begin{equation}
\overline{\lambda}(\{\tilde{t}_j\}) = \frac{\sum_{j=1}^M \lambda_j(\tilde{t}_j)}{M} = \frac{1}{M}\sum_{j=1}^M \frac{\psi_j(\tilde{t}_j)}{\Psi_j(\tilde{t}_j)}.
\label{eq:average instantaneous rate for gen Gillespie}
\end{equation}
The variable $\overline{\lambda}(\{\tilde{t}_j\})$ is
the average instantaneous event rate. 
By instantaneous, we mean that the event rate changes over time even if no event has happened, which contrasts with the situation of the Poisson processes. The variant of a Poisson process in which the event rate varies over time is called the nonhomogeneous Poisson process\index{nonhomogeneous Poisson process}. However, the non-Markovian Gillespie algorithm assumes that the event rate $\overline{\lambda}(\{\tilde{t}_j\})$ stays constant until the next event occurs somewhere in the coupled renewal processes. This is justified because the time to the next event, $\tau$, is small when $M$ is large, and therefore the change in $\overline{\lambda}(\{\tilde{t}_j\})$ should be negligible. See \cite{Legault2019TheorEcol} for an application of the same idea to stochastic population dynamics in ecology when the environment is dynamically changing.

Note that the Taylor expansion given by Eq.~\eqref{eq:nMGA first order approximation 1} assumes that all
$\Psi_j(\tilde{t}_j)$ are analytical at $\tilde{t}_j = 0$. 
This is not always the case in practice, which may cause some terms to diverge in the Taylor expansion of $\Psi_i(\tilde{t}_i + \tau)$, where $i$ is the process that has generated the last event.
To deal with this, the authors proposed to simply remove the renewal process that has generated the last event from the summation in Eq.~\eqref{eq:average instantaneous rate for gen Gillespie}.

We determine the time to the next event by solving 
$\Phi(\tau | \{\tilde{t}_j\}) 
= u$ for $\tau$ using the approximation given by Eq.~\eqref{eq:nMGA first order approximation 2}. 
By doing this, we obtain
\begin{equation}
\tau = - \frac{\ln u} {M\overline{\lambda}(\{\tilde{t}_j\})}.
\label{eq:nMGA tau final}
\end{equation}
Now, the computation of $\tau$ is as fast as that for the original Gillespie algorithm except that the computation of $\overline{\lambda}(\{\tilde{t}_j\})$ may be complicated to some extent.
Because $\tau$ should be small when $M$ is large,
one determines the process that generates this event by setting $\tau=0$ in Eq.~\eqref{eq:Pi(i | tau, set of tildet_j) Gillespie}, i.e.,
\begin{equation}
\Pi_i = \frac{\lambda_i(\tilde{t}_i)}{M\overline{\lambda}(\{\tilde{t}_j\})}.
\label{eq:nMGA Pi_i final}
\end{equation}
Equations~\eqref{eq:nMGA tau final} and \eqref{eq:nMGA Pi_i final}
define the non-Markovian Gillespie algorithm \citep{BogunaLafuerza2014PhysRevE}. For Poisson processes, we have $\lambda_i(\tilde{t}_i) = \lambda_i$, and we recover the original direct method, which is given by Eqs.~\eqref{eq:phi(tau) Gillespie} and \eqref{eq:Pi(i) Gillespie}.
Because the non-Markovian Gillespie algorithm assumes large $M$, its accuracy is considered to be good for large $M$.

By putting together these results we can define an extension of the direct method of Gillespie to simulate coupled renewal processes. The algorithm is described in Box~\ref{box:nMGA}. 
For simplicity, we have assumed so-called ordinary renewal processes, in which all processes have had the last event at $t=0$ \citep{Cox1962book,Masuda2020book}. 

\begin{figure}[t]
\begin{boxedtext}{}
\captionof{floatbox}{\textbf{Non-Markovian Gillespie algorithm.}}
\label{box:nMGA}

\begin{enumerate}[start=0]
   \item Initialization:
      \begin{enumerate}
        \item Define the initial state of the system, and set $t = 0$.
        \item Initialize $\tilde{t}_j=0$ for all $j$
        \item Calculate the rate $\lambda_j(\tilde{t}_j)$ for all $j\in\{1, \ldots, M\}$
        \item Calculate $\overline{\lambda}(\{\tilde{t}_j\}) = \sum_{j=1}^M \lambda_j(\tilde{t}_j)/M$.
      \end{enumerate}
   \item Draw a uniform random variate $u_1$ from $(0, 1]$, and generate the waiting time to the next event by $\tau = - \ln u_1 \big/ [ M\overline{\lambda}(\{\tilde{t}_j\}) ]$.
   \item Draw $u_2$ from a uniform distribution on $(0, M\overline{\lambda}(\{\tilde{t}_j\})]$. 
   Select the event $i$ to occur by iterating over $i=1, \ldots, M$ until we find the $i$ for which $\sum_{j=1}^{i-1} \lambda_j(\tilde{t}_j) < u_2 \le \sum_{j=1}^i \lambda_j(\tilde{t}_j)$. 
   \item Perform the event on reaction channel $i$.
   \item Advance the time according to $t \to t + \tau$. 
   \item 
   \begin{enumerate}
        \item Update the list of times since the last event as $\tilde{t}_j \to \tilde{t}_j + \tau$ for all $j\neq i$, and set $\tilde{t}_i=0$. 
        \item If there are processes $j$ whose distribution of inter-event times has changed upon the event, update $\psi_j(\tau)$ for the affected $j$ values, and reset $\tilde{t}_j=0$ if necessary. 
        \item Update $\lambda_j(\tilde{t}_j)$ for all $j\in\{1, \ldots, M\}$ as well as $\overline{\lambda}(\{\tilde{t}_j\}) = \sum_{j=1}^M \lambda_j(\tilde{t}_j)/M$.
        \end{enumerate}
   \item Return to Step 1.
\end{enumerate}
\end{boxedtext}

\end{figure}

In the context of chemical reaction systems, \cite{Carletti2012ComputMathMethodsMed} developed a second-order variant of the Gillespie algorithm for chemical reactions in a dynamically varying volume, which one can also use to simulate more general non-Markovian processes. The non-Markovian Gillespie algorithm we presented above can be seen as a first-order algorithm in terms of $\tau$ (see Eqs.~\eqref{eq:nMGA first order approximation 1} and \eqref{eq:nMGA first order approximation 2}). 
The idea of the second-order variant is to use the Taylor expansion of $\Psi_j(\tilde{t}_j+\tau)$ up to the second order, i.e.,
\begin{equation}
\Psi_j(\tilde{t}_j+\tau) = \Psi_j(\tilde{t}_j) - \psi_j(\tilde{t}_j)\tau - \frac{\psi_j'(\tilde{t}_j) \tau^2}{2} + O(\tau^3),
\label{eq:nMGA second order approximation 1}
\end{equation}
instead of 
Eq.~\eqref{eq:nMGA first order approximation 1}.
Then, one obtains
\begin{equation}
u = \Phi (\tau | \{\tilde{t}_j\}) \approx
\exp \left[-\tau M \overline{\lambda}(\{\tilde{t}_j\}) + c \tau^2 \right],
\label{eq:nMGA second order approximation 2}
\end{equation}
where $c$ is a constant. We refer to \cite{Carletti2012ComputMathMethodsMed} for its precise form. Therefore, we set $\tau$ by solving the quadratic equation in terms of $\tau$, i.e.,
\begin{equation}
c \tau^2 - M \overline{\lambda}(\{\tilde{t}_j\}) \tau
- \ln u = 0.
\end{equation}

This second order approximation should generally be a more accurate approximation than the first-order one. However, it comes at an increased computational cost for calculating $c$. Furthermore, it still relies on an assumption of large $M$, which breaks down when $M$ is of the order of 1. For contagion processes, for example, $M$ is typically of the order of 1 at the start or near the end of a simulation where it is often the case that only one or a few individuals are infectious.

\subsection{\label{sub:LGA}Laplace Gillespie algorithm}

In spite of the approximations made by the non-Markovian Gillespie algorithm to make it fast enough for practical applications, the algorithm still requires that we update the instantaneous event rates of all processes whenever an event occurs (Step~5(c) in Box~\ref{box:nMGA}). 
This makes its runtime linear in terms of the number of reaction channels $M$. 
Note that we cannot lessen the time complexity by using any of the advanced methods discussed in Section~\ref{sec:computational_complexity}.
This is because these advanced methods only improve the efficiency of selecting the reaction channel in Step~2 in Box~\ref{box:nMGA} and leave the overall time complexity of the entire algorithm to be linear. 
Note that using the binary search tree (Section~\ref{sub:binary_tree}) makes the algorithm less efficient because all nodes in the binary tree then need to be updated after each event, resulting in an algorithm with $\O(M\log M)$ time complexity.

\begin{figure}
\centering
\includegraphics[width=\textwidth]{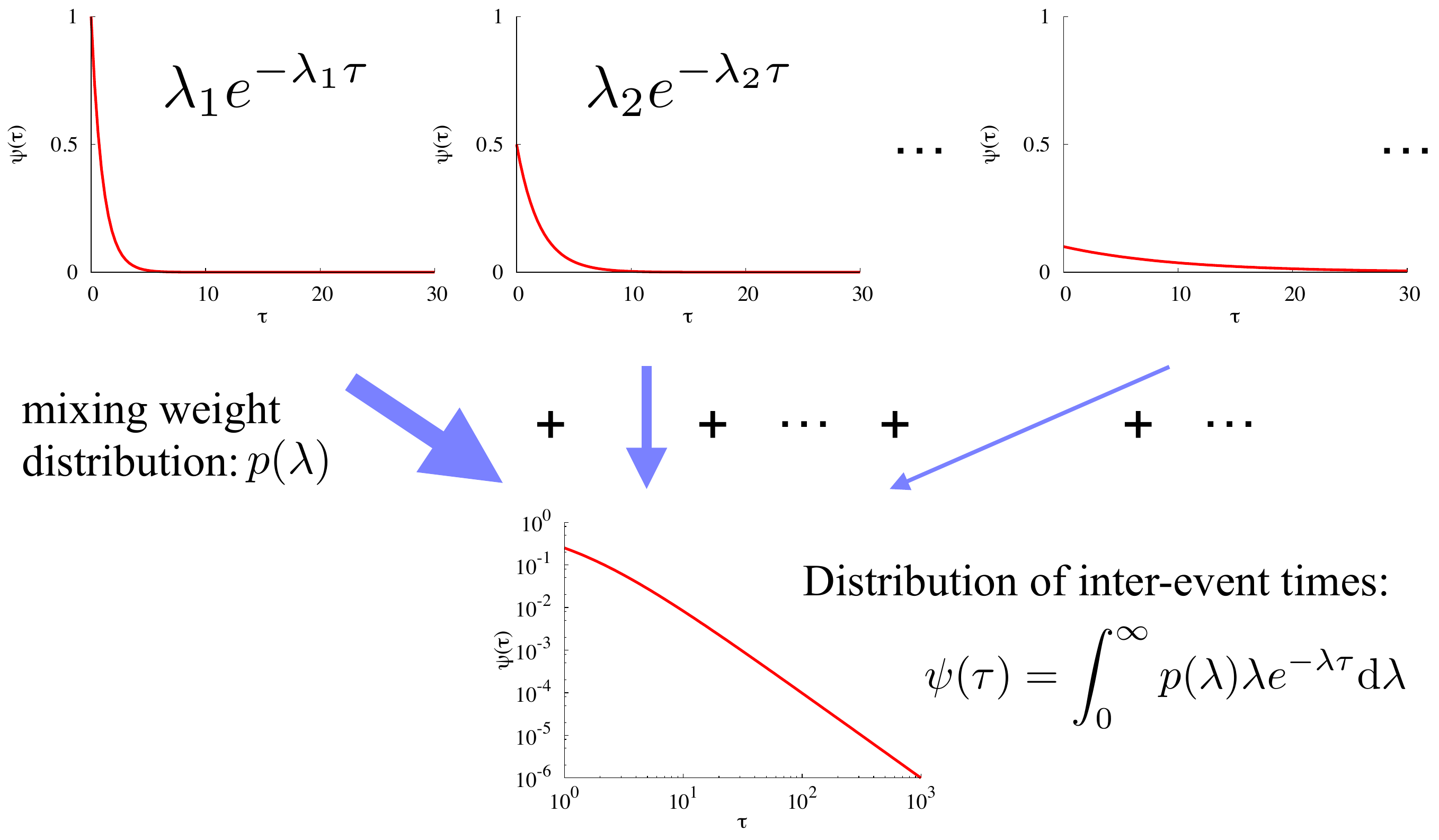}
\caption{Schematic showing a mixture of exponential distributions and the mechanism of the Laplace Gillespie algorithm. One draws $\lambda_1$, $\lambda_2$, and so forth from $p(\lambda)$. The probability density function $p(\lambda)$ is called the mixing weight distribution, representing how probable each value of $\lambda$ is to be drawn. 
Once a $\lambda$ value is drawn, one generates the time to the next time, $\tau$, according to the exponential distribution $\lambda e^{-\lambda \tau}$. As a result, one mixes exponential distributions with mixing weights $p(\lambda)$ to obtain the distribution of inter-event times, $\psi(\tau)$. Although each component distribution is an exponential distribution, the mixture may yield a fat-tailed distribution.}
\label{fig:exponential mixture}
\end{figure}

In this section, we explain an alternative generalization of the direct method to simulate non-Poissonian renewal processes. The algorithm, which we call the \emph{Laplace Gillespie algorithm}\index{Laplace Gillespie algorithm}, exploits mathematical properties of the Laplace transform\index{Laplace transform} \citep{Masuda2018SiamRev} to allow exact and fast simulation of non-Poissonian renewal processes with fat-tailed waiting-time distributions.
It takes advantage of the fact that a fat-tailed distribution often can be expressed as a mixture of exponential distributions. In other words, an appropriately weighted average of $\lambda e^{-\lambda \tau}$ over different values of $\lambda$ approximates a desired fat-tailed distribution well. 
This situation is schematically shown in Fig.~\ref{fig:exponential mixture}.
As nothing ever comes for free, the Laplace Gillespie algorithm does not work for simulating arbitrary renewal processes. It only works for renewal processes whose distribution of inter-event times satisfies a condition known as \emph{complete monotocity}, which we discuss in detail at the end of this section. Luckily, fat-tailed distributions of inter-event times that are ubiquitous in human interaction dynamics are often well modeled by completely monotone functions, so that the Laplace Gillespie algorithm is broadly applicable to social systems. 

To explain the Laplace Gillespie algorithm, we first consider a single renewal process, which has an associated probability density function of inter-event times $\psi(\tau)$. 
Our aim is to (repeatedly) produce inter-event times, $\tau$, that obey the probability density $\psi(\tau)$. 
To this end, we first draw a rate of a Poisson process, denoted by $\lambda$, from a fixed probability density $p(\lambda)$. 
Second, we draw the next value of $\tau$ from
the exponential distribution $\lambda e^{-\lambda \tau}$ as if we were running a Poisson process with rate $\lambda$. 
Third, we advance the clock by $\tau$ and produce the event. Fourth, we repeat the procedure to determine the time to the next event.
In other words, we redraw the rate, which we denote by $\lambda'$ to avoid confusion, from the probability density $p(\lambda)$ and generate the time to the next event from the exponential distribution $\lambda' e^{-\lambda' \tau}$.

If the $\lambda$ value drawn from $p(\lambda)$ happens to be large, then, $\tau$ tends to be small, and vice versa. 
Because there is diversity in the value of $\lambda$, the eventual distribution of inter-event times, $\psi(\tau)$, is more dispersed than a single exponential distribution \citep{Yannaros1994AnnInstStatMath}. 

For a given $p(\lambda)$, the process generated by this algorithm is a renewal process. By construction, $\psi(\tau)$ is the mixture of exponential distributions given by
\begin{equation}
\psi(\tau) = \int_0^{\infty} p(\lambda) \lambda e^{-\lambda \tau} {\rm d}\lambda.
\label{eq:psi(tau) mix}
\end{equation}
For example, if there are only two possible values of $\lambda$, i.e., $\lambda_{\rm low}$ and $\lambda_{\rm high}$ ($> \lambda_{\rm low}$), one obtains
\begin{equation}
p(\lambda) = q \delta (\lambda - \lambda_{\rm low}) + (1-q) \delta (\lambda - \lambda_{\rm high}),
\label{eq:two-peak}
\end{equation}
where $\delta$ is the Dirac delta function. Equation~\eqref{eq:two-peak} just says that
$\lambda = \lambda_{\rm low}$ occurs with probability $q$ and $\lambda = \lambda_{\rm high}$ occurs with probability $1-q$.
Inserting 
Eq.~\eqref{eq:two-peak} in Eq.~\eqref{eq:psi(tau) mix} yields
\begin{equation}
\psi(\tau) = q \lambda_{\rm low}
e^{-\lambda_{\rm low}\tau} + (1-q)
\lambda_{\rm high} e^{-\lambda_{\rm high}\tau},
\end{equation}
i.e., a mixture of two exponential distributions. (See
\cite{Jiang2016JStatMech,Fonsecadosreis2020PhysRevE,Masuda2020PhysRevResearch,Okada2020RSocOpenSci}
for analysis of inter-event times with a mixture of two exponential distributions.)

\begin{figure}
\centering
\includegraphics[width=.8\textwidth]{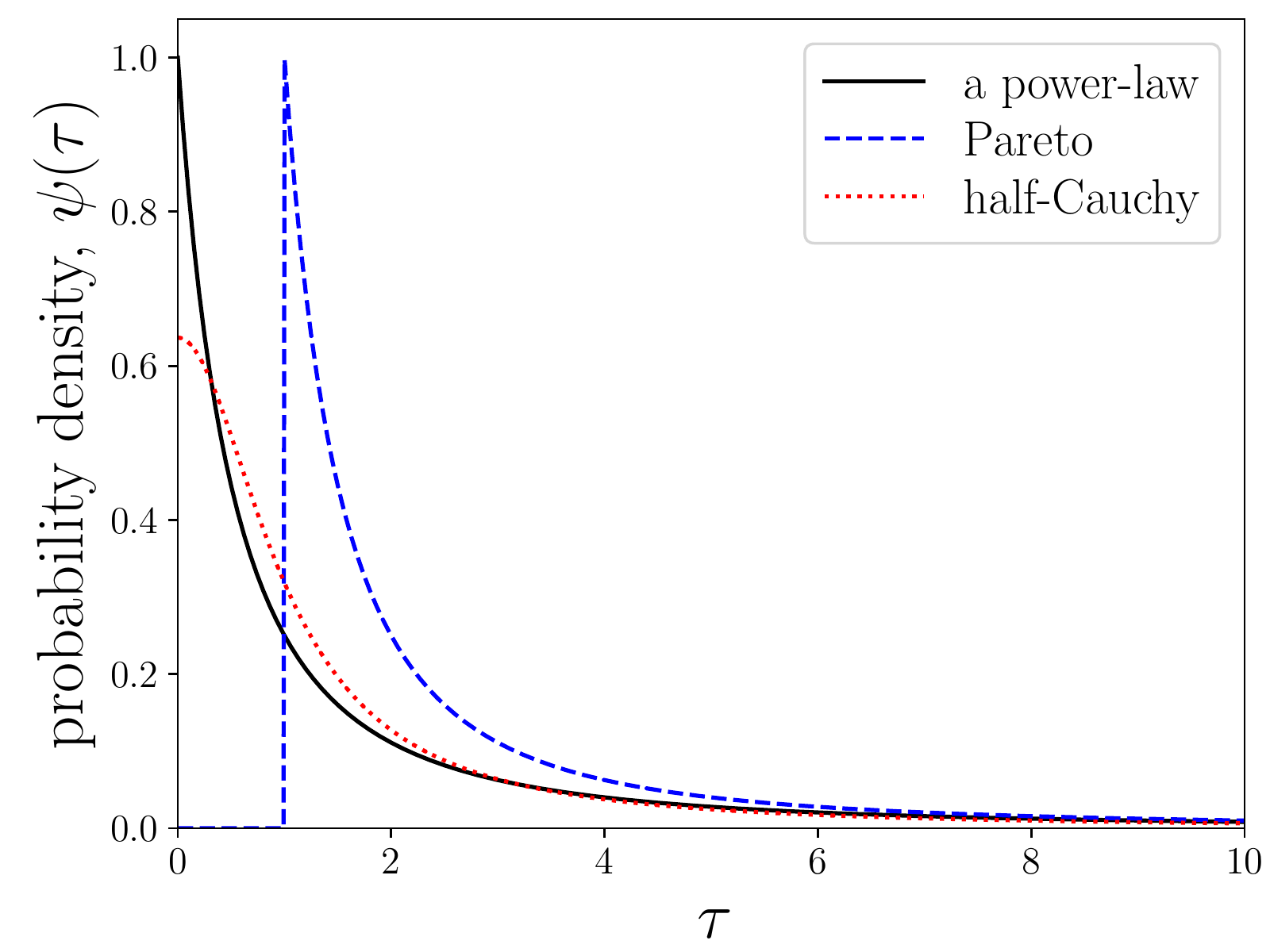}
\caption{Three power-law distributions, i.e., the distribution given by Eq.~\eqref{eq:psi(tau) power law} with $\alpha=1$ and $\kappa=1$, a Pareto distribution with $\alpha=1$ and $\tau_0 = 1$, and a half-Cauchy distribution. 
Note that the three distributions follow the same asymptotic power law, $\psi(\tau)\propto \tau^{-2}$, as $\tau\to\infty$.}
\label{fig:some-psi(tau)}
\end{figure}

As another example, let us consider the gamma distribution for the distribution of mixing weights, i.e.,
\begin{equation}
p(\lambda) = \frac{\lambda^{\alpha-1}e^{-\lambda/\kappa}}{\Gamma(\alpha)\kappa^{\alpha}},
\label{eq:p(lambda) gamma}
\end{equation}
where $\alpha$ and $\kappa$ are the shape and scale parameters of the gamma distribution, respectively, and
\begin{equation}
\Gamma(\alpha) = \int_{0}^{\infty} x^{\alpha-1} e^{-x} \text{d}x
\end{equation}
is the gamma function. 
Inserting
Eq.~\eqref{eq:p(lambda) gamma} in
Eq.~\eqref{eq:psi(tau) mix} yields
\begin{equation}
\psi(\tau) = \frac{\kappa \alpha}{(1+\kappa\tau)^{\alpha+1}},
\label{eq:psi(tau) power law}
\end{equation}
which is a power-law distribution (see the solid line in Fig.~\ref{fig:some-psi(tau)} for an example). 
Crucially, this example shows that one can create a power-law distribution, which is fat-tailed, by appropriately mixing exponential distributions, which are not fat-tailed.

What we want to really simulate is an ensemble of $M$ simultaneously ongoing renewal processes that are governed by given distributions of inter-event times, $\psi_i(\tau)$ ($i=1, \ldots, M$). 
We suppose that we can realize each $\psi_i(\tau)$ as a mixture of exponential distributions by appropriately setting a distribution of mixing weights $p_i(\lambda)$. In other words, we assume that we can find $p_i(\lambda)$ satisfying $\psi_i(\tau) = \int_0^{\infty} p_i(\lambda) \lambda e^{-\lambda \tau} {\rm d}\lambda$.
The Laplace Gillespie algorithm for simulating such a system is described in Box~\ref{box:Laplace}.

\begin{figure}[t]
\begin{boxedtext}{}
\captionof{floatbox}{\textbf{Laplace Gillespie algorithm.}}
\label{box:Laplace}

\begin{enumerate}[start=0]

\item Initialization:
  \begin{enumerate}
    \item Define the initial state of the system, and set $t=0$.
    \item Initialize each of the $M$ renewal processes by drawing the rate $\lambda_j$ of the $j$th Poisson process from $p_j(\lambda_j)$ for all $j\in \{1, \ldots, M\}$.
    \item Calculate the total event rate $\Lambda = \sum_{j=1}^M \lambda_j$.
  \end{enumerate}

\item\label{step:Laplace Gillespie 2} 
Draw a random variate $u$ from a uniform distribution on (0, 1], and generate the waiting time to the next event by $\tau = -\ln u/\Lambda$.

\item Select the process that generates the next event with probability $\Pi_i = \lambda_i/\Lambda$. 

\item Implement the event taking place on the $i$th renewal process.  

\item Advance the clock according to $t \to t + \tau$. 

\item\label{step:Laplace Gillespie 4} 
  \begin{enumerate}
    \item Update $p_i(\lambda_i)$ if it has changed following the event.
    \item  Redraw a rate $\lambda_i$ according to $p_i(\lambda_i)$. 
    \item If there are other processes $j$ whose distribution of inter-event times has changed following the event on the $i$th process, update each affected $p_j(\lambda_j)$, and redraw $\lambda_j$ from the new $p_j(\lambda_j)$. The event rates of the other processes remain unchanged.
    \item Update the total event rate $\Lambda = \sum_{j=1}^M \lambda_j$.
  \end{enumerate}

\item Return to Step~\ref{step:Laplace Gillespie 2}.
\end{enumerate}

\end{boxedtext}
\end{figure}

\begin{figure}
\centering
\includegraphics[width=.49\textwidth]{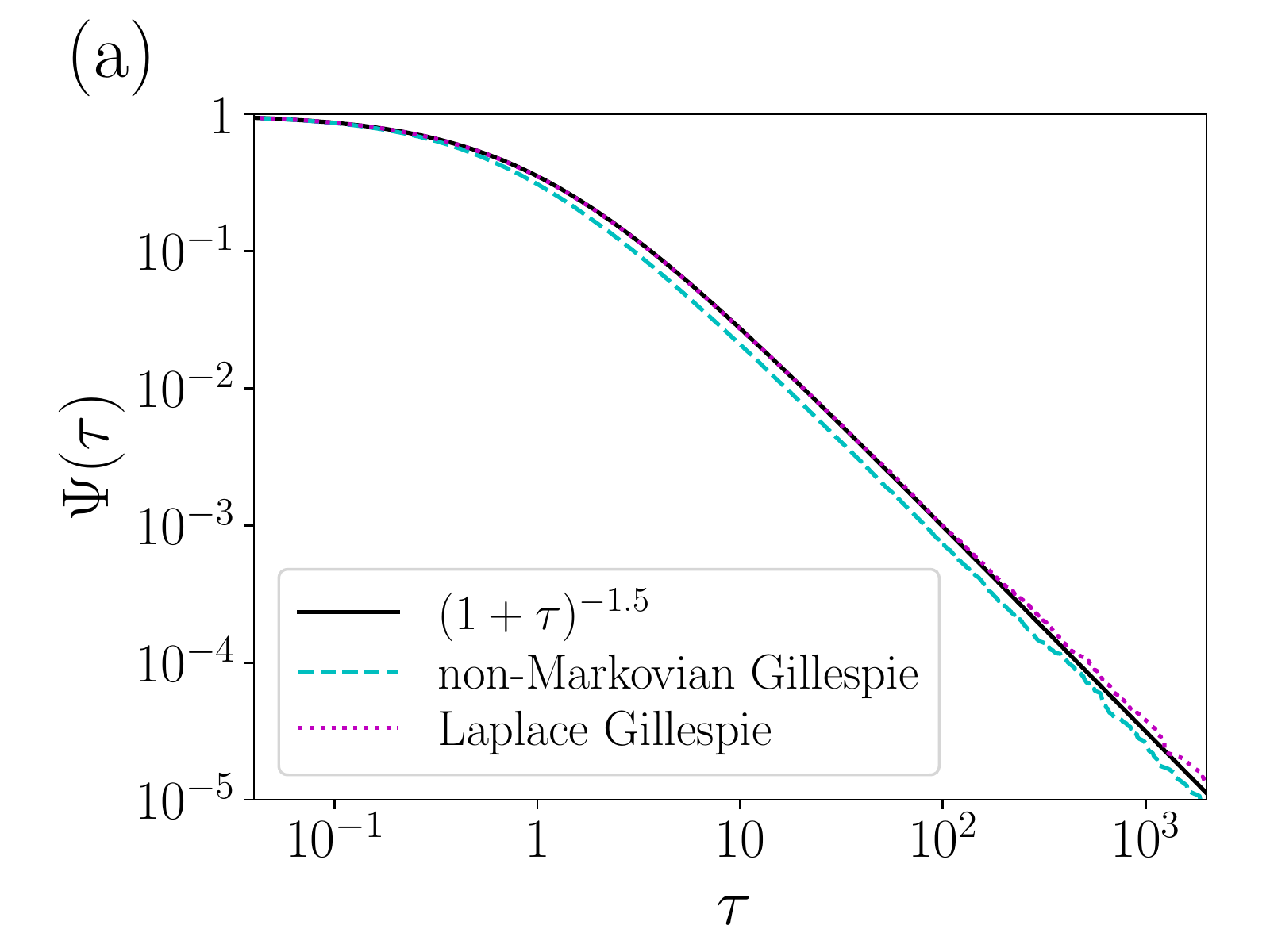}
\includegraphics[width=.49\textwidth]{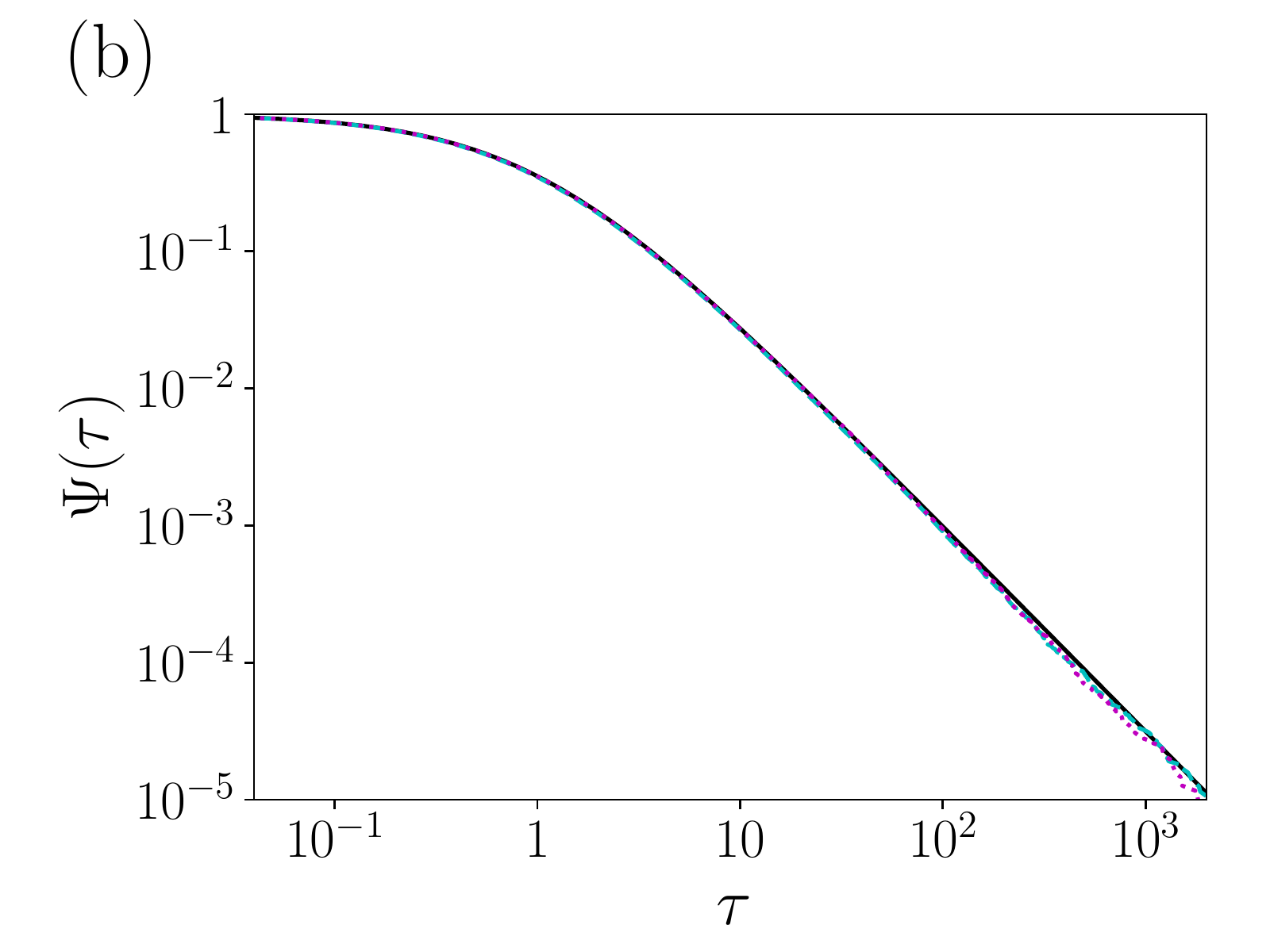}
\caption{Comparison between the non-Markovian Gillespie algorithm and the Laplace Gillespie algorithm. We use a power-law distribution of inter-event times $\psi(\tau) = \alpha/(1+\tau)^{\alpha+1}$ with $\alpha=1.5$. (a) $M=10$. (b) $M=100$. 
For each algorithm, the survival function of the inter-event time distribution is plotted for just one of the $M$ processes and compared against the ground truth, i.e., $\Psi(\tau) = 1/(1+\tau)^{\alpha}$.
}
\label{fig:nMGA vs LGA}
\end{figure}

In contrast to the non-Markovian Gillespie algorithm, the Laplace Gillespie algorithm is exact for arbitrary values of $M$. Figure~\ref{fig:nMGA vs LGA} shows an example in which the Laplace Gillespie algorithm is considerably more accurate than the non-Markovian Gillespie algorithm when $M=10$ (Fig.~\ref{fig:nMGA vs LGA}(a)), whereas both algorithms are sufficiently accurate when $M=100$ (Fig.~\ref{fig:nMGA vs LGA}(b)). 
In addition, the Laplace Gillespie algorithm tends to be faster than the non-Markovian Gillespie algorithm
\citep{Masuda2018SiamRev} because it does not need to update all the $\lambda_i$ values (with $i=1, \ldots, M$) after each event.
Used together with the binary tree structure (Section~\ref{sub:binary_tree}) or the composition and rejection method (Section~\ref{sub:composition-rejection}), the Laplace Gillespie algorithm can thus simulate coupled renewal processes in $\O(\log M)$ or $\O(1)$ time.

Not all functional forms for $\psi_i(\lambda)$ can be generated as a mixture of exponentials.
In these cases we cannot use the Laplace Gillespie algorithm. By contrast, one can  use the non-Markovian Gillespie algorithm for any $\psi_i(\tau)$ in principle. To examine more formally to which cases the Laplace Gillespie algorithm is applicable, we integrate both sides of Eq.~\eqref{eq:psi(tau) mix} to obtain
\begin{equation}
\Psi(\tau) = \int_{\tau}^{\infty} \psi(\tau^{\prime}){\rm d}\tau^{\prime} = \int_0^{\infty} p(\lambda) e^{-\lambda \tau} {\rm d}\lambda.
\label{eq:Psi(tau) mix}
\end{equation}
Equation~\eqref{eq:Psi(tau) mix} indicates that the survival probability of inter-event times, $\Psi(\tau)$, is the Laplace transform of $p(\lambda)$. 
Therefore, the Laplace Gillespie algorithm can simulate a renewal process if and only if $\Psi(\tau)$ is the Laplace transform of a probability density function on non-negative values.

It is mathematically known that a necessary and sufficient condition for the existence of $p(\lambda)$ is that $\Psi(\tau)$ is completely monotone\index{completely monotone} \citep{Feller1971book2} and that $\Psi(0)=1$. 
A function $\Psi(\tau)$ is said to be completely monotone if
\begin{equation}
(-1)^n \frac{{\rm d}^n\Psi(\tau)}{{\rm d}\tau^n} \ge 0\quad (\tau\ge 0, n= 0, 1, \ldots).
\label{eq:completely monotone}
\end{equation}
The condition $\Psi(0) = \int_0^{\infty} \psi(\tau) \text{d}\tau = 1$ is always satisfied since $\Psi$ is a survival function. Equations~\eqref{eq:completely monotone} with $n=0$ and $n=1$ read $\Psi(\tau)\ge 0$ and $\psi(\tau)\ge 0$, respectively. These two inequalities are also always satisfied. Equation~\eqref{eq:completely monotone} 
offers non-trivial conditions when $n\ge 2$. For example, the condition with $n=2$ reads 
\begin{equation}
(-1)^2 \frac{\text{d}^2\Psi(\tau)}{\text{d}\tau^2} = - \frac{\text{d}\psi(\tau)}{\text{d}\tau} \ge 0,
\label{eq:completely monotone n=2}
\end{equation}
i.e., $\text{d}\psi(\tau) / \text{d}\tau \le 0$. Therefore, $\psi(\tau)$ must monotonically decrease with respect to $\tau$. This condition excludes the Pareto distribution, which is a popular form of power-law distribution, i.e.,
\begin{equation}
\psi(\tau)= 
\begin{cases}
\frac{\alpha}{\tau_0}\left(\frac{\tau_0}{\tau}\right)^{\alpha+1} & (\tau\ge \tau_0),\\
0 & (\tau < \tau_0),
\end{cases}
\label{eq:Pareto}
\end{equation}
where $\alpha>0$ and $\tau_0>0$. We show the Pareto distribution with $\alpha=1$ and $\tau_0=1$ by the dashed line in Fig.~\ref{fig:some-psi(tau)}. Note that $\psi(\tau)$ discontinuously increases as $\tau$ increases across $\tau = \tau_0$; note that the Pareto distribution is defined for $\tau \ge 0$ (and $\psi(\tau)=0$ for $0 \le \tau < \tau_0$). Therefore, one cannot use the Laplace Gillespie algorithm when any $\psi_i(\tau)$ is a Pareto distribution.

To show another example of disqualified $\psi(\tau)$, consider Eq.~\eqref{eq:completely monotone}
for $n=3$, i.e., 
$\text{d}\psi^2(\tau)/\text{d}\tau^2 \ge 0$.
The half-Cauchy distribution, which is another form of power-law distribution, defined by 
\begin{equation}
\psi(\tau) = \frac{2}{\pi(\tau^2 + 1)},
\end{equation}
where $\tau \ge 0$, violates this condition. (See the red dotted line in Fig.~\ref{fig:some-psi(tau)} for a plot.) This is because ${\rm d}^2\psi(\tau)/{\rm d}\tau^2 = 4(3\tau^2-1)/\left[\pi(\tau^2+1)^3\right]$, whose sign depends on the value of $\tau$. Specifically, the half-Cauchy distribution has an inflection point at $\tau = 1/\sqrt{3}$. Note that the half-Cauchy distribution satisfies the condition for $n=2$ (Eq.~\eqref{eq:completely monotone n=2}) because
${\rm d}\psi(\tau)/{\rm d}\tau = -4\tau/\left[\pi(\tau^2+1)\right] < 0$.

Complete monotonicity implies that the coefficient of variation (CV)\index{coefficient of variation (CV)}, defined by the standard deviation divided by the mean, of $\psi(\tau)$ is larger than or equal to 1 \citep{Yannaros1994AnnInstStatMath}. 
This is natural because an exponential distribution, $\lambda e^{-\lambda \tau}$,
has a CV equal to one. Because we are mixing exponential distributions with different $\lambda$ values, the CV for the mixture of exponential distributions must be at least 1.
This necessary condition for complete monotonicity excludes some distributions having less dispersion (i.e., standard deviation) than exponential distributions.

We stated various negative scenarios, but there are many distributions of inter-event times, $\psi(\tau)$, for which the Laplace Gillespie algorithm works.
The power-law distribution given by
Eq.~\eqref{eq:psi(tau) power law} is qualified because one can find the corresponding distribution of mixing weights, which is given by
Eq.~\eqref{eq:p(lambda) gamma}.
In fact, using 
Eq.~\eqref{eq:psi(tau) power law}, we obtain
\begin{equation}
\Psi(\tau) = \int_{\tau}^{\infty} \psi(\tau') \text{d}\tau' = \frac{1}{(1+\kappa\tau)^{\alpha}}.
\label{eq:Psi(tau) power law}
\end{equation}
It is easy to verify that this $\Psi(\tau)$ is a completely monotone function.

As a second example, assume that $p(\lambda)$ is a uniform density on $[\lambda_{\min}, \lambda_{\max}]$ \citep{Hidalgo2006PhysicaA}. By Laplace transforming $p(\lambda)$ using Eq.~\eqref{eq:Psi(tau) mix}, we obtain
\begin{equation}
\Psi(\tau) = \frac{e^{-\lambda_{\min}\tau}-e^{-\lambda_{\max}\tau}}{\tau\left(\lambda_{\max}-\lambda_{\min}\right)},
\end{equation}
which yields
\begin{equation}
\psi(\tau) = - \frac{\text{d}\Psi(\tau)}{\text{d}\tau}
=
\frac{\lambda_{\min}e^{-\lambda_{\min}\tau}-\lambda_{\max}e^{-\lambda_{\max}\tau}}{\left(\lambda_{\max}-\lambda_{\min}\right)\tau} + \frac{e^{-\lambda_{\min}\tau}-e^{-\lambda_{\max}\tau}}{\left(\lambda_{\max}-\lambda_{\min}\right)\tau^2}.
\end{equation}
Assume that $\lambda_{\min}\ll \lambda_{\max}$. If $\lambda_{\min}>0$, then $\psi(\tau)\propto e^{-\lambda_{\min}\tau}/\tau$ as $\tau\to\infty$, which is a power-law distribution with an exponential cutoff, often resembling empirical data. If $\lambda_{\min}=0$, then $\psi(\tau)\propto 1/\tau^2$ as $\tau\to\infty$.

A third example is when the inter-event time obeys a gamma distribution, i.e.,
\begin{equation}
\psi(\tau) = \frac{\tau^{\alpha-1}e^{-\tau/\kappa}}{\Gamma(\alpha)\kappa^{\alpha}}.
\label{eq:gamma_event-distribution}
\end{equation}
For this $\psi(\tau)$, we can express $\Psi(\tau)$ as the Laplace transform of $p(\lambda)$ if and only if $0<\alpha\le 1$, and $p(\lambda)$ is given by
\begin{equation}
p(\lambda) = \begin{cases}
0 & (0 < \lambda < \kappa^{-1}),\\
\frac{1}{\Gamma(\alpha)\Gamma(1-\alpha)\lambda(\kappa\lambda-1)^{\alpha}} & (\lambda\ge \kappa^{-1}).
\end{cases}
\end{equation}
It is easy to verify that one obtains the exponential distribution by setting $\alpha=1$ in Eq.~\eqref{eq:gamma_event-distribution}.
We refer to \citet{Masuda2018SiamRev} for more examples of renewal processes that the Laplace Gillespie algorithm can simulate.

\subsection{\label{sub:TGA}Temporal Gillespie algorithm}

The \emph{temporal Gillespie algorithm}\index{temporal Gillespie Algorithm} \citep{Vestergaard2015PlosComputBiol} is an adaptation of the direct method to simulate coupled jump processes taking place on switching temporal networks\index{switching (temporal) network}, i.e., networks whose structure changes discontinuously in discrete points in time (Fig.~\ref{fig:switching}). 
For simplicity in the following presentation, and without loss of generality, we furthermore assume that both the network's dynamics and the dynamical process start at time $t=0$.

The starting point for developing a temporal Gillespie algorithm is a single isolated jump process $i$ which has a time-varying event rate $\lambda_i(t, \tilde{t}_i)$. Note that $\lambda_i(t, \tilde{t}_i)$ may depend on both the ``wall clock'' time, $t$, and the time since the last event, $\tilde{t}_i$.
The explicit dependence on time $t$ is not a property of the renewal processes considered in Sections~\ref{sub:nMGA} and \ref{sub:LGA}, for which the event rate depends only on $\tilde{t}_i$. 
Conversely, if $\lambda_i$ depends on $t$ but not on $\tilde{t}_i$, the process is an nonhomogeneous Poisson process\index{nonhomogeneous Poisson process}. 
In this volume, we will only treat nonhomogeneous Poisson processes because they are simpler than the full problem in which the $\lambda_i$ depends on both $t$ and $\tilde{t}_i$.
The derivation of the temporal Gillespie algorithm for the general case follows the same reasoning as for the algorithm for nonhomogeneous Poisson processes but the mathematics is a bit more involved. 
We thus do not show the details here but refer interested readers to \citet{Vestergaard2015PlosComputBiol}.

Similar to how we calculated the waiting-time distribution for a Poisson process in Section~\ref{sub:waiting_time}, we first consider a discrete-time process and then take the continuous-time limit. 
More precisely, we want to know the probability that the $i$th process does not generate an event in a given time window $[t, t + \tau)$ (i.e., its survival probability), which we denote by $\Psi_i(\tau; t)$.
We approximate $\Psi_i(\tau; t)$ by subdividing the interval into $r$ small time steps of size $\delta t = \tau/r$ as follows:
\begin{equation}
    \Psi_i(\tau; t) \approx  \prod_{r'=0}^{r - 1} \left[ 1 - \lambda_i(t + r'\, \delta t)\delta t \right] .
\end{equation}
Taking the limit $\delta t \to 0$, we find the following exact expression for the survival probability using the exponential identity (Appendix~\ref{app:exp_identity}):
\begin{align}
  \Psi_i(\tau; t) &= \exp\left(-\int_{0}^{\tau} \lambda_i(t + \tau^{\prime})\, \text{d}\tau^{\prime} \right) \nonumber\\
  &= \exp\left(-\int_{t}^{t+\tau} \lambda_i(\tau^{\prime})\, \text{d}\tau^{\prime} \right) .
  \label{eq:S_m-TGA}
\end{align}

Equation~\eqref{eq:S_m-TGA} does not reduce to a simple exponential, except in the special the case of a constant $\lambda_i$. 
It does not even reduce to an analytical expression in general.
For example, in the SIR process on a predefined switching temporal network, the infection rate for a given susceptible node $i$ changes whenever an edge appears or disappears between $i$ and an infectious node.
Say, if a susceptible node $i$ has two infectious neighbors and now gets connected to another infectious neighbor, the rate with which $i$ gets infected changes from $2\beta$ to $3\beta$, where $\beta$ is the infection rate per contact.
This means that one cannot in general solve Eq.~\eqref{eq:S_m-TGA} analytically even for a simple constant-rate SIR process in a temporal network.
Nevertheless, owing to the conditional independence property of the jump processes (see Section~\ref{sub:independence}), we can still find a formal expression for the waiting time for the superposition of a set of $M$ processes.
The survival function for a set of $M$ processes is simply the product of the individual survival functions.
Let $t^{\rm last}$ be the time of the last event amongst all $M$ processes. The survival function for the waiting time $\tau$ until the next event amongst all the processes is
\begin{align}
  \Psi(\tau; t^{\rm last}) 
  &= \prod_{i=1}^M \Psi_i(\tau; t^{\rm last}) \nonumber\\
  &= \prod_{i=1}^M \exp\left({- \int_{t^{\rm last}}^{t^{\rm last}+\tau} \lambda_{i}(\tau^{\prime})\, \text{d}\tau^{\prime}}\right) \nonumber\\
  &= \exp\left({-\int_{t^{\rm last}}^{t^{\rm last}+\tau} \sum_{i=1}^M  \lambda_{i}(\tau^{\prime})\, \text{d}\tau^{\prime}}\right) \nonumber\\
  &= \exp\left({-\int_{t^{\rm last}}^{t^{\rm last}+\tau} \Lambda(\tau^{\prime})\, \text{d}\tau^{\prime}}\right) ,
  \  \label{eq:S-TGA}
\end{align}
where we have defined the \emph{total instantaneous rate} as 
\begin{equation}
   \Lambda(t) \equiv \sum_{i=1}^M  \lambda_{i}(t) .
\end{equation}
Note that $\Lambda(t)$  is simply $M$ times the average instantaneous rate, i.e., $\Lambda(t) = M\,\overline{\lambda}(t)$ (see  Section~\ref{sub:nMGA}).

Due to the lack of an analytic expression for $\Lambda(t)$, we need to numerically integrate Eq.~\eqref{eq:S-TGA} to evaluate it. However, inverting Eq.~\eqref{eq:S-TGA} to directly draw the waiting time is computationally too expensive.
To overcome this, the temporal Gillespie algorithm works instead with unitless, normalized waiting times. 
Given the waiting time, $\tau$, we define the normalized waiting time, denoted by $\overline{\tau}$, as
\begin{equation}
\overline{\tau} = \int_{t^{\rm last}}^{t^{\rm last}+\tau} \Lambda(\tau')\, \text{d}\tau' .
  \label{eq:L(t;t0)}
\end{equation}
The normalized waiting time follows an exponential distribution with an expected value of one. Therefore, it is easy to generate it using inverse sampling, i.e., by $\overline{\tau} = - \ln\,u$, where $u$ is a uniform random variate on $(0,1]$.
The (wall-clock) waiting time $\tau$ is found as the solution to Eq.~\eqref{eq:L(t;t0)} given the $\overline{\tau}$ value that we have generated. 

In fact, all we have done for the moment is to exchange one implicit equation (Eq.~\eqref{eq:S-TGA}) for another (Eq.~\eqref{eq:L(t;t0)}). However, Eq.~\eqref{eq:L(t;t0)} is linear in $\Lambda$, which makes it easier to solve and approximate numerically.
This is in particular the case because we assumed that the temporal network changes only in discrete points in time, as schematically shown in Fig.~\ref{fig:switching}.
We let $t^{\rm net}_0 = 0$, and we denote by $t_1^{\rm net}, t_2^{\rm net}, \ldots$ the subsequent time points at which the temporal network changes. Then, $[t_{n-1}^{\rm net}, t_n^{\rm net})$ is the $n$th interval between network changes. 
Since the temporal network only changes in discrete steps, $\Lambda(t)$ is piecewise constant. 
Therefore, one can solve Eq.~\eqref{eq:L(t;t0)} iteratively as follows:
Suppose that we are given the time of the last event, $t^{\rm last}$, and we want to find the time of the next event denoted by $t^{\rm next} = t^{\rm last} + \tau$. If there is no event yet, and we want to find the time of the first event, we regard that $t^{\rm last} = 0 = t_0^{\rm net}$.
We start from the time interval $[t^{\rm net}_{n^*-1}, t^{\rm net}_{n^*})$ between to successive switches of the network in which the last event took place, i.e., the interval which satisfies $t^{\rm net}_{n^*-1} \le t^{\rm last} < t^{\rm net}_{n^*}$. 
We then sequentially check for each time interval $[t^{\rm last}, t^{\rm net}_{n^*})$, $[t^{\rm net}_{n^*}, t^{\rm net}_{n^*+1})$, $[t^{\rm net}_{n^*+1}, t^{\rm net}_{n^*+2})$, $\ldots$, to determine in which interval $t^{\rm next}$ falls.
In practice, we compare at each step the generated value of $\overline{\tau}$ to the value of the integral $\int_{t^{\rm last}}^{t^{\rm net}_{n}} \Lambda(t)\, \text{d}t$. 
The latter is efficiently calculated as the sum
\begin{equation}
    \int_{t^{\rm last}}^{t^{\rm net}_{n}} \Lambda(t)\, \text{d}t = (t^{\rm net}_{n^*}-t^{\rm last})\Lambda_{n^*}  + \sum_{\ell=n^*+1}^{n} \Delta_{\ell}\Lambda_{\ell} ,
  \label{eq:Ln}
\end{equation}
where $\Delta_{\ell} = t^{\rm net}_{\ell}  - t^{\rm net}_{\ell-1}$ is the length of the $\ell$th interval between successive changes of the network, and $\Lambda_{\ell}$ is the value of $\Lambda(t)$ in this interval.
If $n = n^*$, the sum in the second term on the right-hand side of Eq.~\eqref{eq:Ln} is the empty sum, i.e., it has no summands and thus evaluates to zero, and the equation reduces to $\int_{t^{\rm last}}^{t^{\rm net}_{n^*}} \Lambda(t)\, \text{d}t = (t^{\rm net}_{n^*}-t^{\rm last})\Lambda_{n^*}$.

The smallest value of $n$ that satisfies $\int_{t^{\rm last}}^{t^{\rm net}_{n}} \Lambda(t)\, \text{d}t > \overline{\tau}$ determines the time interval in which the next event takes place. With that $n$ value, the precise time of the next event is given by 
\begin{equation}
t^{\rm next} = t_{n-1} + \frac{\overline{\tau} - \int_{t^{\rm last}}^{t^{\rm net}_{n-1}} \Lambda(t)\, \text{d}t}
{\Lambda_n} .
\end{equation}

Finally, we draw the Poisson process $i$ that produces the event at time $t^{\rm next}$ with probability \begin{equation}
  \Pi_i(t^{\rm next}) = \frac{\lambda_i(t^{\rm next})} {\Lambda(t^{\rm next})} .
  \label{eq:pi_m(t)}
\end{equation}

The steps of the temporal Gillespie algorithm are described in Box~\ref{box:TGA}.
It works by iterating over the list of times at which the network changes. Within each interval between the network's switches, it compares the normalized waiting time, $\overline{\tau}$, to the total instantaneous rate integrated over the time-interval, $\Lambda_n\Delta_n$ (see Step 2).
If $\overline{\tau}$ is larger than or equal to $\Lambda_n\Delta_n$, then nothing happens, and one subtracts $\Lambda_n\Delta_n$ from $\overline{\tau}$ and advances to the next interval, ${n+1}$ (see Step~2(a)).
Alternatively, if $\overline{\tau}$ is smaller than $\Lambda_n\Delta_n$, then an event occurs within the $n$th time window $[t^{\rm net}_{n-1}, t^{\rm net}_n)$ (see Step~2(b)). 
Then, the algorithm determines the timing of the event and selects the reaction channel to produce the event using any of the appropriate selection methods discussed earlier (see Sections~\ref{sub:direct}, \ref{sub:binary_tree}, and \ref{sub:composition-rejection}). 
It then updates the system, draws a new normalized waiting time, and repeats the procedure. 

\begin{figure}[t]
\begin{boxedtext}{}
\captionof{floatbox}{\textbf{Temporal Gillespie algorithm.}}
\label{box:TGA}

\begin{enumerate}[start = 0]
  \item Initialization:
  \begin{enumerate}
      \item Define the initial state of the system, and set $t =0$.
      \item Set $n = 1$ and $\Delta = t^{\rm net}_1 - t^{\rm net}_0$. 
      \item Initialize the rates $\lambda_j$  for all $j=1, \ldots, M$.
      \item Calculate the total rate $\Lambda = \sum_{j=1}^M \lambda_j$.
     \end{enumerate}
  \item Draw a normalized waiting time $\overline{\tau} = - \ln u$, where $u$ is a uniform random variate on $(0,1]$.
  \item Compare $\Lambda\Delta$ to $\overline{\tau}$:
    \begin{enumerate}
      \item If $\Lambda\Delta \leq \overline{\tau}$, then no reaction takes place in the $n$th time window.
        \begin{enumerate}
          \item Set $\overline{\tau} \to \overline{\tau} - \Lambda\Delta $. 
          \item Advance to the next time window by setting $t \to t^{\rm net}_{n}$ and $\Delta \to t^{\rm net}_{n+1} - t^{\rm net}_n$;
          update $n \to n+1$. 
          \item Update all $\lambda_j$ affected by changes in the temporal network, and update $\Lambda$ accordingly. 
          \item Return to Step~2.
        \end{enumerate}
      \item If $\Lambda\Delta > \overline{\tau}$, then an event takes place at time $t^{\rm next} = t + \overline{\tau}/\Lambda$. 
        \begin{enumerate}
          \item Select the reaction channel $i$ that produces the event with probability $\Pi_{i} = \lambda_{i}/\Lambda$.
          \item Update the time as $t \to t^{\rm next}$. Also update the remaining length of the present time window as $\Delta \to \Delta - \overline{\tau}/\Lambda$. 
          \item Update the rates $\lambda_j$ that are affected by the event, and update $\Lambda$ accordingly.
          \item Return to Step~1.
        \end{enumerate}
    \end{enumerate}
\end{enumerate}

\end{boxedtext}
\end{figure}

\cite{Vestergaard2015PlosComputBiol} furthermore proposed to adapt the temporal Gillespie algorithm to simulate non-Markovian processes in temporal networks.
To make the algorithm computationally efficient, they proposed two approximations to solve Eq.~\eqref{eq:L(t;t0)} by simply iterating over the times $t^{\rm net}_n$ at which the network changes, as we did for nonhomogeneous Poisson processes.
These approximations avoid having to use numerical integration to solve the implicit equation, which would make the algorithm slow for large systems.

The first approximation is to regard the total instantaneous rate $\Lambda(t,\{\tilde{t}_j\})$, which in the non-Markovian case can depend on the times since the last events for all $M$ processes, as constant during each interval $[t^{\rm net}_{n-1}, t^{\rm net}_{n})$ between the consecutive changes in the network. 
This approximation is accurate when the network changes much faster than the total rate $\Lambda(t,\{\tilde{t}_j\})$ does, i.e., when 
\begin{equation}
  \frac{ \Lambda(t^{\rm net}_{n+1},\{\tilde{t}_j\}) - \Lambda(t^{\rm net}_n,\{\tilde{t}_j\}) } { \Lambda(t^{\rm net}_n,\{\tilde{t}_j\}) } \ll 1 ,
\end{equation}
where $\Lambda(t^{\rm net}_{n+1},\{\tilde{t}_j\}) - \Lambda(t^{\rm net}_n,\{\tilde{t}_j\})$ is the change of $\Lambda(t,\{\tilde{t}_j\})$ between two successive intervals.
When simulating spreading processes in temporal networks, the network dynamics is often much faster than the spreading dynamics in practice. 
For example, the time scale of recordings of physical proximity networks is typically of the order of seconds to minutes while the infection and recovery of flu-like diseases occur in the order of hours to days~\citep{Vestergaard2015PlosComputBiol}.
With this first approximation, one can directly apply Eq.~\eqref{eq:Ln} and use the same procedure as for the Poissonian case. 

The second approximation is to use a first-order cumulant expansion, similar to the non-Markovian Gillespie algorithm (Section~\ref{sub:nMGA}), in addition to the first approximation. 
It amounts to assuming that each $\lambda_i$ is constant as long as no event takes place and no change of the network that directly affects the $\lambda_i$ value takes place. 
One thus avoids having to update all the $\lambda_i$ values each time we go to the next time window (i.e., from $[t^{\rm net}_{n-1}, t^{\rm net}_n)$ to $[t^{\rm net}_n, t^{\rm net}_{n+1})$), and the algorithm runs much faster.
To increase the accuracy of the algorithm when the number of reaction channels $M$ is small and the cumulant expansion is not accurate (e.g., at the start or near the end of an SIR process where only a few nodes are typically infectious), they proposed a heuristic approach, in which one updates $\lambda_i$ only if the time elapsed since the last update of $\lambda_i$ exceeds a given threshold $\delta$.
They proposed to choose the value of $\delta$ as a given fraction of the expected waiting time of a single reaction channel. 
Therefore, when $M$ is large, the waiting time between events will almost never exceed $\delta$, and the algorithm will be similar to the non-Markovian Gillespie algorithm.
When $M$ is small, the algorithm updates the $\lambda_i$ more frequently, making it more accurate at an added computational cost.

With the above approximations, the application of the temporal Gillespie algorithm to general non-Markovian processes only slightly changes the implementation from that for the nonhomogeneous Poisson processes described in Box~\ref{box:TGA}. We refer interested readers to~\citet{Vestergaard2015PlosComputBiol} for details.

\subsection{Event-driven simulation of the SIR process\label{sub:event-data-driven_simulation}}

Holme proposed another efficient event-based algorithm, related to the first reaction method, when the time-stamped contact events are given as data \citep{Holme2021PlosOne}. 
Although the timing of the event is no longer stochastic, the overall dynamics is still stochastic. This is because, in the SIR model for example, infection upon each contact event occurs with a certain probability and recovery occurs as a Poisson process with rate $\mu$. The efficiency of the Holme's algorithm comes from multiple factors. Suppose that the $i$th node is infected and its neighboring node, $j$, is susceptible. First, the algorithm tactically avoids searching all the contact events between $i$ and $j$ when determining the event with which $i$ successfully infects $j$. Second, it uses the binary heap to maintain a carefully limited set of times of the events with which infection may occur between pairs of nodes. The corresponding code for simulating the SIR model, implemented in C with a Python wrapper, is available on Github \citep{Holme2021PlosOne}.

\section{Conclusions}

The aim of this article has been two-fold: to provide a tutorial of the standard Gillespie algorithms and to review recent Gillespie algorithms that improve upon their computational efficiency and extend their scope. 
While our emphasis and examples lean towards social multi-agent dynamics in populations and networks, the applicability of the Gillespie algorithms and their variants is extensive. We believe that the present article is useful for students and researchers in various fields, such as epidemiology, ecology, control theory, artificial life, complexity sciences, and so on.

In fact, many models of adaptive networks, where the network change is induced by the change of the status of e.g. nodes, have been mostly described by ODEs and assume that the interaction strength between pairs of nodes vary in response to changes in individuals' behavior \citep{GrossBlasius2008JRSocInterface,GrossSayama2009book,Wang2015PhysLifeRev}. If such changes occur in an event-driven manner, Gillespie algorithms are readily applicable. How to deploy and develop Gillespie algorithms and their variants to adaptive network scenarios is a practical concern.

We briefly discussed simulations on empirical time-stamped contact event data (Sections~\ref{sub:TGA} and \ref{sub:event-data-driven_simulation}). In this setting, it is the given data that determines the times and edges (i.e., node pairs) of the events, which is contrary to the assumption of the Gillespie algorithms that jump process models generate events. Despite the increasing demand of simulations on the given time-stamped contact event data, this is still an underexplored area of research. 
\cite{Vestergaard2015PlosComputBiol} and \cite{Holme2021PlosOne} showed that ideas and techniques from the Gillespie algorithms are useful for such simulations although the developed algorithms are distinct from the historical Gillespie algorithms. This is another interesting area of future research.

\appendix  



\section*{Acknowledgments}

We thank the SocioPatterns collaboration (see \url{http://www.sociopatterns.org}) for providing the experimental data set. 
N.M.\ thanks the financial support by AFOSR European Office (under Grant No. FA9550-19-1-7024), National Science Foundation (under Grant No. DMS-2052720), the Nakatani Foundation, the Sumitomo Foundation, and the Japan Science and Technology Agency (under Grant No. JPMJMS2021).
C.L.V.\ was supported in part by the the French government under management of Agence Nationale de la Recherche as part of the ``Investissements d’avenir" program, reference ANR-19-P3IA-0001 (PRAIRIE 3IA Institute).

\section{\label{app:exp_identity}Exponential identity}

In this appendix, we prove the identity
\begin{equation}
    \lim_{x\to 0} (1 + x)^{1/x} = e .
    \label{eq:exp_identity}
\end{equation}
Because $e^x$ is continuous in $x$, we obtain
\begin{equation}
    \lim_{x\to 0} (1 + x)^{1/x} = e^{\lim_{x\to0} \ln(1+x)/x} .
\end{equation}
Thus, we can prove Eq.~\eqref{eq:exp_identity} by showing that $\lim_{x\to0} \ln(1+x)/x = 1$. 
We do this using l'H\^opital's rule as follows: 
\begin{equation}
     \lim_{x\to0} \frac{\ln(1+x)}{x} = \frac{\lim_{x\to0}\frac{1}{1+x}}{\lim_{x\to0}1} = 1 .
\end{equation}

\bibliography{citations}

\end{document}